\documentclass[10pt]{article}
\usepackage{colordvi}
\usepackage{epsfig}
\usepackage{axodraw}
\usepackage{epsfig}
\usepackage{graphicx}
\usepackage{rotate}
\usepackage{latexsym}
\usepackage{amssymb}
\usepackage{amsmath}
\newcommand\pubnumber{}
\newcommand\pubdate{\today}
\newcommand\hepnumber{hep-ph/0612124}

\def\csuma{Deutsches Electronen - Synchrotron, DESY, Platanenallee 6, 15738
Zeuthen, Germany}
\def\csumb{Dipartimento di Fisica Teorica, Universit\`a di Torino, Italy\\
INFN, Sezione di Torino, Italy}
\def\support{\footnote{Work supported by MIUR under contract
2001023713$\_$006 and by  the European Community's Marie-Curie Research 
Training Network under contract MRTN-CT-2006-035505
`Tools and Precision Calculations for Physics Discoveries at Colliders'.}}
\def\Title#1{\begin{center} {\Large\bf #1 } \end{center}}

\def\Author#1{\begin{center}{ \sc #1} \end{center}}
\def\Address#1{\begin{center}{ \it #1} \end{center}}

\newcommand\pubblock{\rightline{\begin{tabular}{l} \pubnumber\\
         \pubdate\\ \hepnumber \\  DESY 06-225 \\  SFB/CPP-06-56\end{tabular}}}
\newenvironment{Abstract}{\begin{quotation}  }{\end{quotation}}

\def\Acknowledgments{\bigskip  \bigskip \begin{center}
          \large\bf Acknowledgments\end{center}}
\def\email#1{\footnote{#1}}
\makeatletter
\def\section{\@startsection{section}{0}{\z@}{5.5ex plus .5ex minus
 1.5ex}{2.3ex plus .2ex}{\large\bf}}
\def\subsection{\@startsection{subsection}{1}{\z@}{3.5ex plus .5ex minus
 1.5ex}{1.3ex plus .2ex}{\normalsize\bf}}
\def\subsubsection{\@startsection{subsubsection}{2}{\z@}{-3.5ex plus
-1ex minus  -.2ex}{2.3ex plus .2ex}{\normalsize\sl}}

\renewcommand{\@makecaption}[2]{%
   \vskip 10pt
   \setbox\@tempboxa\hbox{\small #1: #2}
   \ifdim \wd\@tempboxa >\hsize     
       \small #1: #2\par          
     \else                        
       \hbox to\hsize{\hfil\box\@tempboxa\hfil}
   \fi}
 \def\citenum#1{{\def\@cite##1##2{##1}\cite{#1}}}
\def\citea#1{\@cite{#1}{}}
\newcount\@tempcntc
\def\@citex[#1]#2{\if@filesw\immediate\write\@auxout{\string\citation{#2}}\fi
  \@tempcnta\z@\@tempcntb\m@ne\def\@citea{}\@cite{\@for\@citeb:=#2\do
    {\@ifundefined
       {b@\@citeb}{\@citeo\@tempcntb\m@ne\@citea\def\@citea{,}{\bf }\@warning
       {Citation `\@citeb' on page \thepage \space undefined}}%
    {\setbox\z@\hbox{\global\@tempcntc0\csname b@\@citeb\endcsname\relax}%
     \ifnum\@tempcntc=\z@ \@citeo\@tempcntb\m@ne
       \@citea\def\@citea{,}\hbox{\csname b@\@citeb\endcsname}%
     \else
      \advance\@tempcntb\@ne
      \ifnum\@tempcntb=\@tempcntc
      \else\advance\@tempcntb\m@ne\@citeo
      \@tempcnta\@tempcntc\@tempcntb\@tempcntc\fi\fi}}\@citeo}{#1}}
\def\@citeo{\ifnum\@tempcnta>\@tempcntb\else\@citea\def\@citea{,}%
  \ifnum\@tempcnta=\@tempcntb\the\@tempcnta\else
  {\advance\@tempcnta\@ne\ifnum\@tempcnta=\@tempcntb \else\def\@citea{--}\fi
    \advance\@tempcnta\m@ne\the\@tempcnta\@citea\the\@tempcntb}\fi\fi}
\makeatother
\newcommand{\nl}{\nonumber\\}

\newcommand{\lpar}{\left(}                            
\newcommand{\rpar}{\right)}

\newcommand{\bq}{\begin{equation}}                    
\newcommand{\eq}{\end{equation}}
\newcommand{\bqa}{\arraycolsep 0.14em\begin{eqnarray}}
\newcommand{\eqa}{\end{eqnarray}}
\newcommand{\ba}[1]{\begin{array}{#1}}
\newcommand{\ea}{\end{array}}
\newcommand{\ben}{\begin{enumerate}}
\newcommand{\een}{\end{enumerate}}
\newcommand{\bei}{\begin{itemize}}
\newcommand{\eei}{\end{itemize}}
\newcommand{\eqn}[1]{Eq.(\ref{#1})}
\newcommand{\eqns}[2]{Eqs.(\ref{#1})--(\ref{#2})}

\newcommand{\eqnsc}[2]{Eqs.(\ref{#1}) and (\ref{#2})}

\newcommand{\tabn}[1]{Tab.~\ref{#1}}

\newcommand{\tabns}[2]{Tabs.~\ref{#1}--\ref{#2}}

\newcommand{\fig}[1]{Fig.~\ref{#1}}
\newcommand{\figs}[2]{Figs.~\ref{#1}--\ref{#2}}
\newcommand{\sect}[1]{Section~\ref{#1}}

\newcommand{\subsect}[1]{Subsection~\ref{#1}}

\newcommand{\GeV}{\mathrm{GeV}}

\def\Re{\mathop{\operator@font Re}\nolimits}
\def\Im{\mathop{\operator@font Im}\nolimits}
\newcommand{\ord}[1]{{\cal O}\lpar#1\rpar}

\newcommand{\wb}{W}

\newcommand{\zb}{Z}

\newcommand{\hb}{H}

\newcommand{\fbpsi}{{\overline{\psi}}}

\newcommand{\fbe}{{\overline{e}}}
\newcommand{\fbu}{{\overline{u}}}
\newcommand{\fbv}{{\overline{v}}}

\newcommand{\fbnu}{{\overline{\nu}}}

\newcommand{\mV}{M_{\ssV}}
\newcommand{\mw}{M_{_W}}

\newcommand{\mz}{M_{_Z}}

\newcommand{\mh}{M_{_H}}

\newcommand{\mt}{m_t}

\newcommand{\mws}{M^2_{_W}}

\newcommand{\mzs}{M^2_{_Z}}

\newcommand{\mhs}{M^2_{_H}}

\newcommand{\shat}{{\hat s}}

\newcommand{\gf}{G_{\ssF}}

\newcommand{\stw}{s_{\theta}}             
\newcommand{\ctw}{c_{\theta}}
\newcommand{\stws}{s_{\theta}^2}

\newcommand{\ctws}{c_{\theta}^2}

\newcommand{\shs}{{\hat s}^2}
\newcommand{\chs}{{\hat c}^2}

\newcommand{\sla}[1]{/\!\!\!#1}

\newcommand{\gfd}{\gamma_5}                    
\newcommand{\gap}{\lpar 1+\gamma_5\rpar}

\newcommand{\gadu}[1]{\gamma_{#1}}

\newcommand{\intfx}[1]{\int_{\scriptstyle 0}^{\scriptstyle 1}\,d#1}

\newcommand{\sgh}{{\hat\Sigma}}

\newcommand{\MSB}{\overline{MS}}

\newcommand{\ep}{\epsilon}

\newcommand{\Reb}{{\rm{Re}}}
\newcommand{\Imb}{{\rm{Im}}}
\newcommand{\upar}[1]{u}

\newcommand{\ssA}{{\scriptscriptstyle{A}}}
\newcommand{\ssB}{{\scriptscriptstyle{B}}}

\newcommand{\ssD}{{\scriptscriptstyle{D}}}
\newcommand{\ssE}{{\scriptscriptstyle{E}}}
\newcommand{\ssF}{{\scriptscriptstyle{F}}}
\newcommand{\ssG}{{\scriptscriptstyle{G}}}
\newcommand{\ssH}{{\scriptscriptstyle{H}}}
\newcommand{\ssI}{{\scriptscriptstyle{I}}}

\newcommand{\ssL}{{\scriptscriptstyle{L}}}
\newcommand{\ssM}{{\scriptscriptstyle{M}}}

\newcommand{\ssO}{{\scriptscriptstyle{O}}}
\newcommand{\ssP}{{\scriptscriptstyle{P}}}
\newcommand{\ssQ}{{\scriptscriptstyle{Q}}}
\newcommand{\ssR}{{\scriptscriptstyle{R}}}
\newcommand{\ssS}{{\scriptscriptstyle{S}}}
\newcommand{\ssT}{{\scriptscriptstyle{T}}}
\newcommand{\ssU}{{\scriptscriptstyle{U}}}
\newcommand{\ssV}{{\scriptscriptstyle{V}}}
\newcommand{\ssW}{{\scriptscriptstyle{W}}}

\newcommand{\ssZ}{{\scriptscriptstyle{Z}}}

\newcommand{\QED}{\rm{\scriptscriptstyle{QED}}}

\newcommand{\bqas}{\begin{eqnarray*}}
\newcommand{\eqas}{\end{eqnarray*}}

\def\app#1#2 {{\it Acta. Phys. Pol.} {\bf#1},#2}
\def\cpc#1#2 {{\it Computer Phys. Comm.} {\bf#1},#2}
\def\np#1#2 {{\it Nucl. Phys.} {\bf#1},#2}
\def\pl#1#2 {{\it Phys. Lett.} {\bf#1},#2}
\def\prep#1#2 {{\it Phys. Rep.} {\bf#1},#2}
\def\prev#1#2 {{\it Phys. Rev.} {\bf#1},#2}
\def\prl#1#2 {{\it Phys. Rev. Lett.} {\bf#1},#2}
\def\zp#1#2 {{\it Zeit. Phys.} {\bf#1},#2}
\def\sptp#1#2 {{\it Suppl. Prog. Theor. Phys.} {\bf#1},#2}
\def\mpl#1#2 {{\it Modern Phys. Lett.} {\bf#1},#2}
\def\jetp#1#2 {{\it Sov. Phys. JETP} {\bf#1},#2}
\def\fpj#1#2 {{\it Fortschr. Phys.} {\bf#1},#2}
\def\afp#1#2 {{\it Acta.Phys. Polon.} {\bf#1},#2}
\def\err#1#2 {{\it Erratum} {\bf#1},#2}
\def\ijmp#1#2 {{\it Int. J. Mod. Phys} {\bf#1},#2}
\def\nc#1#2 {{\it Nuovo Cimento} {\bf#1},#2}
\def\ap#1#2 {{\it Ann. Phys.} {\bf#1},#2}
\def\cmp#1#2 {{\it Comm. Math. Phys.} {\bf#1},#2}
\def\el#1#2 {{\it Europhys. Lett.} {\bf#1},#2}
\def\hpa#1#2 {{\it Helv. Phys. Acta} {\bf#1},#2}
\def\yf#1#2 {{\it Yad. Fiz.} {\bf#1},#2}
\def\nim#1#2 {{\it Nucl. Instrum. Meth.} {\bf#1},#2}
\def\spz#1#2 {{\it Sov. Pisma Zhetf} {\bf#1},#2}
\def\jetpl#1#2 {{\it JETP Lett.} {\bf#1},#2}
\def\sjnp#1#2 {{\it Sov. J. Nucl. Phys.} {\bf#1},#2}
\def\ptp#1#2 {{\it Progr. Theor. Phys. (Kyoto)} {\bf#1},#2}
\def\rmp#1#2  {{\it Rev. Mod. Phys.} {\bf#1},#2}
\def\zhetf#1#2 {{\it ZhETF} {\bf#1},#2}
\def\prs#1#2 {{\it Proc. Roy. Soc.} {\bf#1},#2}
\def\phys#1#2 {{\it Physica} {\bf#1},#2}

\def\bfi{\begin{figure}}
\def\efi{\end{figure}}

\newcommand{\hdel}{{\hat\delta}}
\newcommand{\bdel}{{\bar\delta}}
\newcommand{\hgam}{{\hat\gamma}}
\newcommand{\bgam}{{\bar\gamma}}
\newcommand{\hmu}{{\hat{\mu}}}
\newcommand{\hnu}{{\hat{\nu}}}

\newcommand{\htp}{{\hat{p}}}
\newcommand{\brp}{{\bar{p}}}

\newcommand{\LB}{{\cal L}oop{\cal B}ack}
\newcommand{\GS}{{\cal G}raph{\cal S}hot}

\newcommand{\bmid}{\Bigr|}

\newcommand{\DUV}{{\Delta}_{\ssU\ssV}}
\newcommand{\ghat}{{\hat g}}
\newcommand{\Mhat}{{\hat M}}
\newcommand{\shq}{{\hat s}^4}
\newcommand{\shsix}{{\hat s}^6}
\newcommand{\cpw}{s_{\ssW}}
\newcommand{\cpz}{s_{\ssZ}}
\newcommand{\cpv}{s_{\ssV}}
\newcommand{\ext}{\,;\,{\rm ext}}
\newcommand{\rpw}{\mu^2_{\ssW}}
\newcommand{\rpz}{\mu^2_{\ssZ}}

\newcommand{\cph}{s_{\ssH}}
\newcommand{\Cph}{S_{\ssH}}
\newcommand{\rph}{\mu^2_{\ssH}}

\newcommand{\xWs}{x^2_{\ssW}}

\newcommand{\xLs}{x^2_{\ssL}}
\newcommand{\xBs}{x^2_{\ssB}}
\newcommand{\xTs}{x^2_{\ssT}}

\newcommand{\xW}{x_{\ssW}}
\newcommand{\xH}{x_{\ssH}}
\newcommand{\xL}{x_{\ssL}}
\newcommand{\xB}{x_{\ssB}}
\newcommand{\xT}{x_{\ssT}}

\newcommand{\cL}{{\cal L}}
\newcommand{\cM}{{\cal M}}

\newcommand{\WEAK}{\rm{\scriptscriptstyle{WEAK}}}
\newcommand{\FT}{\rm{\scriptscriptstyle{FT}}}
\newcommand{\REST}{\rm{\scriptscriptstyle{REST}}}
\newcommand{\BOX}{\rm{\scriptscriptstyle{BOX}}}

\newcommand{\ssQQ}{{\scriptscriptstyle{Q}\scriptscriptstyle{Q}}}
\newcommand{\ssVV}{{\scriptscriptstyle{V}\scriptscriptstyle{V}}}
\newcommand{\ssWW}{{\scriptscriptstyle{W}\scriptscriptstyle{W}}}
\newcommand{\ssHH}{{\scriptscriptstyle{H}\scriptscriptstyle{H}}}

\newcommand{\OS}{{\scriptscriptstyle{OS}}}
\newcommand{\oSigma}{{\overline \Sigma}}
\newcommand{\oF}{{\overline F}}
\newcommand{\tF}{{\tilde F}}
\newcommand{\oD}{{\overline \Delta}}

\begin{document}
\begin{titlepage}
\pubblock
\vfill
\def\thefootnote{\fnsymbol{footnote}}
\Title{Two-Loop Renormalization in the Standard Model\\[5mm]
Part III: Renormalization Equations and their Solutions\support}
\vfill
\Author{Stefano Actis\email{Stefano.Actis@desy.de}}
\Address{\csuma}
\Author{Giampiero Passarino\email{giampiero@to.infn.it}}
\Address{\csumb}
\vfill
\begin{Abstract}
\noindent 
In part I and II of this series of papers all elements have been introduced 
to extend, to two loops, the set of renormalization procedures which are 
needed in describing the properties of a spontaneously broken gauge theory. 
In this paper, the final step is undertaken and finite renormalization is 
discussed. Two-loop renormalization equations are introduced and their 
solutions discussed within the context of the minimal standard model of 
fundamental interactions. These equations relate renormalized Lagrangian 
parameters (couplings and masses) to some input parameter set containing 
physical (pseudo-)observables. Complex poles for unstable gauge and Higgs 
bosons are used and a consistent setup is constructed for extending the 
predictivity of the theory from the Lep1 $Z$-boson scale (or the Lep2 $WW$ 
scale) to regions of interest for LHC and ILC physics.
\end{Abstract}
\vfill
\begin{center}
\end{center}
\end{titlepage}
\def\thefootnote{\arabic{footnote}}
\setcounter{footnote}{0}
\small
\thispagestyle{empty}
\tableofcontents
\normalsize
\clearpage
\setcounter{page}{1}
\section{Introduction \label{intro}}
The end of the Lep period represented a moment of transition in the 
development of techniques designed for producing high precision results for
collider physics.

It is a well known fact that the advent of a new hadronic machine, the LHC in
this case, gives a privileged role to QCD but what is the real implication
of this fact?

QCD is a theory devoted to studying strong interactions by means of
perturbative methods, a road which has been made possible by the discovery
of asymptotic freedom. It is a theory with very few scales and therefore its 
perturbative aspects are technically simpler.
For this reason most of the new ideas are, first of all, tested in QCD
where we have an extensive literature of explicit results up to four-loop
Feynman diagrams. The very recent twistor spinoff for collider 
physics~\cite{Dixon:2005cf} has its immediate target in the realm of QCD
going beyond the traditional field-theoretic point of view.

Generally speaking we are now witnessing the development of three parallel
roads: techniques (mostly Monte Carlo driven~\cite{Mangano:2002ea} or twistor 
inspired~\cite{Cachazo:2004kj}) to deal with tree level processes with many 
particles in the final state; complete one-loop calculations for $2 \to 4$ 
processes with a particular emphasis on the proper treatment of intermediate 
unstable particles~\cite{Denner:2005fg} and genuine electroweak two-loop 
calculations of physical observables (\cite{Aglietti:2006yd}
and~\cite{Hollik:2005ns}), including supersymmetric 
effects~\cite{Haestier:2005tx}.
The first item in the list is deeply linked to the LHC physic programme and has
already collected a sizable number of important results. Extension of one-loop
calculations to four particles in the final state represents a bridge between
LHC and LC physics while two-loop electroweak physics is, at the moment, in
some early stage of development. It is worth mentioning, however, that past
history has told us about the importance of a complete calculation when
dealing with the final analysis of the experimental data.

Lep1 physics has ben mostly dealing with $2 \to 2$ processes, where one has
been able to assemble the most complete set of predictions in the whole
history of radiative corrections. Most notable is the fact that the technology
needed for the operation has been pushed well beyond what was expected at the 
beginning of the period, very much as in the case of the experimental 
analysis. The key ingredients have been

\begin{itemize}

\item[--] the construction of a complete one-loop renormalization procedure 
for the standard model~\cite{Bardin:1997xq}, with the inclusion of several 
leading and next-to-leading higher order effects;

\item[--] the development of libraries for assembling the relevant one-loop
diagrams, including reduction of tensor integrals to scalar ones and 
analytical evaluation of the latter~\cite{Montagna:1998kp};

\item[--] inclusion of higher-order, real and virtual, QED 
effects~\cite{Bardin:1999ak};

\item[--] development of fitting procedures to deal with pseudo and realistic
experimental observables.

\end{itemize}

At the very end of the Lep period it became evident that these procedures
should be generalized if we want to have a full two-loop interpretation of 
the data. Furthermore, it became obvious that a step forward is necessary for 
treating the technological elements which are mandatory in obtaining a 
satisfactory solution to all the items of our wish-list. There are 
several aspects which deserve a specific comment: the main one can be 
summarized by saying that assembling a two-loop package is not, anymore, a 
one-man-show as it was in the past but requires instead a dedicated 
involvement of a well coordinated group. 

There are many new aspects in the project that put an unprecedented challenge: 
for any multi-scale process we have to abandon the fully analytical way and 
new, numerical, algorithms must be developed to handle the complexity of
Feynman diagram evaluation with cutting-edge calculations. Generally speaking 
we are referring here to the evaluation of master integrals, a minimal set 
of diagrams to which all other diagrams are reduced. The techniques for 
reduction should be general enough to handle three or more external legs in 
a fully massive world.
On top of that one should not forget to develop a comprehensive
two-loop renormalization procedure, flexible enough to be implemented in
practice.

What we are describing here is a multi-step program and only one of the steps 
will be presented in this paper; to fully understand the results, however, it 
is important to mention that everything has been developed in parallel
and what we are going to illustrate is already fitting in the general
layout of our program. The general strategy for handling multi-loop, multi-leg
Feynman diagrams was designed in~\cite{Passarino:2001wv} and the collection of 
results necessary for evaluating two-loop, two-point integrals can be found
in ref.~\cite{Passarino:2001jd}.
Next the calculation of two-loop three-point scalar integrals: 
infrared-convergent configurations are discussed 
in~\cite{Ferroglia:2003yj} and infrared- and collinear-divergent ones are 
analyzed in~\cite{Passarino:2006gv}.
Finally, our method for dealing with two-loop tensor integrals can be found 
in~\cite{Actis:2004bp} and results for one-loop multi-leg integrals are 
shown in~\cite{Ferroglia:2002mz}.

First of all, we needed a technique for reducing diagrams to some set of master
integrals. Here, we are referring to a class of two-loop calculation with
few external legs, avoiding the exponential complexity that would be
encountered with many legs.
This step requires our ability in treating all relevant limits, when
masses are negligible or when external momenta must be set to zero.
We should mention that the real world is massive, with all masses different,
but most of them negligibly small. To mention an important aspect we may say
that any numerical treatment that does not extract collinear and Sudakov
logarithms from the very beginning is doomed to failure.
Furthermore, all the results must be stored in the most appropriate form,
since our final numerical integration must be absolutely stable.

Next, a full two-loop renormalization had to be designed and we have to
make clear the meaning of this sentence. Diagrams have to be generated with
some automatic procedure; in our project, we did not want to rely on any black 
box, so we have constructed our own set of procedures, creating the $\GS$ 
package~\cite{GraphShot}.

After assembling diagrams we wanted to perform all sort of canonical tests
which, essentially, amounts to check all possible unrenormalized WST 
identities. After checking that everything has been properly assembled
we have been moving to the next logical step, removing all ultraviolet
infinities. At the very end one has to admit that the removal is a simple
business but we pretend having done it in the most rigorous way. To mention an 
example we quote the problem of overlapping divergencies; non-local residues
of ultraviolet poles must cancel in the total if unitarity of the theory has 
to be preserved: it is highly non trivial but they indeed do cancel. After that
one has to check renormalized WST identities.

Removing ultraviolet poles means trading bare parameters for renormalized ones 
and, after that, renormalization equations must be written and solved which is 
equivalent to express theoretical predictions in terms of an input parameter 
set. In this paper, we illustrate the role of the running e.m. coupling 
constant and give a full account of the calculation of the Fermi coupling 
constant. 

Another, non trivial, aspect of our work is related to proving that what we 
expect to vanish is indeed zero; for instance one can prove that the standard 
model can be made, to some extent, QED-like: at the level of S-matrix elements 
we expect that vertices, with the inclusion of wave-function factors, do not 
contribute to the renormalization of the electric charge. To prove this 
property at two loops is far from trivial.

It is by now clear that one cannot produce all the results in a two-loop
calculation without
having a software package, $\GS$~\cite{GraphShot} in our case; $\GS$ collects 
all the relevant algorithms. At this stage the package is far from being {\em
user friendly} (for instance we even miss a user's guide) and it is not
even clear if somebody else will use it but writing the code has been
an essential step without which none of the results presented in this paper
would have been achieved. 
Once more, past experience is telling us that with the present level of 
complexity there will be no time, in a few years, to go back and to allow for
extensions which were not foreseen from the very beginning.

To summarize, we have performed several steps towards complete two-loop 
renormalization in the standard model, steps that could be easily generalized, 
with few minor adjustments, to an arbitrary renormalizable quantum field 
theory. The introductory elements have been given in~\cite{IofIII} (hereafter 
I) and in~\cite{IIofIII} (hereafter II).

As it is well known, finite $S$-matrix elements can be obtained without the
explicit use of counterterms, a fact that has been fully described in the
one-loop renormalization~\cite{Bardin:1999ak}. 
However, most of the people are more familiar with the language of counterterms
and, for this reason, we have decided to adapt our approach. In part II we 
have described the strategy for making the one- and two-loop Green functions 
of the theory ultraviolet finite; this amounts to rewrite the Lagrangian in 
terms of renormalized quantities.

The next step in our procedure will be to express renormalized parameters in 
terms of physical observables belonging to some input parameter set 
(hereafter, IPS).
In any one-loop calculation one tries to improve upon the accuracy of the
result by including leading (and even next-to-leading) higher order effects 
(which are often available in analytical form) and by performing
resummations which, quite often, cannot be fully justified. Here we focus
on the issue of setting up a complete two-loop calculus, postponing the 
question of building improved two-loop resummations.

At two loops we have a new feature, as described in details by many 
authors: the use of on-shell mass renormalization is not allowed anymore and 
complex poles are the only meaningful quantities: a property of the 
$S\,$-matrix, as it is often stated~\cite{Stuart:1991xk}. A better argument
is that complex poles are gauge parameter independent to all orders, as shown
by using Nielsen identities~\cite{Gambino:1999ai}.

Renormalization with complex poles should not be confused with a simple
recipe for the replacement of running widths with constant widths; there are 
many more ingredients in the scheme. Actually, this scheme allows for the 
introduction of a beautiful language, the one of effective (complex) 
couplings. 
The whole organization of loop corrections is most conveniently organized in 
terms of running couplings, as illustrated by the development of the fermion 
loop approximation. Unfortunately gauge invariance -- the ingredient at the 
basis of the successful proof of renormalization -- prevents us from fully
extending the use of running, resummed, couplings to the bosonic sector of the
theory (as a matter of fact even the concept of a fermionic sector is 
meaningless abeyond one loop).
Despite this caveat we will organize the presentation of our results 
according to the language of running couplings; in a way this language is much 
easier and one should only remember that final results must be expanded in 
perturbation theory up to second order. Partial resummations, i.e. the attempt 
of isolating gauge parameter independent blocks that can be moved freely from
numerators to denominators is beyond the scope of this paper.

Neglecting the fine points of the procedure we may say that the transition
from renormalized parameters to physical quantities depends on the choice 
of the latter. Quantities like $\gf$ and $\alpha$, the Fermi coupling constant 
and the fine structure constant, will always be included in our choice of the 
IPS. For most of the cases under consideration fermion masses and the Higgs 
boson mass are only needed at one loop; here on-mass-shell masses 
can be used. For gauge bosons instead, we have to extract from the data some 
information about the position of the corresponding complex poles. The fact 
that we still indulge in presenting theoretical prediction for, say $\mw$, is a
consequence of an established attitude that every pseudo-observable can
be defined and derived from data, although its theoretical degree of
purity is vanishing small. To an even larger extent the situation with the
Higgs boson {\em mass} (as used in any LHC Monte Carlo program) is fully
unclear.

Having adopted this strategy, we may say that the final step in renormalization
can still be seen as the moment where we write a system of (essentially)
three coupled equations, relating $g, \stw$ and $M$ (all renormalized 
quantities) to $\gf, \alpha,\,\dots$.

In this paper, we shall follow the same notations as defined in our companion 
papers I and II. We therefore refer the reader to that papers for notations;
in particular, $M$ stands for the bare (renormalized) $W$ boson mass
(we do not distinguish unless strictly needed), $M_0 = M/\ctw$ where 
$\ctw$ is the bare (renormalized) cosine of the weak-mixing angle; $\mw$ is 
the on-shell $W$ boson mass and $\mz = \mw/\ctw$.

The outline of the paper is as follows: in \sect{two} we summarize our 
procedure, and in \sect{FEqGF} we present a 
renormalization equation based on the use of the Fermi coupling constant;
aspects of the calculation connected with the proper definition of $\gamma^5$
are discussed in \sect{folwg5}. A second renormalization equation connected
with the fine structure constant is analyzed in \sect{FEqalpha}.
The running of $\alpha$ beyond one-loop fermion terms is discussed in
\sect{alpharun}. Complex poles are introduced in \sect{CPoles} and
solutions of the renormalization equations in \sect{solREe}.
Loop diagrams with dressed propagators are introduced in \sect{LPDP},
unitarity, gauge parameter independence and WST identities in \sect{UgpIWI}.
A suggestion on how to improve the complex mass scheme~\cite{Denner:2005fg} 
is finally introduced in \sect{giveitatry}.
\section{Outline of the calculation}
\label{two}
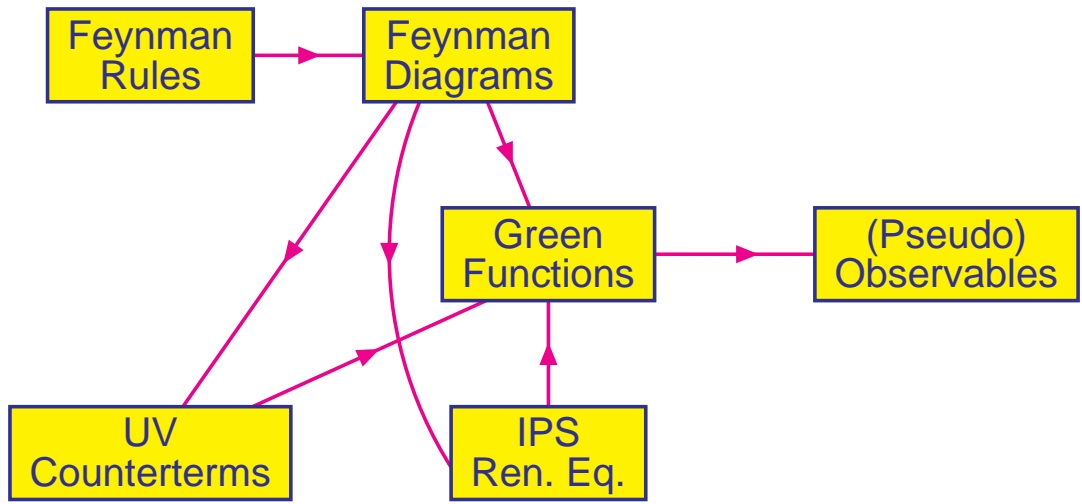
\begin{figure}[th]
\begin{center} \begin{picture}(320,320)(0,0)
\SetPFont{Helvetica}{10}
\SetScale{1.5}
\SetColor{Magenta}
\SetWidth{0.9}
\ArrowLine(0,200)(80,200)
\ArrowLine(80,200)(100,150)
\ArrowLine(0,100)(110,150)
\ArrowLine(100,100)(100,150)
\ArrowLine(100,150)(200,150)
\ArrowLine(70,200)(0,100)
\ArrowArc(160,150)(100,145,215)
\IfColor{\C2Text(0,200){Blue}{Yellow}{Feynman}{Rules}
}{\G2Text(0,200){0.9}{Feynman}{Rules}}
\IfColor{\C2Text(80,200){Blue}{Yellow}{Feynman}{Diagrams}
}{\G2Text(80,200){0.9}{Feynman}{Diagrams}}
\IfColor{\C2Text(0,100){Blue}{Yellow}{UV}{Counterterms}
}{\G2Text(0,100){0.9}{UV}{Counterterms}}
\IfColor{\C2Text(100,100){Blue}{Yellow}{IPS}{Ren. Eq.}
}{\G2Text(100,100){0.9}{IPS}{Ren. Eq.}}
\IfColor{\C2Text(100,150){Blue}{Yellow}{Green}{Functions}
}{\G2Text(100,150){0.9}{Green}{Functions}}
\IfColor{\C2Text(200,150){Blue}{Yellow}{(Pseudo)}{Observables}
}{\G2Text(200,150){0.9}{(Pseudo)}{Observables}}
\end{picture} 
\end{center}
\vspace{-3.4cm}
\caption[]{Renormalization - flowchart. Feynman rules define the theory,
renormalizability guarantees that ultraviolet poles have polynomial residues
and can be subtracted. Any input parameter set allows us to replace 
renormalized quantities with experimental data. A prediction follows.}
\label{Rfc}
\end{figure}
The whole renormalization procedure has been summarized in the flowchart of
\fig{Rfc} where IPS stands for Input Parameter Set. Object of this paper is 
to introduce renormalization equations (RE) and to solve them, therefore 
undertaking the task of writing any (pseudo-)observable, not in the IPS, in 
terms of the quantities of the IPS. The minimal standard model is essentially
a three-parameter theory and therefore we seek for a system of three 
(coupled) REs whose unknowns are $g, M$ and $\stw$, all renormalized 
parameters. 
In \sect{FEqGF} we discuss the first RE, related to the Fermi coupling 
constant, $\gf$. In \sect{FEqalpha} we present the second RE, related to fine
structure constant $\alpha$. In \sect{CPoles} we introduce the third RE, based
on the notion of complex pole for unstable gauge bosons.
The solution of our RE is discussed in \sect{solREe} where we present two
choices for the IPS. 
\section{The Fermi-coupling constant \label{FEqGF}}
In this section we present the results for $\gf$. A critical observation is 
that writing a renormalization equation for the Fermi-coupling constant should 
not be confused with predicting the muon lifetime or the Fermi coupling 
constant itself (also in an effective field-theory approach, see i.e. 
Ref.~\cite{Czakon:2003wg}).

To proceed further, we illustrate our method for relating the 
$\MSB$-renormalized parameters of the Standard Model to the Fermi coupling 
constant at two loops and we construct our first renormalization equation.

The Fermi coupling constant, $\gf$, is defined by
\bq\label{ferdef}
\frac{1}{\tau_\mu} \,= \, 
\frac{\gf^2\, m_\mu^5}{192\,\pi^3}\, ( \, 1\, +\, \Delta q \, ) ,
\eq
where $\tau_\mu$ and $m_\mu$ are the observed muon lifetime and mass.
The parameter $\Delta q$ summarizes both real and virtual corrections to $
\tau_\mu$, at leading order in $\gf$ and to all orders in the 
fine-structure constant $\alpha$. The corrections are generated by the 
effective Lagrangian
\bq\label{lagF}
\cL_{\QED\times\FT} \,=\,
\cL_{\FT} \, +\, \cL_{\QED},
\qquad
\cL_{_{\FT}} \,=\, 
\frac{\gf}{\sqrt{2}}\, \,\,\,
\left[\, \fbnu_\mu\, \gadu{\alpha}\, \gap\,\mu \,\right]\,\,\cdot \,\,
\left[\, \fbe\, \gadu{\alpha}\, \gap\,\nu_e\,\right] .
\eq
Here $\cL_{\FT}$ describes the contact-interaction Fermi theory (FT), 
$\cL_{\QED}$ is the usual QED Lagrangian and $\mu$, $e$, $\nu_\mu$ and 
$\nu_e$ are the spinor fields for the muon, the electron and their related 
neutrinos.

The inclusion of electromagnetic effects allows to determine the numerical 
value of $\gf$ in terms of the measurable quantities $\tau_\mu$, $\alpha$, 
$m_\mu$ and the electron mass $m_e$.
The combination of the one-loop \cite{Kinoshita:1958ru} 
and two-loop \cite{vanRitbergen:1998yd} QED$\,\times\,$FT contributions 
with the non-perturbative hadronic components \cite{vanRitbergen:1998hn}
leads to estimate
\bq
\gf\,=\, (1.16637 \, \pm\, 0.00001)\, \times 10^{-5} \, \GeV^{-2}.
\eq
Unfortunately, another definition is often employed in the literature
(see e.g.\ Ref.~\cite{PDG}),
\bq\label{ferdef2}
\frac{1}{\tau_\mu} \,=\, 
\frac{\gf^2\, m_\mu^5}{192\,\pi^3}\, ( \, 1\, +\, \Delta q \, ) \,\,
F\left(\frac{m_e^2}{m_\mu^2}\right)\,\,
\left(\, 1\, + \, \frac{3}{5} \, \frac{m_\mu^2}{M_{\ssW}^2} \, \right ),
\eq
where $M_{\ssW}$ is the $W$-boson mass and $F(m_e^2\slash m_\mu^2)$ follows 
from the phase-space integration at lowest order in perturbation theory,
\bq\label{PhS}
F(x)\,=\, 1 - 8 \, x - 12\, x^2\, \ln x  + 8\, x^3 -x^4.
\eq
However, the factor $(1 + 3m_\mu^2)\slash (5M_{\ssW}^2)$ is the tree-level 
$W$-propagator effect and is not generated by the Fermi-contact interaction.
Moreover, as pointed out by the authors of Ref.~\cite{vanRitbergen:1998yd}, 
the function $F$ does not factorize in the same way at higher orders.
Therefore, in order to avoid unnecessary ambiguities, we will use the 
definition of \eqn{ferdef}.

In the context of our renormalization procedure, $\gf$ is an input data
and we have to derive the appropriate renormalization equation.
First we express $\tau_\mu$ through the SM renormalized parameters. Next, we 
match our result with the definition of \eqn{ferdef} getting a relation 
between a measurable quantity, $\gf$, and the SM renormalized parameters,
\bq\label{deg}
\frac{\gf}{\sqrt{2}}\, = \,
\frac{g^2}{8\, M^2} \, (\, 1\, +\, \Delta g\, ).
\eq
Here $g$ and $M$ are the renormalized weak-coupling constant 
and $W$-boson mass and the quantity $\Delta g$ is constructed order-by-order
through the purely-weak corrections to the muon lifetime.
Electromagnetic components are already included in \eqn{ferdef} and they are 
discarded through the matching procedure.

It is worth noting that the answer for $\Delta g$ will contain renormalized 
parameters and counterterms. The choice of the renormalization scheme 
determines the explicit expressions for the counterterms and the final result 
for $\Delta g$.
A popular strategy, followed in recent two-loop calculations of the muon-decay
width \cite{Freitas:2000gg,Awramik:2002wn}, is to employ the on-mass-shell 
(OMS) scheme and to define renormalized parameters by means of measurable 
quantities. At lowest order one has simple relations,
\bq\label{OMS1}
e^2 = g^2 \stws = 4\pi\, \alpha,\qquad
\ctws = \frac{\mws}{\mzs},
\qquad M = \mw.
\eq
Here $e$ is the renormalized electric charge, $\mw$ and $\mz$ are the 
on-shell masses of the $W$ and $Z$ bosons and $c_\theta$($s_\theta$) is the 
renormalized cosine (sine) of the weak-mixing angle.
A replacement of the OMS renormalization prescription into \eqn{deg}
leads to the traditional parametrization employed to describe the 
interdependence between the masses of the vector bosons introduced in 
Ref.~\cite{Sirlin:1980nh}.

Rather that defining exactly what the renormalized parameters are, like in 
\eqn{OMS1}, we prefer to follow a minimal-subtraction scheme.
Let us prescribe the values of the counterterms as an intermediate step
to remove ultraviolet poles and regularization-dependent factors.
Relations among renormalized parameters and physical quantities are not 
imposed by hand and represent the solution of the chosen set of 
renormalization equations. 
Therefore, our result for $\Delta g$ should not be confused with 
$\Delta r$ as reported by the authors of 
Ref.~\cite{Freitas:2000gg,Awramik:2002wn}.

The basic prerequisite to derive the renormalization equation for $\gf$ is the 
extraction of the QED$\,\times\,$FT components from the full SM calculation of 
the muon lifetime. In \subsect{fact} we discuss our method and detail the
result with some relevant diagrammatic examples.
Note that the definition of $\gf$ relegates the infrared structure of muon 
decay to the \emph{soft} electromagnetic effects summarized by $\Delta q$ in
\eqn{ferdef} and does not affect the \emph{hard} weak remainder, $\Delta g$.

Moreover, in our approach, external legs will be provided with appropriate 
wave-function renormalization (WFR) factors; since most of the existing 
literature deals with the OMS scheme, where WFR factors are usually replaced 
by field counterterms, we devote \subsect{WFR} to discuss their role. 
Of course, the connection between field renormalization and wave-function 
factors is well understood, they are simply connected by a field 
transformation.

Finally, in \subsect{RESULTS} and \subsect{Gopt}, we give our results for 
$\Delta g$ at one loop (consistency check) and at two loops.
\subsection{Extraction of the electromagnetic components}
\label{fact}
In this subsection we consider the Standard Model tree-level amplitude 
for muon-decay and fix our notations an conventions. The process is
\bq
\mu(p_1)\,\to\,\nu_\mu(p_2)\,+\,e(p_3)\,+\,\fbnu_e(p_4).
\eq
Neglecting the electron mass and the squared momentum carried by the 
intermediate $W$-boson propagator, the result in 't Hooft-Feynman gauge reads 
as
\bq\label{tree}
\cM^0\,=\,
- \,(2 \pi)^4 \, i \, \frac{g^2}{8\, M^2} \,\,\, 
\left[\, \fbu_2\,\gadu{\alpha}\,\gap\,u_1\,\right] \,\cdot \,
\left[\, \fbu_3\,\gadu{\alpha}\, \gap\,v_4 \right],
\eq
where we used the short-hand notation for spinors $u_i = u(p_i)$ 
and $v_i = v(p_i)$.

The key observation is that the QED corrections in the context of the 
Fermi-contact interaction can be systematically identified and removed from 
the full SM amplitude. The one-loop matrix element factorizes as
\bq
\cM^{1L}\,=\,\cM^{0}\,\,
\left(\, 1\, +\, \Delta \cM^{1L}_{\QED\times\FT}\,
 \right)\,\,
\left(\, 1\, +\, \Delta \cM^{1L}_{\WEAK}\, \right)\, 
+ \ord{g^6},
\eq
where $\Delta \cM^{1L}_{\QED\times\FT}$ is a scalar function which 
summarizes all the QED$\,\times\,$FT soft contributions and the purely-weak 
remainder is entirely relegated to the hard component 
$\Delta\cM^{1L}_{\WEAK}$.
At two loops the amplitude can be decomposed as
\bq\label{FACT2}
\cM^{2L}\,=\,\cM^{0}\,\,
\left(\, 1\, +\, 
\Delta \cM^{1L}_{\QED\times\FT}\, + \, 
\Delta \cM^{2L}_{\QED\times\FT}\, \right)\,\,
\left(\, 1\, +\, \Delta \cM^{1L}_{\WEAK}\, + \, 
\Delta \cM^{2L}_{\WEAK}\right)\, + \ord{g^8},
\eq
where $\Delta \cM^{2L}_{\QED\times\FT}$ and 
$\Delta \cM^{2L}_{\WEAK}$ denote two-loop soft and hard corrections.
The class of soft effects generates the parameter $\Delta q$ of 
\eqn{ferdef}. We discard them and we identify the hard terms with
the quantity $\Delta g$ introduced in \eqn{deg},
\bq\label{deltag}
\Delta g\, =\, \Delta \cM^{1L}_{\WEAK}\, 
+ \, \Delta \cM^{2L}_{\WEAK}\, + \ord{g^6}.
\eq
%
\begin{figure}
\begin{center}
\includegraphics[scale=0.7]{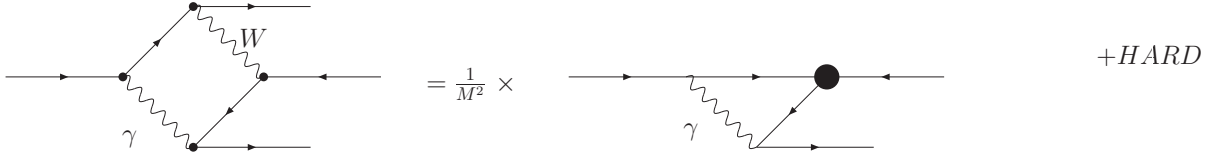}
\end{center}
\caption{Diagrammatic interpretation of \eqn{sdec}.
The first graph is a box diagram in the full SM context.
The one-photon vertex diagram is a QED correction
in the Fermi-contact interaction, denoted by the black circle.}
\label{box1sum}
\end{figure}
%

A simple one-loop example shows how the QED$\,\times\,$FT components 
factorize in the SM amplitude. We apply the procedure introduced in 
Ref.~\cite{Green:1980bd} to separate off the soft electromagnetic corrections 
from the hard remainder in the SM box diagram of \fig{box1sum}.
If $q$ denotes the momentum of the virtual photon and $q+p$ the momentum 
flowing through the $W$-boson line, we decompose the integrand in the large 
$W$-mass limit,
\bq
\label{sdec}
\frac{1}{q^2 \, [\, (q+p)^2+M^2\, ]} = \frac{1}{M^2}\,\,
\Bigl(\,
\frac{1}{q^2} - \frac{1}{q^2+M^2}\, \Bigr)\,
\Bigl(\, 1 \, -\,  \frac{2 q\cdot p + p^2}{q^2+M^2}+\ldots\, \Bigr).
\eq
Here the first term generates the soft QED-vertex correction to the local 
Fermi interaction of \fig{box1sum}, which can be obtained replacing 
the $W$-boson propagator by $1\slash M^2$.
We discard this component and we evaluate the hard weak remainder in the 
large vector-boson mass limit, where the external momenta and the lepton 
masses can be safely neglected,
\bq\label{Boxd1}
{\cM}^{\ssW\gamma}_{\WEAK}
= \frac{g^4\, \stws}{8\, M^2}\,\mu^{4-n}\,
\int d^n q\,
\frac{q_\mu\, q_\nu }{(q^2)^2(q^2+M^2)} \, \Gamma_{\mu\nu}^{{\ssW\gamma}}.
\eq
Here $\mu$ is the 't Hooft unit of mass and the spinor chain reads as
\bq \label{stringA}
\Gamma^{{\ssW\gamma}}_{\mu\nu}\,
 \equiv\, \left[\, \fbu_2\,\gamma_{\alpha} \,\gamma_+\,
\gamma_{\mu}\,\gamma_{\beta} \,u_1 \, \right]\,\, \cdot \,\, 
\left[\, \fbu_3\,\gamma_{\beta} \,\gamma_{\nu}\,
\gamma_\alpha \,\gamma_+\,v_4\, \right] ,
\eq
where $\gamma_{\pm} = 1 \pm \gamma^5$.
Note that the purely-weak components can be obviously obtained starting from 
the full SM amplitude and subtracting the QED$\,\times\,$FT terms.
Therefore, instead of extracting the hard correction from the box 
diagram of \fig{box1sum}, we can use an alternative strategy:
\vspace{0.1cm}

\noindent
I) We construct the difference of the box diagram and the QED$\,\times\,$FT 
component of \fig{box1sum}. This can be evaluated nullifying the 
external momenta and the lepton masses.
\vspace{0.1cm}

\noindent
II) Here the soft electromagnetic term is a massless tadpole which vanishes in 
dimensional regularization. With a non-zero electron mass, of course, it 
develops infrared and collinear singularities.
Therefore, any information about the infrared structure is lost but
these effects are already included in $\Delta q$ of \eqn{ferdef} and 
are not relevant for $\Delta g$.
\vspace{0.1cm}

\noindent
III) As a result, the hard part follows directly from the complete one-loop 
diagram by nullifying the lepton masses and the external momenta, 
\bq
\label{BoxD2}
{\cM}^{\ssW\gamma}_{\WEAK} = - \frac{g^4\, \stws}{8}\,
\mu^{4-n}\,\int d^n q\,
\frac{q_\mu q_\nu}{(q^2)^3(q^2+M^2)} \, \Gamma_{\mu\nu}^{\ssW\gamma}.
\eq
Using the fact that in dimensional regularization an integral without scales
is zero it is easy to see that the two representations of \eqn{Boxd1} and 
\eqn{BoxD2} are completely equivalent.
Since QED factorization should not be proven by assumption, and one has 
anyway to separate off the QED$\,\times\,$FT components, the choice between 
the two procedures is just a matter of taste.
However, the subtraction method of \eqn{BoxD2}, introduced in 
Ref.~\cite{Awramik:2002wn}, appears more appropriate in the context of an 
automatic approach, where all the Feynman diagrams are generated and 
evaluated by neglecting the soft scales from the very beginning.

The matching on the Fermi-theory spinor-chain configuration can be finally 
completed when we introduce the projector,
\bq
\label{proj}
{\cal P}\, =\,
\sum_{\rm{spins}}\, \left[ \, \fbv_4\,\gadu{\lambda}\, \gamma_+\,
u_2\, \right]\,\, \cdot\, \,
\left[\, \fbu_1\,\gadu{\lambda}\, \gamma_+\,
u_3 \, \right]
\eq
which will act on the matrix element for muon decay.
For example, for the hard part of \eqn{Boxd1} we can write
\bq
\label{deltaM}
\cM^{\ssW\gamma}_{\WEAK} =  \cM^{0} 
\, \,\cdot \,\, \Delta \cM^{\ssW\gamma}_{\WEAK},
\quad \mbox{with} \quad
\Delta \cM^{\ssW\gamma}_{\WEAK} = 
\frac{ {\cal P}  \cM^{\ssW\gamma}_{\WEAK}}{
       {\cal P}  \cM^{0}}.
\eq

\begin{figure}
\begin{center}
\includegraphics[scale=0.7]{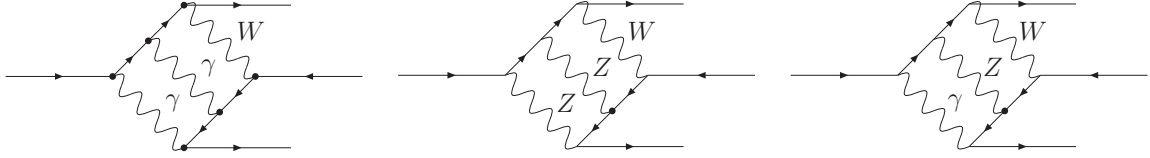}
\end{center}
\caption{Two-loop box diagrams for muon decay.}
\label{box2list}
\end{figure}

\begin{figure}
\begin{center}
\includegraphics[scale=0.5]{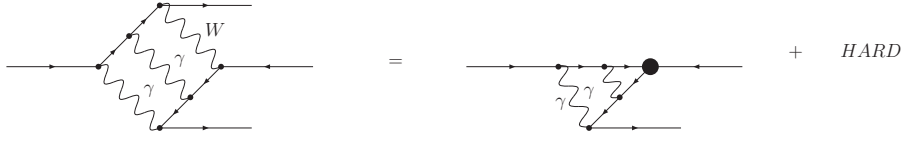}
\end{center}
\caption{Two-loop soft $\otimes$ soft splitting.}
\label{split2}
\end{figure}

Two-loop virtual corrections to the muon-decay amplitude can be classified
according to \fig{box2list}.
\noindent
Here diagram a) contains two photons and one hard scale and can be treated 
in complete analogy with the one-loop case. After shrinking the heavy line 
to a point like in \fig{split2}, the two-photon graph in the local 
Fermi theory is discarded. The hard weak remainder is obtained by nullifying
the soft scales in the full two-loop box diagram.
%
\noindent
Diagram b) includes just heavy components, and does not require any 
subtraction. We will evaluate it in the soft limit obtaining a finite answer.

\begin{figure}
\begin{center}
\includegraphics[scale=0.5]{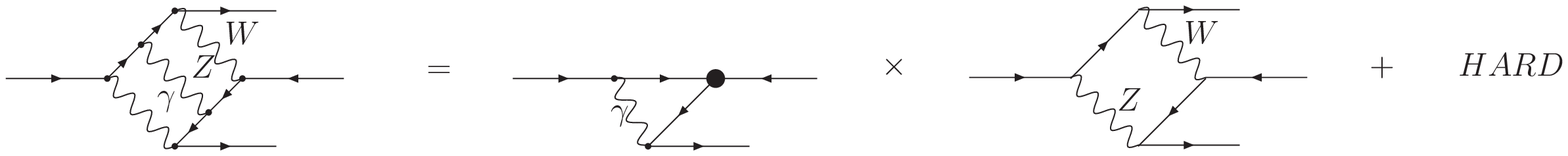}
\end{center}
\caption{Two-loop soft $\otimes$ hard splitting.}
\label{splitMIX}
\end{figure}

Since in diagram c) soft and hard components are entangled, we discuss it 
in more detail.
We consider the subloop with a photon and a $Z$ boson and we decompose it
in the large $Z$-boson mass limit. If $q_1$ is the momentum flowing along 
the photon line and $q_1-q_2$ is the momentum of the $Z$ propagator,
we can write
\bq
\label{decgzw}
\frac{1}{q_1^2 \, [\, (q_1-q_2)^2+M_0^2\, ]} =
\frac{1}{q_1^2}\,
\frac{1}{q_2^2+M_0^2}\,\,
\,
\Bigl(\, 1 \, +\,  \frac{2 q_1\cdot q_2 - q_1^2}{q_2^2+M_0^2}+\ldots\, \Bigr),
\eq
where $M_0= M\slash c_\theta$.
The first term gives the product of a one-loop QED$\,\times\,$FT vertex
and a one-loop weak remainder, as shown in \fig{splitMIX}.
%
The rest is a purely-weak two-loop component which will be evaluated after
nullification of external momenta and lepton masses.

The soft$\,\times\,$hard term is crucial in proving the two-loop factorization 
property of \eqn{FACT2}, because it generates the 
$\Delta \cM^{1L}_{\QED\times\FT} \times \Delta \cM^{1L}_{\WEAK}$ part.
To evaluate the hard remainder, we start again from the complete 
representation for the SM diagram, where we nullify the soft scales,
\bq
\cM^{\ssW\ssZ\gamma}_{\WEAK} \, = \,
- \, \frac{1}{(2 \pi)^4\, i}\,
\frac{1}{32}\, \frac{ g^2\, \stws}{\ctws}\,
\mu^{2(4-n)}\,\int d^nq_1 d^nq_2\,
\frac{q_{1\delta}\, q_{2\nu}\, q_{1\rho}\, q_{2\sigma}}
{(q_1^2)^3\, (q_2^2)^2\, (q_2^2+M^2)\, [(q_1-q_2)^2+M_0^2]}\,\,
\Gamma^{\ssW\ssZ\gamma}_{\delta\nu\rho\sigma}.
\eq
Here the spinor chain reads as
\bq
\Gamma^{\ssW\ssZ\gamma}_{\mu\nu\rho\sigma} \, = \,  
\left[ \, 
\fbu_2 \, \gamma_{\alpha} \, \gamma_+ \, \gamma_{\nu} \, 
\gamma_\beta \, (v_\mu + a_\mu \gamma_5)\,
\gamma_{\delta} \, \gamma_\gamma \, u_1 
\, \right] \,\,\cdot \, \, \left[ \,
\fbu_3\,
\gamma_\gamma\, \gamma_{\rho}\, \gamma_\beta\, (v_e + a_e \gamma_5)\,
\gamma_{\sigma} \, \gamma_{\alpha}\, \gamma_+ \, v_4
\, \right]
\eq
and $v_f= I_f^{(3)}-2 Q_f^2 s_\theta^2$ and $a_f= I_f^{(3)}$, where $I_f^{(3)}$
is the weak-isospin third component for the fermion $f$ and $Q_f$ is the 
related electric-charge quantum number.
Finally, a correct description of muon decay involves also real radiative 
effects. This is essential in showing that the QED$\,\times\,$FT corrections 
are finite \cite{Kinoshita:1958ru,vanRitbergen:1998yd} and  allows to define 
the Fermi-coupling constant through \eqn{ferdef}.
Nevertheless, one can easily prove that real soft-radiation diagrams can be 
systematically neglected from the calculation.
\subsection{Wave-function renormalization factors}
\label{WFR}
Wave-function renormalization factors for external legs (hereafter WFR)
contain derivatives of two-point Green's functions which are infrared divergent
and deserve a special discussion. 
In order to derive explicit representations for the muon-decay case we 
consider the fermion one-particle irreducible Green's function,
\bq
\Sigma(p) \,=\, (2\pi)^4\, i\,\, \left\{\,  a_{\ssS}(p^2) 
\,+\, \left[\, a_{\ssV}(p^2) \,-\,a_{\ssA}(p^2)\,\gamma_5
\,\right]\,i\, \sla{p} \, \right\},
\eq
where we introduced a scalar, a vector and an axial form factor (the SM 
pseudo-scalar component obviously vanishes). The fermion Dyson-resummed 
propagator reads as
\bq\label{Dys}
S(p) \,=\, \Bigl\{ (2\pi)^4\, i \,\, \Bigl[
\Bigl(\,  1\, -\, a_{\ssV}(p^2)\, +\, a_{\ssA}(p^2)\, \gamma_5 
\, \Bigr)  \, i \sla{p} \, +\,  m\,  -\, a_{\ssS}(p^2) \, 
\Bigr] \, \Bigr\}^{-1},
\eq
where $m$ is the renormalized fermion mass. Let us accordingly define 
appropriate WFR factors using a second representation for $S(p)$ around 
the on-shell fermion mass, $m_{\OS}$,
\bq\label{WFRDys}
S(p) \, =\,  \Bigl\{ (2\pi)^4\, i\, \Bigl[
\Bigl( \, 1 \, -\, W_{\ssV}\, +\, W_{\ssA}\,\gfd \,\Bigr) \,
\Bigl( i \sla{p} + m_{\OS} \,
+\ord{(i \sla{p} + m_{\OS})^2}\Bigr)
\, \Bigl( \, 1 - W_{\ssV} - W_{\ssA}\,\gfd \, \Bigr) \Bigr] \Bigr\}^{-1}.
\eq
Here $W_{\ssV}$ and $W_{\ssA}$ are the vector and axial WFR factors, 
whose expressions can be obtained through fermion-mass renormalization 
followed by a straightforward matching procedure of \eqn{WFRDys} with 
\eqn{Dys}.
\vspace{0.2cm}

\noindent
\underline{\emph{Fermion-mass renormalization}}. We trade the renormalized 
mass for the on-shell one writing
\bq
\begin{split}
S(p)\,D(p^2)\,&=\,- \frac{1}{(2\pi)^4i}\,\,\Bigl[\, 
1\, -\, a_{\ssV}(p^2)\,  +\, a_{\ssA}(p^2)\, \gamma_5 
\, \Bigr] \, i\, \sla{p}\, - \, m\, +\, a_{\ssS}(p^2),\\
&\\
D(p^2)\,&=\,\left[\,
1 \,- \,2 a_{\ssV}(p^2)\, +\, a_{\ssV}^2(p^2)\, -\, a_{\ssA}^2(p^2)
\, \right]\, p^2 + \left[\, m - a_{\ssS}(p^2)\, \right]^2
\end{split}
\eq
and imposing fermion-mass renormalization, $D(-m_{\OS}^2) = 0$.
We obtain a perturbative solution expanding $a_{\ssS}$, $a_{\ssV}$ and 
$a_{\ssA}$ through powers of the weak-coupling constant $g$,
\bq
a_{\ssI}(p^2) \,=\, 
\sum_{j=1}^{\infty}\, g^{2j}\, a_{\ssI}^{jL}(p^2),\qquad
I=S,V,A ,
\eq
and performing a second expansion around the mass shell,
\bq
a_{\ssI}^{jL}(p^2)\,=\, A_{\ssI}^{jL}\,
+\,
2\, m_{\OS}\, \left(\, i\, \sla{p}\, +\, m_{\OS}\, \right)\, 
B_{\ssI}^{jL}\,
+\, \ord{\left(\, i\, \sla{p}\, +\, m_{\OS}\, \right)^2}.
\eq
Here we introduced short-hand notations for the form factors evaluated 
on the fermion mass shell and their derivatives respect to the squared 
external momentum,
\bq
A_{\ssI}^{jL}\,=\, a_{\ssI}^{jL}(-m^2_{\OS}),
\qquad
B_{\ssI}^{jL}\,=\,
\frac{\partial a_{\ssI}^{jL}(p^2)}{\partial p^2}\vert_{p^2=-m^2_{\OS}}.
\eq
The solution for fermion-mass renormalization reads as
\bq\label{massDEF}
m \, =\,m_{\OS}\,+\,
g^2 \Bigl(
A_{\ssS}^{1L} \, 
- \, m_{\OS} \, A_{\ssV}^{1L}\Bigr)\\
\,+\,g^4 \Bigl\{
A_{\ssS}^{2L} \, - \, m_{\OS} \Bigl[ A_{\ssV}^{2L}
\, + \, \frac{1}{2} \Bigl( A_{\ssA}^{1L}\Bigr)^2 \Bigr]
\Bigr\}\, + \ord{g^6}
\eq
and removes from \eqn{Dys} the renormalized mass.
Note that one-loop fermion-mass renormalization involves just the axial and 
scalar form factors evaluated on the fermion mass shell. At two loops, 
instead, we are left to consider an additional term given by the square of the 
one-loop axial component.
\vspace{0.2cm}

\noindent
\underline{\emph{Wave-function renormalization}}. Expand the WFR factors 
introduced in \eqn{WFRDys} order-by-order in perturbation theory,
\bq
W_{\ssI}\,=\, \sum_{j=1}^{\infty}\, g^{2j}\, W_{\ssI}^{jL},\qquad
I=V,A.
\eq
We wish to determine the explicit expressions of the WFR factors at one and 
two loops by matching \eqn{Dys} with \eqn{WFRDys},
\bqa
\label{Wfacts}
W^{1L}_{\ssV}\,&=&\,\frac{1}{2}\, A^{1L}_{\ssV}\,
+\, m_{\OS}\, B^{1L}_{\ssS}\,-\, m^2_{\OS}\, B^{1L}_{\ssV},
\qquad
W^{1L}_{\ssA}\,=\,\frac{1}{2}\, A^{1L}_{\ssA},
\nl
W^{2L}_{\ssV}\,&=&\,
\frac{1}{2}\, \Bigl\{\,A^{2L}_{\ssV}\,+\, \frac{1}{4}\,
\Bigl[\,\Bigl(\, 
A^{1L}_{\ssV}\, \Bigr)^2 \, +\, \Bigl(\, A^{1L}_{\ssA} \, 
\Bigr)^2\,\Bigr]\, \Bigr\}
+\, m_{\OS}\,
\Bigl(\,B^{2L}_{\ssS}\, + \, \frac{1}{2}\, A^{1L}_{\ssV}\, 
B^{1L}_{\ssS}\, \Bigr)
\nl 
&-&\,  m_{\OS}^2\,
\Bigl\{\,
 B^{2L}_{\ssV}\, + \, \frac{1}{2}\, 
\Bigl[\,A^{1L}_{\ssV}\, B^{1L}_{\ssV}\,-\,\Bigl(\, B^{1L}_{\ssS} \, \Bigr)^2
\, \Bigr]\, \Bigr\} 
-\,  m_{\OS}^3\,B^{1L}_{\ssS}\,B^{1L}_{\ssV}\,
+\,  \frac{1}{2}\, m_{\OS}^4\,
\Bigl(\, B^{1L}_{\ssV} \, \Bigr)^2,
\nl
W^{2L}_{\ssA}\,&=&\,\frac{1}{2}\, \Bigl(\,A^{2L}_{\ssA}\,
+\, \frac{1}{2}\, A^{1L}_{\ssV}\,A^{1L}_{\ssA}\,\Bigl)\,
+\,  \frac{1}{2}\, m_{\OS}\,A^{1L}_{\ssA}\,
B^{1L}_{\ssS}\,-\,  \frac{1}{2}\, m_{\OS}^2\,
A^{1L}_{\ssA}\,B^{1L}_{\ssV}.
\eqa

Beyond the tree level we provide the external spinors with the WFR factors.
The one-loop matrix element includes a component given by tree level 
diagrams where each external leg gets separately a WFR factor at $\ord{g^2}$,
\bq
\label{1WFRspin}
u \, \Rightarrow \,g^2\,\,\, \left(\, 
W^{1L}_{\ssV}+W^{1L}_{\ssA}\gamma_5\,\right)\, u,
\qquad
\fbu\, \Rightarrow \,g^2\,\,\, \fbu\,  
\left(\, W^{1L}_{\ssV}-W^{1L}_{\ssA}\gamma_5\,\right).
\eq
The two-loop amplitude requires a more careful treatment, and involves tree 
classes of diagrams containing WFR factors:

\bei
\item[--] tree level diagrams where three external legs are not corrected,
and one includes a WFR factor at $\ord{g^4}$,
\bq\label{2WFRspin}
\begin{split}
u           \, \Rightarrow& \,
g^4\,\, \left[\, W^{2L}_{\ssV}
+ \sum_{\ssI=\ssV,\ssA}\,\left( W^{1L}_{\ssI} \right)^2
+\Bigl( W^{2L}_{\ssA}
 + 2\, W^{1L}_{\ssV}W^{1L}_{\ssA}\Bigr)\gamma_5
\,\right]\, u,\\
\fbu\, \Rightarrow& \,
g^4\,\, \fbu\,  \left[\, W^{2L}_{\ssV}
+ \sum_{\ssI=\ssV,\ssA}\,\left( W^{1L}_{\ssI} \right)^2
-\Bigl(W^{2L}_{\ssA}
 + 2\, W^{1L}_{\ssV}W^{1L}_{\ssA}\Bigr)\gamma_5
\,\right].
\end{split}
\eq
\item[--] Tree level diagrams where two external legs are simultaneously 
providedwith a WFR factor at $\ord{g^2}$ as in \eqn{1WFRspin}.

\item[--] One-loop diagrams where the external legs are corrected 
one-by-one at $\ord{g^2}$ through the WFR components of \eqn{1WFRspin}.

\eei

Concerning the muon-decay case, we are dealing with an external positron.
Relations for antiparticle spinors follow trivially from 
Eqs.~\eqref{1WFRspin}-\eqref{2WFRspin} replacing $u$ and $\fbu$ by 
$v$ and $\fbv$.

\eqn{Wfacts} contains derivatives of the 1PI Green's function.
These terms are suppressed by positive powers of the fermion mass,
and in QED they survive with double and single infrared poles and logarithmic 
mass singularities.
In the computation of the purely-weak component, instead, we can neglect the 
fermion masses in \eqn{Wfacts} and compute WFR factors through the 
scalar, vector and axial form factors evaluated at zero-momentum transfer.
\subsection{Two-Loop Corrections to $\Delta g$}
\label{RESULTS}
In this section we will extend to the two-loop amplitude; in order to 
evaluate $\Delta g$ we follow three steps:

\bei
\item[--] We generate the one-loop and two-loop matrix elements $\cM^{1L}$ and 
$\cM^{2L}$ for muon decay by nullifying the external momenta and the 
light-fermion masses.
As discussed in \sect{fact}, this implies an automatic subtraction of the 
soft QED$\,\times\,$FT corrections.
Therefore, our expressions for the one-loop and two-loop amplitude only 
contain the weak contributions.

\item[-] We project $\cM^{1L}$ and $\cM^{2L}$ on the tree level amplitude 
through the projector defined in \eqn{proj} and we extract the hard components 
$\Delta\cM^{1L}_{\WEAK}$ and $\Delta\cM^{2L}_{\WEAK}$.
The sum over the spins of the external particles leads to traces over Dirac 
matrices which have to be evaluated in $n$ dimensions.

\item[--] At this stage, we have to compute a large set of massive tadpole 
diagrams. Using integration-by-part identities they are reduced to one master 
tadpole integral, as shown in appendix~A of II.

\eei

Before showing the two-loop result, we review the one-loop corrections.
The one-loop hard contributions can then be 
conveniently written as
\bq
\Delta\cM^{1L}_{\WEAK}\,=\,\Delta\cM^{1L}_{\WEAK,\ssW}
+ \Delta\cM^{1L}_{\WEAK,\REST},
\eq
where $\Delta\cM^{1L}_{\WEAK,\ssW}$ denotes universal $W$-boson self-energy 
corrections, and $\Delta\cM^{1L}_{\WEAK,\REST}$ represents process-dependent 
vertex, box and wave-function components.
The result for the latter reads as
\bq
\Delta\cM^{1L}_{\WEAK,\REST}\,= \frac{g^2}{16\, \pi^2}
\, \Bigl[ 6 + \frac{7-4 s_\theta^2}{2 s_\theta^2} \ln c_\theta^2\Bigr].
\eq
Since we removed all ultraviolet poles and regularization-dependent factors
from one-loop diagrams, the quantities $\Delta\cM^{1L}_{\WEAK,\ssW}$ 
and $\Delta\cM^{1L}_{\WEAK,\REST}$ do not show any ultraviolet-divergent 
component. However, as explained in ~\cite{Passarino:1990xx}, 
$\Delta\cM^{1L}_{_{\WEAK,\REST}}$ is ultraviolet finite also in the bare 
theory, owing to the introduction of the parameter $\Gamma_1$ (see I).

Two-loop corrections to the muon-decay amplitude can be organized 
through a universal component, represented by $W$-boson self-energy reducible 
and irreducible diagrams, and a process-dependent part containing irreducible 
two-loop box, vertex and WFR diagrams and reducible ones,
\bq
\Delta\cM^{2L}_{\WEAK} \,=\,\Delta\cM^{2L}_{\WEAK,\ssW}
+ \Delta\cM^{2L}_{\WEAK,\REST}.
\eq
Irreducible two-loop box diagrams give ($x_i= m^2_i/M^2$)
\bqa\label{Box2}
\Delta\cM^{2L}_{\WEAK, \BOX}\vert_{UV}&=&
\frac{g^4}{(16 \pi^2)^2}\,
\Bigl\{
\frac{2}{\ep}
\Bigl(\Delta_{\ssU\ssV} - \frac{1}{\ep}\Bigr)\,s^4_\theta
+ \frac{1}{\ep}
\Bigl[
 - \frac{157}{12}
 +\frac{3}{4}\frac{1}{c_\theta^2}
 +\frac{157}{6} c_\theta^2
 -\frac{83}{6} c_\theta^4 \nonumber\\
{}\nl{}
&-& \frac{3}{2} s_\theta^2 x_t
- \frac{s^2_\theta}{x_\ssH} \Bigl( \frac{1}{c_\theta^4} -
  12  x_t^2 
 + 2  + 12 x_t^2 \ln x_t
  + \frac{3}{c_\theta^4} \ln c_\theta^2
                \Bigr) \nonumber \\
{}\nl{}
&+& \Bigl( \frac{3}{2} \ln x_\ssH - \frac{7}{4} \Bigr) x_\ssH s_\theta^2
-  \Bigl( 6 - \frac{5}{2} \frac{1}{c_\theta^2} - 
4 c_\theta^2 + 2 c_\theta^4 \Bigr)\ln c_\theta^2
 \nonumber \\
{}\nl{}
&-& \frac{3}{2} s_\theta^2  
 \frac{x_\ssH}{x_\ssH-1} \ln x_\ssH + 3  x_t s_\theta^2 \ln x_t
\Bigr]
\Bigr\},
\eqa
and the sum of the other process-dependent components is
\bqa\label{Ver2}
\Delta\cM^{2L}_{\WEAK, other}\bmid_{\ssU\ssV}&=&
-\, \Delta\cM^{2L}_{\WEAK, \BOX}\bmid_{\ssU\ssV},
\eqa
where we have included one-loop diagrams with a one-loop counterterm 
insertion but no two-loop counterterms, since one can show that 
-- exactly like for the one-loop case -- their contribution cancels,
and the sum of Eqs.~\eqref{Box2}-\eqref{Ver2} is an ultraviolet-finite 
quantity by itself.
Therefore, a one-loop renormalization is enough to cancel the ultraviolet 
poles for $\Delta\cM^{2L}_{\WEAK,\REST}$ and to construct the 
process-dependent component of the Fermi-coupling constant by neglecting 
two-loop counterterms.
Note that a crucial role here is played by the parameter $\Gamma_2$.
The inpact of two-loop corrections on the finite quantity $\delta_{\ssG}$
is shown in \tabn{nideltag}. The complete expressions of all the components 
of $\Delta g$ are too lengthy to be presented here and have been stored in
http://www.to.infn.it/\~{}giampier/REN/GF.log.
\subsection{Process independent, resummed, Fermi constant \label{Gopt}}
If we neglect, for the moment, issues related to gauge parameter independence
it is convenient to define a $G$ constant that is totally process independent,
\bq
\Delta g = \delta_{\ssG} + \Delta g^{\ssS}, \qquad
G = \frac{\gf}{\sqrt{2}}\,\lpar 1 - \frac{g^2}{8\,M^2}\,\delta_{\ssG}\rpar,
\quad
\delta_{\ssG} = \sum_{n=1}\,\lpar \frac{g^2}{16\,\pi^2}\rpar^n\,
\delta^{(n)}_{\ssG}.
\label{defDG}
\eq
Alternatively, but always neglecting issues related to gauge parameter 
independence, we could resum $\delta_{\ssG}$ by defining
$\sqrt{2}\,G_{\ssR} = \gf/(1+\delta_{\ssG})$. In one case we obtain
\bq
G = \frac{g^2}{8\,M^2}\,\Bigl[ 1 - 
\frac{g^2}{16\,\pi^2\,M^2}\,\Sigma_{\ssWW}(0)\Bigr]^{-1},
\qquad
\Sigma_{\ssWW}(0) = \Sigma^{(1)}_{\ssWW}(0) + \frac{g^2}{16\,\pi^2}\,
\Sigma^{(2)}_{\ssWW}(0),
\label{defGuniv}
\eq
where $\Sigma_{\ssWW}$ is the $\wb$ self-energy, whereas with resummation
we get
\bq
G_{\ssR} = \frac{g^2}{8\,M^2}\,\Bigl[ 1 - 
\frac{g^2}{16\,\pi^2\,M^2}\,\oSigma_{\ssWW}(0)\Bigr]^{-1},
\quad
\oSigma_{\ssWW}(0) = \Sigma^{(1)}_{\ssWW}(0) 
+ \frac{g^2}{16\,\pi^2}\,\Bigl[
\Sigma^{(2)}_{\ssWW}(0) - \Sigma^{(2)}_{\ssWW}(0)\,\delta^{(1)}_{\ssG}
\Bigr].
\eq
In \sect{solREe} we will show how solutions for the renormalization equation
of the SM are simpler when written in terms of $G$.
\section{The $\gamma^5$ problem \label{folwg5}}
To compute two-loop corrections to $\mu$ decay we must specify how to treat
$\gamma^5$. The {\em naive} scheme is defined by
\bq
\{\gamma^{\mu}\,,\,\gamma^5\} = 0,
\qquad
\mbox{Tr}\, \Bigl( \gamma^5\,\gamma^{\mu}\,\gamma^{\nu}\,
\gamma^{\alpha}\,\gamma^{\beta}\Bigr) = 4\,\ep^{\mu\nu\alpha\beta},
\label{naive}
\eq
in $n-$dimensions. 
A complete set of calculational rules with algebraic consistency requires to
consider $\gamma^{\mu}, \gamma^5$ and $\ep_{\mu\nu\alpha\beta}$
as formal objects\cite{Breitenlohner:hr} with the following properties:
\bq
\{\gamma^{\mu},\gamma^{\nu}\} = 2\,\delta^{\mu\nu}\,I, \qquad
{\rm Tr}\,I = 4,
\nonumber
\eq
\bq
\delta_{\mu\nu} = \hdel_{\mu\nu} + \bdel_{\mu\nu},
\quad
\bdel_{\mu\alpha}\,\hdel_{\alpha\nu} = 0,
\quad 
\bdel_{\mu\mu} = n-4,
\quad
\hdel_{\mu\mu} = 4,
\nonumber
\eq
\bq
\gamma^5 = \frac{1}{4!}\,\ep^{\mu\nu\alpha\beta}\,{\rm Tr}\,
\Bigl( \gamma_{\mu}\gamma_{\nu}\gamma_{\alpha}\gamma_{\beta} \Bigr)=
\frac{1}{4!}\,\ep^{{\hat\mu}{\hat\nu}{\hat\alpha}{\hat\beta}}\,{\rm Tr}\,
\Bigl( \gamma_{\hat\mu}\gamma_{\hat\nu}\gamma_{\hat\alpha}\gamma_{\hat\beta} \Bigr),
\nonumber
\label{HVBM}
\eq
\bq
\{\hgam^{\mu},\gamma^5\} = 0, \qquad [\bgam^{\mu},\gamma^5] = 0 
\Rightarrow \Delta^{\mu} = \{\gamma^{\mu},\gamma^5\} = 2\,\bgam^{\mu}\gamma^5,
\nonumber
\eq
\bq
{\rm Tr}\,\gamma^5 = 
{\rm Tr}\, \Bigl( \gamma^5\,\gamma_{\mu}\gamma_{\nu} \Bigr)=
{\rm Tr}\, \Bigl( \gamma^5\,\gamma_{\mu_1}\dots\gamma_{\mu_{2p+1}} \Bigr)= 0,
\nonumber
\eq
\bq
{\rm Tr}\,
\Bigl( \gamma^5\gamma^{\mu}\gamma^{\nu}\gamma^{\alpha}\gamma^{\beta} \Bigr)=
4\,\ep^{\mu\nu\alpha\beta}\, \qquad
(\gamma^5)^2 = 1.
\eq
\eqn{HVBM} defines the 't Hooft - Veltman - Breitenlohner - Mason scheme
(hereafter HVBM).

The HVBM scheme breaks all WST identities (so-called spurious or avoidable
violations) which can be restored afterwards by introducing suitable 
ultraviolet finite counterterms. The procedure, however is lengthy and 
cumbersome. 

\begin{figure}[th]
\vspace{1.3cm}
\bqas  
{}&{}&
  \vcenter{\hbox{
  \begin{picture}(0,0)(100,0)
  \SetScale{0.7}
  \SetWidth{1.2}
  \SetColor{Red}
  \ArrowLine(50,0)(100,25)
  \ArrowLine(100,25)(100,-25)
  \ArrowLine(100,-25)(50,0)
  \SetColor{Black}
  \Line(0,0)(50,0)
  \Line(100,25)(150,50)
  \Line(100,-25)(150,-50)
  \ArrowLine(150,50)(150,-50)
  \ArrowLine(175,65)(150,50)
  \ArrowLine(150,-50)(175,-65)
  \Text(0,5)[cb]{$W$}
  \Text(80,35)[cb]{$W$}
  \Text(80,-40)[cb]{$Z$}
  \end{picture}}}
\qquad\qquad 
\qquad\qquad
\qquad\quad
  \vcenter{\hbox{
  \begin{picture}(0,0)(40,0)
  \SetWidth{1.2}
  \SetScale{0.7}
  \SetColor{Red}
  \ArrowLine(100,25)(50,0)
  \ArrowLine(100,-25)(100,25)
  \ArrowLine(50,0)(100,-25)
  \SetColor{Black}
  \Line(0,0)(50,0)
  \Line(100,25)(150,50)
  \Line(100,-25)(150,-50)
  \ArrowLine(150,50)(150,-50)
  \ArrowLine(175,65)(150,50)
  \ArrowLine(150,-50)(175,-65)
  \Text(0,5)[cb]{$W$}
  \Text(80,35)[cb]{$W$}
  \Text(80,-40)[cb]{$Z$}
  \end{picture}}}
\eqas
\vspace{0.9cm}
\caption[]{Example of two-loop diagrams contributing to the anomaly.}
\label{anom}
\end{figure}
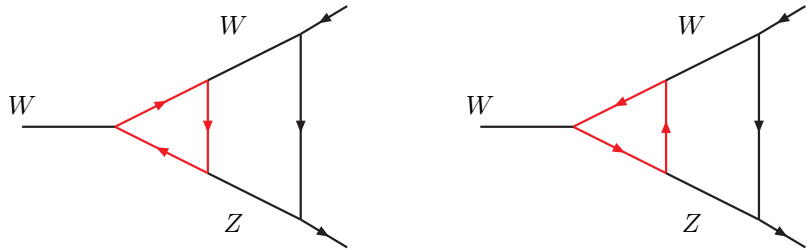

The usual statement that we find in the literature is: consider the two 
diagrams of \fig{anom}, they are the only place where the difference
between the two schemes is relevant.
If we can prove that the fermion triangles inside the diagrams of \fig{anom} 
in the HVBM scheme give
\bq
\Gamma^{\alpha\mu\nu}_{ssW}\,\bmid_{\ssH\ssV\ssB\ssM} + \,\mbox{c.t.} =
\Gamma^{\alpha\mu\nu}_{ssW}\,\bmid_{{\rm ac}\,\gamma^5} + \ord{n-4},
\label{epequal}
\eq
(and for both schemes the anomaly is correctly reproduced) then 
the difference ($\ord{n-4}$) is irrelevant since the remaining integration in 
the two-loop integrals is ultraviolet finite; the latter follows in
general from renormalizability and it is confirmed by our explicit calculation
which shows that all $\ep-$tensor terms are ultraviolet finite. 
However, Feynman rules are derived from the Lagrangian and one cannot change
the former but not the latter. The relevant analysis has been performed by
F.~Jegerlehner~\cite{Jegerlehner:2000dz} and we simply repeat the argument; 
in constructing the SM Lagrangian we use chiral fields and the relation
\bq
\fbpsi_{\ssL,\ssR}\,\gamma^{\mu}\,\psi_{\ssR,\ssL} =
\frac{1}{2}\,\fbpsi\,\gamma^{\mu}\,\gamma_-\,\psi,
\eq
is valid only if $\{\gamma^{\mu}\,,\,\gamma^5\} = 0$. The consequence is that
we miss a chirally invariant dimensional regularization, i.e. we cannot
regularize fermion-loop integrals by continuation in $n$.
If we insist on the trace condition then gauge invariance must be broken in
order to obtain a pseudo-regularization, i.e.\ we use an $n$-dimensional
fermion propagator
\bq
S_{\ssF} = \frac{-i\,\sla{\hat p} + m}{p^2 + m^2 -i\,\delta},
\qquad
p^2= {\hat p}^2 + {\bar p}^2.
\eq
Pseudo-regularization introduces spurious (i.e. avoidable) violations of WST 
identities which must be restored afterwards by introducing suitable 
ultraviolet finite counterterms. The procedure is lengthy because all
Green functions containing a fermion loop are now anomalous and not only the 
fermion triangles inside the diagrams of \fig{anom}.
A simple example comes from the difference between one-loop naive and
one-loop pseudo-regularized vector - vector, vector - scalar and 
scalar - scalar transitions; define finite counterterms in the 
pseudo-regularized formulation that make this difference zero.
We will get a set of equations,
\bqa
{}&{}&
Z_{\ssW}\,\Bigl[
- (\htp^2 + Z_{\ssM}\,M^2)\,\delta^{\hmu\hnu} +
\lpar 1 - \frac{1}{\xi_{\ssW}^2\,Z_{\xi_{\ssW}}^2}\rpar\,
\htp^{\hmu}\,\htp^{\hnu}\Bigr] +
\sum_f\,N^c_f\Delta^{\hmu\hnu}_{f\,;\,\ssWW} = 0.
\nl
{}&{}&
\Delta^{\hmu\hnu}_{f\,;\,\ssWW} =
\frac{g^2}{32\,\pi^2}\,\delta^{\hmu\hnu}\,\Bigl[
\frac{1}{3}\,( 1 + 2\,{\cal E}_p )\,\htp^2 + (m_{uf}^2 + m_{df}^2)\Bigr],
\eqa
etc, where $N^c_f$ is the fermion color factor and where we have introduced 
the evanescent operator
\bq
\brp^2 = (4-n)\,{\cal E}_p\,\htp^2.
\eq
If we expand finite counterterms,
\bq
Z_{\ssW} = 1 + \frac{g^2}{16\,\pi^2}\,\delta Z_{\ssW},
\eq
etc, it is easy to derive a solution,
\bq
\delta Z_{\ssW} = \frac{2}{3}\,\lpar 1 + 2\,{\cal E}_p\rpar,
\quad
\delta Z_{\ssZ} =
\frac{2}{9}\,\lpar \frac{5}{\ctws} - 8\,\stws - 2\rpar\,
\lpar 1 + 2\,{\cal E}_p\rpar,
\eq
etc. Although a solution for self-energies can be obtained it is clear that 
the number of anomalous WSTI is greater than the number of Lagrangian
counterterms, i.e. counterterms specifically associated to parameters and
fields. Even this fact does not pose a serious problem since we are talking
about finite counterterms and, in principle, we can associate an ad hoc
counterterm to each avoidable anomaly. Actually there is more; if all fermion 
loops induce anomalies then the argument of \eqn{epequal} is violated and we 
must introduce counterterms at $\ord{n-4}$.

Clearly it is not an ideal solution. After considerable wrangling one is lead
to the conclusion that the only sensible solution is the one proposed by
Jegerlehner~\cite{Jegerlehner:2000dz}: $n$-dimensional $\gamma$-algebra with 
strictly anti-commuting $\gamma^5$ together with $4$-dimensional treatment of 
the hard anomalies.
\section{The fine structure constant \label{FEqalpha}}
To discuss the effect of radiative corrections on $\alpha$ and its interplay 
with renormalization we need the photon propagator, including two-loop 
contributions; using the results obtained in Sect.~5 of I, we write 
\bq
D_{\ssA\ssA}(p^2) = \stws\,\sum_{n=1,2}\,\lpar \frac{g^2}{16\,\pi^2}\rpar^n\,
\Pi^{(n)}_{\ssQQ\ext}(p^2)\,p^2,
\eq
where the external LQ decomposition of self-energies in the neutral sector
has been introduced and discussed in Sect.~6 of I (Eqs.(123)--(125)). Using 
$e^2= g^2\,\stws$ we write
\bq
\Pi_{\ssQQ}(p^2) = \frac{e^2}{16\,\pi^2}\,\Bigl[
\Pi^{(1)}_{\ssQQ\ext}(p^2\,,\,\{m\}) +
\frac{e^4}{16\,\pi^2\,\stws}\,
\Pi^{(2)}_{\ssQQ\ext}(p^2\,,\,\{m\}\,,\,\stws)\Bigr],
\eq
where $\{m\}$ denotes the full set of masses, including bosons;
note that the two-loop SM contribution to the photon self-energy is not 
proportional to $e^4$; as a consequence, charge renormalization does not
decouple from the remaining REs.
Finally, we construct $\Pi_{\ssQQ}(0)$, as illustrated in II; for this 
operation several steps are needed: although one could start from the very 
beginning with $p^2 = 0$ where all diagrams are vacuum bubbles that can be 
reduced by integration-by-part techniques~\cite{Tkachov:1981wb} (but see also 
ref.~\cite{ftes}) to one master integral or products of one-loop functions, we 
prefer to perform the limit $p^2 \to 0$ of the output of $\GS$ which 
represents a highly non-trivial test of the procedure. It is useful to define
\bq
{\hat \Pi}^{(2)}_{\ssQQ\ext} = \frac{1}{\stws}\,\Pi^{(2)}_{\ssQQ\ext},
\eq
and our renormalization equation reads as follows:
\bq
{\hat \Pi}^{(2)}_{\ssQQ}(0)\,\frac{e^4}{256\,\pi^2} +
\Bigl[ \frac{\Pi^{(1)}_{\ssQQ}(0)}{16\,\pi^2} +
\frac{1}{4\,\pi\,\alpha}\Bigr]\,e^2 - 1 = 0,
\label{FSCeq}
\eq
where $\alpha$ is the fine structure constant. Strictly speaking, our 
renormalization equations form a set of coupled equations; at one loop
there is no residual dependence on the weak-mixing angle once we write
$e^2 = g^2 \stws$, but the two-loop contribution modifies this simple 
structure.
In this case, however, we should simply insert the lowest order result
for $\stws$-dependent terms of $\ord{e^4}$ in \eqn{FSCeq}. What to choose 
depends on the IPS that we select; postponing, for a moment, this question we
obtain
\bq
\frac{e^2}{4\,\pi\,\alpha} = 1 - 
\frac{\alpha}{4\,\pi}\,\Pi^{(1)}_{\ssQQ}(0) +
\frac{\alpha^2}{16\,\pi^2}\,\Bigl\{ \Bigl[ \Pi^{(1)}_{\ssQQ}(0)\Bigr]^2 -
\Pi^{(2)}_{\ssQQ\ext}(0,{\bar s}^2)\Bigr\}.
\label{FSCrenEq}
\eq
The value of ${\bar s}^2$ in the argument of the two-loop corrections is fixed
by the corresponding lowest order solution, for which we will need the 
remaining two renormalization equations.

The total contribution to vacuum polarization is split into several components:
\bq
\Pi_{\ssQQ}(0) =  \Pi^{\rm bos}_{\ssQQ}(0) +
\Pi^{\rm lep}_{\ssQQ}(0) +
\Pi^{\rm per}_{\ssQQ}(0) +
\Pi^{\rm had}_{\ssQQ}(0),
\eq
where the fermionic part contains three lepton generations, a perturbative
quark contribution and a non-perturbative one. The latter, associated to
diagrams where a light quark couple to a photon, is related to
$\Delta \alpha^{5}_{\rm had}(\mzs)$~\cite{Eidelman:1995ny}. 
The top contribution at one loop and two loops can be computed in 
perturbation theory due to the large scale of the top mass; always at 
two loops, diagrams where quarks are coupled internally to vector bosons 
are also computed perturbatively. However, QED and QCD contributions to the 
light-quark part are always subtracted.

There are important tests on the result once the limit $p^2 \to 0$ has been
taken. QED is always included in the leptonic sector and collinear logarithms 
are present in the final answer. However, as it is well-known, they 
are of sub-leading nature, i.e. the correction factor is proportional to 
$g^4\,\ln m^2_l$ and not to $g^4\,\ln^2 m^2_l$; indeed, the leading logarithms 
are controlled by the renormalization group equations and are related to Dyson
re-summation of one-loop diagrams.

For the quark sector we introduce a perturbative contribution and a        
non-perturbative one associated to diagrams where a light quark couple to a 
photon. Consider a doublet of light quarks, we subtract the QED part from the
total and give the perturbative component; this means that for each up and 
down quark we neglect the two loop-diagrams which are built with all possible 
insertions of a photon line in a loop of light quarks and we also neglect the 
QED component in the light quark mass renormalization. 
It is worth noting that the subtracted terms are gauge invariant. Since QED 
has been subtracted we do not expect collinear logarithms.

We also have a perturbative heavy-light contribution where the up quark is the
top. In this case only the QED component of the $b$-quark is subtracted and we 
do not neglect the top mass. Note that logarithms of the mass of the $b$ 
quark are expected to disappear from the answer, as for any other light quark.
In the following section we sketch how the result is constructed in QED.
\subsection{The QED case}
To understand renormalization at the two-loop level we consider first the case
of pure QED where we have
\bq
\Pi_{\ssQ\ssE\ssD}(s,m) = \frac{e^2}{16\,\pi^2}\,\Pi^{(1)}_{\ssQQ}(s,m) + 
\frac{e^4}{256\,\pi^4}\,\Pi^{(2)}_{\ssQQ}(s,m),
\eq
where $p^2= -s$ and where we have indicated a dependence of the result on 
the (bare) electron mass. Suppose that we compute the two-loop contribution
($3$ diagrams) in the limit $m = 0$. The result is
\bq
\Pi^{(2)}_{\ssQQ}(s,0)= -\,\frac{4}{\ep} + \ord{1},
\eq
where $n= 4 - \ep$. This is a well-known result which shows the cancellation of
the double ultraviolet pole as well as of any non-local residue. The latter 
result is related to the fact that the four one-loop diagrams with one-loop
counterterms cancel due to a Ward identity. Let us repeat the calculation with
a non-zero electron mass; after scalarization of the result we consider
the ultraviolet divergent parts of the various diagrams. Collecting all the 
terms we obtain
\bq
\Pi^{(2)}_{\ssQQ}(s,m) = - \frac{1}{\ep}\,\Bigl[
4\,\lpar 1 + 24\,\frac{m^2}{s} \rpar +
192\,\frac{m^4}{s^2}\,\frac{1}{\beta(\mu^2)}\,
\ln\frac{\beta(\mu^2) + 1}{\beta(\mu^2) - 1}\Bigr] + \ord{1}.
\label{scaledep}
\eq
Note that the $m$ dependent part is not only finite but also zero in the 
limit $s \to 0$; indeed, in the limit $s \to 0$ and with $\mu^2 = m^2/s -
i\,\delta$ we have
\bq
\beta = \Bigl( 1 - 4\,\mu^2\Bigr)^{1/2} = 
2\,i\,\mu - \frac{i}{2\,\mu} + \ord{\mu^{-2}},
\qquad 
\frac{1}{\beta}\,\ln\frac{\beta+1}{\beta-1} = -\,\frac{1}{2\,\mu^2} +
\ord{mu^{-3}},
\eq
so that
\bq
\Pi^{(2)}_{\ssQQ}(0,m) = -\,\frac{4}{\ep} + 
\Pi^{(2)}_{\ssQQ\,;\,\rm fin}(0,m).
\label{fite}
\eq
\eqn{fite} is the main ingredient to build our renormalization equation
and contains only bare parameters, respecting the true spirit of the 
renormalization equations that express a measurable input ($\alpha$ in this 
case) in terms of bare parameters ($e$ and $m$ in this case) and of 
ultraviolet singularities (this last aspect can be avoided by introducing 
counterterms).

To make a prediction, the running of $\alpha$ in this case, is a different 
issue and, actually, does not depend on the introduction of counterterms: the 
scattering of two charged particles is proportional to
\bq
\frac{e^2}{1 - f(s)} = e^2\,\Bigl[ 1 + f(s) + f^2(s) + \,\cdots\,\Bigr],
\qquad
f(s) =  \frac{e^2}{16\,\pi^2}\,\Pi^{(1)}_{\ssQQ}(s) +
\frac{e^4}{(16\,\pi^2)^2}\,\Pi^{(2)}_{\ssQQ}(s) + \ord{e^6}.
\eq
Renormalization amounts to substituting
\bq
e^2 = 4\,\pi\,\alpha - \alpha^2\,\Pi^{(1)}_{\ssQQ}(0) +
\frac{\alpha^3}{4\,\pi}\,\Bigl\{\Bigl[ \Pi^{(1)}_{\ssQQ}(0)\Bigr]^2 - 
\Pi^{(2)}_{\ssQQ}(0) \Bigr\} + \ord{\alpha^4},
\eq
with the following result:
\bqa
\frac{e^2}{1 - f(s)} &=&  4\,\pi\,\alpha\,\Bigl\{ 1 +
\frac{\alpha}{4\,\pi}\,\Pi^{(1)}_{\ssR}(s) +
\lpar \frac{\alpha}{4\,\pi} \rpar^2\,\Bigl[
\Pi^{(1)}_{\ssR}(s)\,\Pi^{(1)}_{\ssR}(s) + \Pi^{(2)}_{\ssR}(s) +
\ord{\alpha^3}\Bigr\},
\nl
\Pi^{(n)}_{\ssR}(s) &=& \Pi^{(n)}_{\ssQQ}(s) - \Pi^{(n)}_{\ssQQ}(0).
\eqa
To have an ultraviolet finite result we also need that the poles in 
$\Pi^{(n)}_{\ssQQ}(s)$ should not depend on the scale $s$. This is 
obviously true for the one-loop result but what is the origin of the 
scale-dependent extra term in \eqn{scaledep}? One should take into account 
that
\bqa
\Pi^{(1)}_{\ssQQ}(s,m) &=& -\,\frac{8}{3}\,\frac{1}{\ep}      
+ \frac{4}{3}\,\Bigl[ \ln\frac{m^2}{M^2} + ( 1 + 2\,\frac{m^2}{s})\,
\beta(\frac{m^2}{s})\,\ln\frac{\beta(m^2/s)+1}{\beta(m^2/s)-1}\Bigr] 
- \frac{20}{9} + \frac{4}{3}\,\DUV 
- \frac{16}{3}\,\frac{m^2}{s},
\eqa
and that $m$ is the bare electron mass. To proceed further we introduce
a renormalized electron mass (not to be confused with the physical one) which 
is given by
\bq
m = m_{\ssR}\,\Bigl[ 1 + \frac{e^2}{16\,\pi^2}\,\lpar -\,\frac{6}{\ep} +
\mbox{finite part} \rpar\Bigr].
\label{notyet}
\eq
If we write $m^2 = m^2_{\ssR}\,(1 + \delta)$ then 
\bqa
\beta(\frac{m^2}{s}) &=& \beta(\frac{m^2_{\ssR}}{s}) - 
2\,\frac{m^2_{\ssR}}{\beta(\frac{m^2_{\ssR}}{s})\,s}\,\delta + \ord{\delta^2},
\nl
\ln\frac{\beta(m^2/s)+1}{\beta(m^2/s)-1} &=& 
\ln\frac{\beta(m^2_{\ssR}/s)+1}{\beta(m^2_{\ssR}/s)-1} - 
\frac{\delta}{\beta(m_{\ssR})} + \ord{\delta^2}.
\eqa
Inserting this expansion into our results we obtain
\bqa
\Pi_{\ssQ\ssE\ssD}(s,m_{\ssR}) &=& 
\frac{e^2}{\pi^2}\,\Bigl[
 \frac{1}{12}\,\Bigl( \DUV - \frac{2}{\ep}\Bigr)
+ \frac{1}{12}\,\ln\frac{m^2_{\ssR}}{M^2}
+ \frac{1}{3}\,\lpar \frac{1}{4} - \frac{1}{2}\,\frac{m^2_{\ssR}}{s} -
2\,\frac{m^4_{\ssR}}{s^2}\rpar\,\frac{1}{\beta(m^2_{\ssR}/s)}\,
\ln\frac{\beta(m^2_{\ssR}/s)+1}{\beta(m^2_{\ssR}/s)-1} - 
\nl
{}&-& \frac{5}{36}  - 
\frac{1}{3}\,\frac{m^2_{\ssR}}{s}\Bigr]
+ \frac{e^4}{\pi^4}\,\Bigl[ - \frac{1}{64\,\ep} +
\frac{1}{256}\,\Pi^{(2)}_{\rm fin}(s,m_{\ssR})\Bigr],
\eqa
showing cancellation of the ultraviolet poles in 
$\Pi^{(n)}_{\ssR}(s,m_{\ssR})$ with $n = 1,2$.
\subsection{The standard model case}
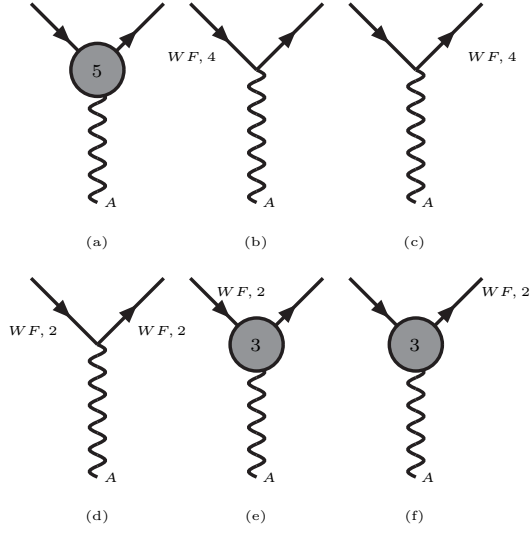
\begin{figure}[ht]
\begin{center}
\SetScale{1}
\SetWidth{1.3}
\begin{eqnarray*}
\hbox{\begin{picture}(180,100)(0,0)
\ArrowLine(0,100)(25,75)
\ArrowLine(25,75)(50,100)
\Photon(25,75)(25,25){3}{6}
\CCirc(25,75){10}{Black}{Gray}
\Text(25,75)[]{\scriptsize{$5$}}
\Text(30,25)[]{\tiny{$A$}}
\Text(25,10)[]{\tiny{(a)}}
\ArrowLine(60,100)(85,75)
\ArrowLine(85,75)(110,100)
\Photon(85,75)(85,25){3}{6}
\Text(70,80)[r]{\tiny{$WF,4$}}
\Text(90,25)[]{\tiny{$A$}}
\Text(85,10)[]{\tiny{(b)}}
\ArrowLine(120,100)(145,75)
\ArrowLine(145,75)(170,100)
\Photon(145,75)(145,25){3}{6}
\Text(165,80)[l]{\tiny{$WF,4$}}
\Text(150,25)[]{\tiny{$A$}}
\Text(145,10)[]{\tiny{(c)}}
\end{picture}}\\
\hbox{\begin{picture}(180,100)(0,0)
\ArrowLine(0,100)(25,75)
\ArrowLine(25,75)(50,100)
\Photon(25,75)(25,25){3}{6}
\Text(10,80)[r]{\tiny{$WF,2$}}
\Text(40,80)[l]{\tiny{$WF,2$}}
\Text(30,25)[]{\tiny{$A$}}
\Text(25,10)[]{\tiny{(d)}}
\ArrowLine(60,100)(85,75)
\ArrowLine(85,75)(110,100)
\Photon(85,75)(85,25){3}{6}
\CCirc(85,75){10}{Black}{Gray}
\Text(85,75)[]{\scriptsize{$3$}}
\Text(70,95)[l]{\tiny{$WF,2$}}
\Text(90,25)[]{\tiny{$A$}}
\Text(85,10)[]{\tiny{(e)}}
\ArrowLine(120,100)(145,75)
\ArrowLine(145,75)(170,100)
\Photon(145,75)(145,25){3}{6}
\CCirc(145,75){10}{Black}{Gray}
\Text(145,75)[]{\scriptsize{$3$}}
\Text(170,95)[l]{\tiny{$WF,2$}}
\Text(150,25)[]{\tiny{$A$}}
\Text(145,10)[]{\tiny{(f)}}
\end{picture}}
\end{eqnarray*}
\end{center}
\caption[]{
Two-loop process-dependent corrections to Rutherford scattering not involving 
propagator corrections.
}
\label{fig:alpha:twoloop1}
\end{figure}

Quite often \eqn{FSCrenEq} is presented without additional theoretical
support. The correct statement is that $\alpha$ is defined from Thompson
scattering at zero momentum transfer. At two loops there are four
classes of diagrams contributing to the process:
\begin{itemize}

\item[I] Irreducible two-loop vertices and wave-function factors, product
of one-loop corrected vertices with one-loop wave-function factors;

\item[II] one-loop vacuum polarization $\,\otimes\,$ one-loop vertices or
one-loop wave-function factors;

\item[III] irreducible two-loop $AA, AZ, A\phi^0$ transitions;

\item[IV] reducible two-loop $AA, AZ, A\phi^0$ transitions.

\end{itemize}
The various contributions are depicted in 
\figs{fig:alpha:twoloop1}{fig:alpha:twoloop4}, where we have made a 
distinction between process-dependent and universal corrections.

Using $\GS$ we have generated the whole set of corrections, I - IV, at 
$p^2 = 0$ including all special vertices (see appendices A -- C of I) . 
By algebraic methods. i.e. full reduction of tensor structures without using 
the explicit expressions for the scalar integrals, we have verified that 
\bei
\item {\em the non-vanishing contribution originates from III and IV only 
and, within these terms, only the reducible and irreducible $AA$ transition 
survives. For this result the role of $\Gamma$ (see Sect.~8 of II) is vital.}
\eei
\begin{figure}[h]
\begin{center}
\SetScale{1}
\SetWidth{1.3}
\begin{picture}(190,130)(-10,-10)
\ArrowLine(0,100)(25,75)
\ArrowLine(25,75)(50,100)
\Photon(25,50)(25,25){3}{3}
\Line(25,50)(25,75)
\CCirc(25,75){8}{Black}{Gray}
\CCirc(25,50){8}{Black}{Gray}
\Text(25,75)[]{\scriptsize{$3$}}
\Text(25,50)[]{\scriptsize{$2$}}
\Text(30,25)[]{\tiny{$A$}}
\Text(25,10)[]{\tiny{(g)}}
\Text(65,105)[b]{\tiny{$WF,2$}}
\ArrowLine(60,100)(85,75)
\ArrowLine(85,75)(110,100)
\Photon(85,50)(85,25){3}{3}
\Line(85,50)(85,75)
\CCirc(85,50){8}{Black}{Gray}
\Text(85,50)[]{\scriptsize{$2$}}
\Text(90,25)[]{\tiny{$A$}}
\Text(85,10)[]{\tiny{(h)}}
\ArrowLine(120,100)(145,75)
\ArrowLine(145,75)(170,100)
\Text(170,105)[b]{\tiny{$WF,2$}}
\Photon(145,50)(145,25){3}{3}
\Line(145,50)(145,75)
\CCirc(145,50){8}{Black}{Gray}
\Text(145,50)[]{\scriptsize{$2$}}
\Text(150,25)[]{\tiny{$A$}}
\Text(145,10)[]{\tiny{(i)}}
%
\end{picture}
\end{center}
\caption[]{Mixed corrections to two-loop fine-structure-constant 
renormalization equation involving propagator, vertex and wave-function 
corrections. Solid lines stand for a photon, a $Z$ boson or a Higgs-Kibble 
scalar.}
\label{fig:alpha:twoloop2}
\end{figure}
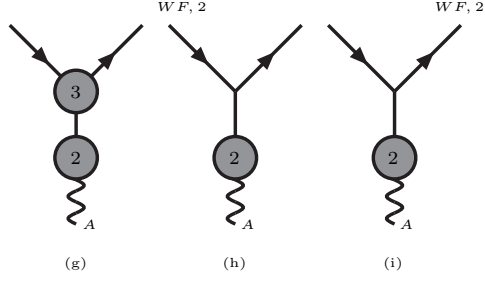
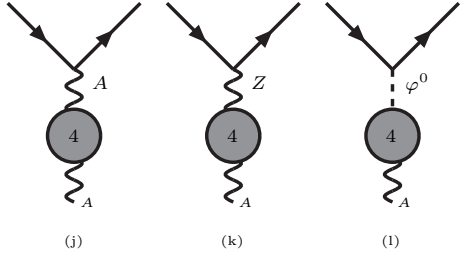
\begin{figure}[ht]
\begin{center}
\SetScale{1}
\SetWidth{1.3}
\begin{picture}(180,120)(0,0)
\ArrowLine(0,100)(25,75)
\ArrowLine(25,75)(50,100)
\Photon(25,75)(25,25){3}{6}
\CCirc(25,50){10}{Black}{Gray}
\Text(30,25)[]{\tiny{$A$}}
\Text(25,10)[]{\tiny{(j)}}
\ArrowLine(60,100)(85,75)
\ArrowLine(85,75)(110,100)
\Photon(85,75)(85,25){3}{6}
\CCirc(85,50){10}{Black}{Gray}
\Text(90,25)[]{\tiny{$A$}}
\Text(85,10)[]{\tiny{(k)}}
\ArrowLine(120,100)(145,75)
\ArrowLine(145,75)(170,100)
\DashLine(145,75)(145,50){3}
\Photon(145,25)(145,50){3}{3}
\CCirc(145,50){10}{Black}{Gray}
\Text(150,25)[]{\tiny{$A$}}
\Text(145,10)[]{\tiny{(l)}}
\Text(25,50)[]{\scriptsize{$4$}}
\Text(85,50)[]{\scriptsize{$4$}}
\Text(145,50)[]{\scriptsize{$4$}}
\Text(35,70)[]{\scriptsize{$A$}}
\Text(95,70)[]{\scriptsize{$Z$}}
\Text(155,70)[]{\scriptsize{$\varphi^0$}}
%
\end{picture}
\end{center}
\caption[]{Process-independent irreducible corrections contributing 
to the two-loop renormalization equation for the fine-structure constant.}
\label{fig:alpha:twoloop3}
\end{figure}
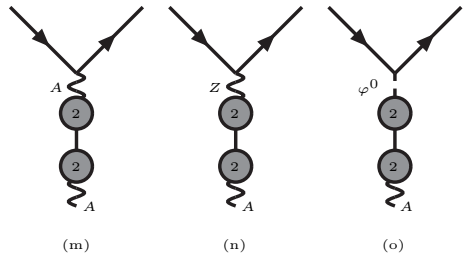
\begin{figure}[b]
\begin{center}
\SetScale{1}
\SetWidth{1.3}
\begin{picture}(200,120)(-20,0)
\ArrowLine(0,100)(25,75)
\ArrowLine(25,75)(50,100)
\Photon(25,25)(25,45){3}{3}
\Photon(25,75)(25,55){3}{3}
\Line(25,55)(25,45)
\ArrowLine(60,100)(85,75)
\ArrowLine(85,75)(110,100)
\Photon(85,25)(85,45){3}{3}
\Photon(85,75)(85,55){3}{3}
\Line(85,55)(85,45)
\ArrowLine(120,100)(145,75)
\ArrowLine(145,75)(170,100)
\Photon(145,25)(145,45){3}{3}
\DashLine(145,75)(145,55){3}
\Line(145,55)(145,45)
%
\CCirc(25,40){6}{Black}{Gray}
\CCirc(25,60){6}{Black}{Gray}
\CCirc(85,40){6}{Black}{Gray}
\CCirc(85,60){6}{Black}{Gray}
\CCirc(145,40){6}{Black}{Gray}
\CCirc(145,60){6}{Black}{Gray}
\Text(25,40)[]{\tiny{$2$}}
\Text(85,40)[]{\tiny{$2$}}
\Text(145,40)[]{\tiny{$2$}}
\Text(25,60)[]{\tiny{$2$}}
\Text(85,60)[]{\tiny{$2$}}
\Text(145,60)[]{\tiny{$2$}}
%
%
\Text(20,70)[r]{\tiny{$A$}}
\Text(80,70)[r]{\tiny{$Z$}}
\Text(140,70)[r]{\tiny{$\varphi^0$}}
\Text(30,25)[]{\tiny{$A$}}
\Text(25,10)[]{\tiny{(m)}}
\Text(90,25)[]{\tiny{$A$}}
\Text(85,10)[]{\tiny{(n)}}
\Text(150,25)[]{\tiny{$A$}}
\Text(145,10)[]{\tiny{(o)}}
\end{picture}
\end{center}
\caption[]{Process-independent reducible corrections contributing 
to the two-loop renormalization equation for the fine-structure constant.}
\label{fig:alpha:twoloop4}
\end{figure}
To present results in the full standard model we introduce auxiliary
quantities,
\bq
\xW = \frac{\mws}{s}, \qquad x_i = \frac{m^2_i}{\mws}, 
\qquad f^1_{-1}(M)= \frac{1}{\ep} - \frac{1}{2}\,\DUV(M^2),
\qquad L_{\beta} = \ln\frac{\beta+1}{\beta-1}, 
\eq
and obtain the following results:

\noindent
$-$ bosonic part
\bqa
\Pi^{(1)}_{\rm bos} &=&
    6\,f^1_{-1}(M) + 8\,\xW -
    (3 + 4\,\xW)\,\beta(\xW)\,L_{\beta}(\xW),
\eqa

\noindent
$-$ leptonic part
\bqa
\Pi^{(1)}_l &=&
  -\,\frac{8}{3}\,f^1_{-1}(M) 
  - \frac{20}{9}\,z_l
  + \frac{4}{3}\,\ln x_l
  + \frac{4}{3}\,( 1 + 2\,z_l )\,\beta(z_l)\,
    L_{\beta}(z_l),
\eqa

\noindent
$-$ top - bottom contribution
\bqa
\Pi^{(1)}_{tb} &=&
  - \frac{40}{9}\,f^1_{-1}(M)
  - \frac{100}{27} 
  + 4\,\sum_{f=b,t}\,Q^2_f\,\Bigl[ \ln x_f - z_f
+ ( 1 + 2\,z_f)\,\beta(z_f)\,L_{\beta}(z_f)\Bigr], 
\eqa
with $z_i = m^2_i/s$. In the limit $s \to 0$ the vacuum polarization becomes
\bq
\Pi^{(1)}_{\rm bos}(0) = 6\,f^1_{-1}(M),
\eq
\bq
\Pi^{(1)}_l(0) = -\,\frac{8}{3}\,f^1_{-1}(M) + \frac{4}{9} + 
\frac{4}{3}\,\ln x_l,
\quad
\Pi^{(1)}_{tb}(0) = -\,\frac{40}{9}\,f^1_{-1}(M)
 + \frac{20}{27} + 4\,\sum_{f=b,t}\,Q^2_f\,\ln x_f.
\eq
First we consider fermion mass renormalization, obtaining
\bq
m^2_f= m^2_{f\,\ssR}\,
\Bigl( 1 + 2\,\frac{g^2}{16\,\pi^2}\,\frac{\delta Z^f_m}{\ep}\Bigr),
\eq
with renormalization constants given in Sect.~5 of II.
Consider the fermionic part of $\Pi^{(1)}$ relative to one fermion generation
($\nu_l, l, t$ and $b$) and perform fermion mass renormalization; we obtain
\bq
\Pi^{(1)}_{\rm fer} \to \Pi^{(1)}_{\rm ferm} + 
\frac{g^2}{\pi^2\,\ep}\,\Delta \Pi^{(1)}_{\rm ferm}.
\eq
When we add the two-loop result we obtain 
\bq
\frac{g^2}{16\,\pi^2}\,\Pi^{(1)}_{\rm fer} +
\frac{g^4}{(16\,\pi^2)^2}\,\Pi^{(2)} =
\mbox{one loop} + 
\frac{g^4}{\pi^4}\,\Bigl[ R^{(2)}\,\ep^{-2} +
R^{(1)}\,\ep^{-1} + \Pi_{\rm fin} \Bigr].
\eq
The two residues are given by
\bq
R^{(2)} = -\,\frac{11}{256},
\eq
\bqa
R^{(1)} &=&
\frac{11}{256}\,\DUV
          + \frac{407}{27648}
          + \frac{9}{64} \ctw^{-4} \frac{\xW}{\xH}
          - \frac{9}{128} \ctw^{-2} \xW
          - \frac{131}{6912} \ctw^{-2}
\nl
{}&+& \frac{3}{64}\,\xW\,\lpar 
            \xL
          - 4 \frac{\xLs}{\xH}
          + 3 \xB
          - 12 \frac{\xBs}{\xH}\rpar
+ \frac{9}{128} \xW\,\lpar 
            2 \xT
          - 8 \frac{\xTs}{\xH}
          + 4 \frac{1}{\xH}
          + \xH\rpar
\nl
{}&+&     \frac{1}{32} \xW
          + \frac{3}{512} \xL
          + \frac{7}{1536} \xB
          + \frac{13}{1536} \xT
+ \beta^{-1}(\xW) L_{\beta}(\xW)\,\Bigl[
          - \frac{11}{768}
          + \frac{3}{64} \ctw^{-4} \frac{\xW}{\xH}\,(1 + 6\,\xW)
\nl
{}&-& \frac{1}{128}\, \xW\,\lpar  
            4 \ctw^{-2} 
          + 18 \ctw^{-2} \xW
          - 3 \xL
          + 8 \frac{\xLs}{\xH}
          - 9 \xB\rpar
- \frac{3}{128}\,\xW\, \lpar 
            8 \frac{\xBs}{\xH}
          - 3 \xT
          + 8 \frac{\xTs}{\xH}
          - 4 \frac{1}{\xH}\rpar
\nl
{}&+& \frac{1}{384}\,\xW\,\lpar 
            9 \xH
          - 13
          + 36 \xW \xL
          - 144 \xW \frac{\xLs}{\xH}\rpar
+ \frac{9}{32}\,\xWs\,\lpar 
           \xB
          - 4 \frac{\xBs}{\xH}
          + \xT
          - 4 \frac{\xTs}{\xH}\rpar
\nl
{}&+& \frac{1}{64}\,\xWs\lpar 
            \frac{36}{\xH}
          + 9 \xH
          + 4 \rpar \Bigr].
\eqa
Therefore mass renormalization has removed all logarithms in the residue
of the simple ultraviolet pole for the fermionic part while a non-local
residue remains in the bosonic part.

If we work in the 't Hooft - Feynman gauge a simple procedure of $\wb$ mass 
renormalization is not enough to get rid of logarithmic residues in the 
bosonic component and the reason is that in a bosonic loop we may have 
three different fields, the $\wb$, the $\phi$ and the charged ghosts and 
only one mass is available, $M$.  

The situation is illustrated in \fig{wmren} where the cross denotes insertion
of a counterterm $\delta Z_{\ssM}$; the latter is fixed to remove the 
ultraviolet pole in the $\wb$ self-energy and one easily verifies that the 
total in the second and third line of \fig{wmren} ($\phi$ and $X$ 
self-energies, respectively) is not ultraviolet finite.

The procedure has to be changed if we want to make the result in the bosonic 
sector as similar as possible to the one in the fermionic sector.
Also with this goal in mind we have introduced the $R_{\xi\xi}$ gauge in
sect.~3 of II.

Collecting all diagrams, renormalizing the $\wb$ mass and inserting the 
solution for the renormalization constants in the $R_{\xi\xi}$ gauge we find 
a convenient expression for the bosonic, one-loop, $AA$ self-energy:
\bq
\Pi^{(1)}_{\rm bos} \to
  6\,f^1_{-1}(M)
  + 6
  + 8 \xW
- \beta^{-1}(\xW) L_{\beta}(\xW)   (
           3
          - 8 \xW
          - 16 \xWs )
+ \frac{g^2}{\pi^2}\,\sum_{l=1,2}\,\Delta \Pi^{(1,l)}_{\rm bos}\,\ep^{-l},
\eq
with a correction given by
\bq
\Delta \Pi^{(1,2)}_{\rm bos} = - \frac{11}{24},
\eq
\bqa
\Delta \Pi^{(1,1)}_{\rm bos} &=&
            \frac{11}{24}
          + \frac{11}{48} \DUV
          - \frac{9}{4} \ctw^{-4} \frac{\xW}{\xH}
          + \frac{9}{8} \ctw^{-2} \xW
\nl
{}&+& 3\,\xW\,\lpar 
            \frac{\xLs}{\xH}
          + 3 \frac{\xTs+\xBs}{\xH}
          - \frac{3}{2} \frac{1}{\xH}
          - \frac{3}{8} \xH
          - \frac{1}{4} \xL \rpar
          - \frac{9}{4}\,\xW\,\lpar 
             \xT + \xB + \frac{2}{9} \rpar
\nl
{}&+& \beta^{-1}(\xW) L_{\beta}(\xW)\,\Bigl[
          \frac{11}{48}
          - \frac{3}{2}\,\xW\,\lpar
            \frac{1}{2} \ctw^{-4} \frac{1}{\xH}
          + 3 \ctw^{-4} \frac{\xW}{\xH}
          - 3 \ctw^{-2} 
          - \frac{3}{2} \ctw^{-2} \xW\rpar
\nl
{}&+& \xW\,\lpar 
              \frac{\xLs}{\xH}
          + 3 \frac{\xTs+\xBs}{\xH}
          - \frac{3}{2} \frac{1}{\xH}
          - \frac{3}{8} (\xH+\xL)\rpar
\nl
{}&+& \xW\,\lpar 
            \frac{13}{24} 
          - \frac{9}{8} (\xT+\xB)
          + 6 \frac{\xW \xLs}{\xH}
          + 18 \frac{\xW(\xTs+\xBs)}{\xH}\rpar
\nl
{}&-& 3\,\xWs\,\lpar 
            \frac{3}{\xH}
          + \frac{3}{4} \xH
          + \frac{1}{2} \xL
          + \frac{3}{2} (\xT+\xB)
          + 3 \rpar
          \Bigr]
\eqa
Including both components and taking into account the additional contribution 
arising from renormalization we finally get residues for the ultraviolet poles 
which show the expected properties:
\bqa
R^{(2)} &=& -\,\frac{55}{768},
\nl
R^{(1)} &=&
\frac{11}{192}\,\DUV
 + \frac{1199}{27648}
 - \frac{131}{6912} \ctw^{-2}
 + \frac{3}{512} \xL
 + \frac{13}{1536} \xT
 + \frac{7}{1536} \xB.
\label{residues}
\eqa
\eqn{residues} shows complete cancellation of poles with a logarithmic residue;
furthermore the two residues in \eqn{residues} are scale independent and
cancel in the difference $\Pi(p^2) - \Pi(0)$. 

A final comment concerns the $\zb$-photon transition which is not zero, at 
$p^2 = 0$, in any gauge where $\xi \not= 1$ even after the $\Gamma_1$ 
re-diagonalization procedure. However, in our case, the non-zero
result shows up only due to a different renormalization of the two bare gauge 
parameters and it is, therefore, of $\ord{g^4}$; it can be absorbed into 
$\Gamma_2$ which does not modify our result for $\Pi$ since there are no 
$\Gamma_2$-dependent terms in the $AA$ transition (only $\Gamma^2_1$ appears).

It is important to stress that in computing $\Pi$ one-loop field and
charge counterterms are irrelevant and only mass and gauge parameter
renormalization matters.

\eqn{notyet} is not yet a true renormalization equation since the latter 
should contain the physical electron mass $m_e$ and not the intermediate 
parameter $m_{\ssR}$ but the relation between the two is ultraviolet finite. 
All of this is telling us that a renormalization equation has the structure
\bq
p_{\rm phys} = f \lpar \frac{1}{\ep}\,,\,p_{\rm bare}\rpar,
\eq
where the residue of the ultraviolet poles must be local. A prediction,
\bq
O\lpar \frac{1}{\ep}\,,\,p_{\rm bare} \rpar \equiv O( p_{\rm phys} ),
\eq
gives a finite quantity that can be computed in terms of some input parameter
set. 

Note that we could follow an altogether different approach, avoiding 
the explicit introduction of counterterms; instead of the chain
bare $\,\to\,$ renormalized $\,\to\,$ physical we simply drop the second
step, bare $\,\to\,$ physical.
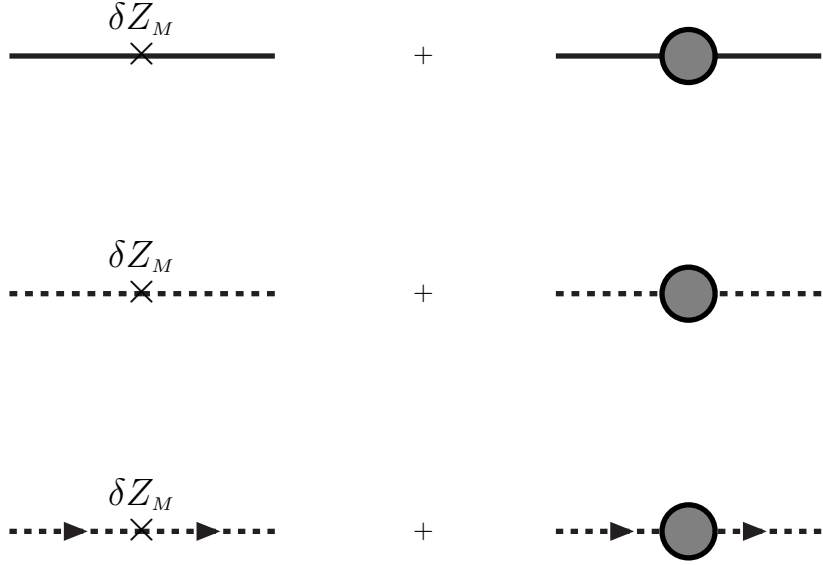
\begin{figure}[th]
\vspace{2.5cm}
\bqas  
  \vcenter{\hbox{
  \begin{picture}(150,0)(0,0)
  \SetScale{1.0}
  \SetWidth{2.}
  \Line(0,0)(50,0)
  \Line(50,0)(100,0)
  \Text(50,8)[cb]{\Large $\delta Z_{\ssM}$}
  \Text(50,-5)[cb]{\LARGE $\times$}
  \end{picture}}}
&+\qquad\qquad&
  \vcenter{\hbox{
  \begin{picture}(150,0)(0,0)
  \SetScale{1.0}
  \SetWidth{2.}
  \Line(0,0)(50,0)
  \Line(50,0)(100,0)
  \GCirc(50,0){10}{0.5}
  \end{picture}}}
\nl\nl\nl\nl\nl\nl
  \vcenter{\hbox{
  \begin{picture}(150,0)(0,0)
  \SetScale{1.0}
  \SetWidth{2.}
  \DashLine(0,0)(50,0){3.}
  \DashLine(50,0)(100,0){3.}
  \Text(50,8)[cb]{\Large $\delta Z_{\ssM}$}
  \Text(50,-5)[cb]{\LARGE $\times$}
  \end{picture}}}
&+\qquad\qquad&
  \vcenter{\hbox{
  \begin{picture}(150,0)(0,0)
  \SetScale{1.0}
  \SetWidth{2.}
  \DashLine(0,0)(50,0){3.}
  \DashLine(50,0)(100,0){3.}
  \GCirc(50,0){10}{0.5}
  \end{picture}}}
\nl\nl\nl\nl\nl\nl
  \vcenter{\hbox{
  \begin{picture}(150,0)(0,0)
  \SetScale{1.0}
  \SetWidth{2.}
  \DashArrowLine(0,0)(50,0){3.}
  \DashArrowLine(50,0)(100,0){3.}
  \Text(50,8)[cb]{\Large $\delta Z_{\ssM}$}
  \Text(50,-5)[cb]{\LARGE $\times$}
  \end{picture}}}
&+\qquad\qquad&
  \vcenter{\hbox{
  \begin{picture}(150,0)(0,0)
  \SetScale{1.0}
  \SetWidth{2.}
  \DashArrowLine(0,0)(50,0){3.}
  \DashArrowLine(50,0)(100,0){3.}
  \GCirc(50,0){10}{0.5}
  \end{picture}}}
\eqas
\vspace{0.5cm}
\caption[]{$\wb$ mass counterterm insertion for charged transitions in
the 't Hooft - Feynman gauge. While the $\wb\wb$ one is ultraviolet finite the 
same is not true for $\phi\phi$ and ghost-ghost transitions.}
\label{wmren}
\end{figure}
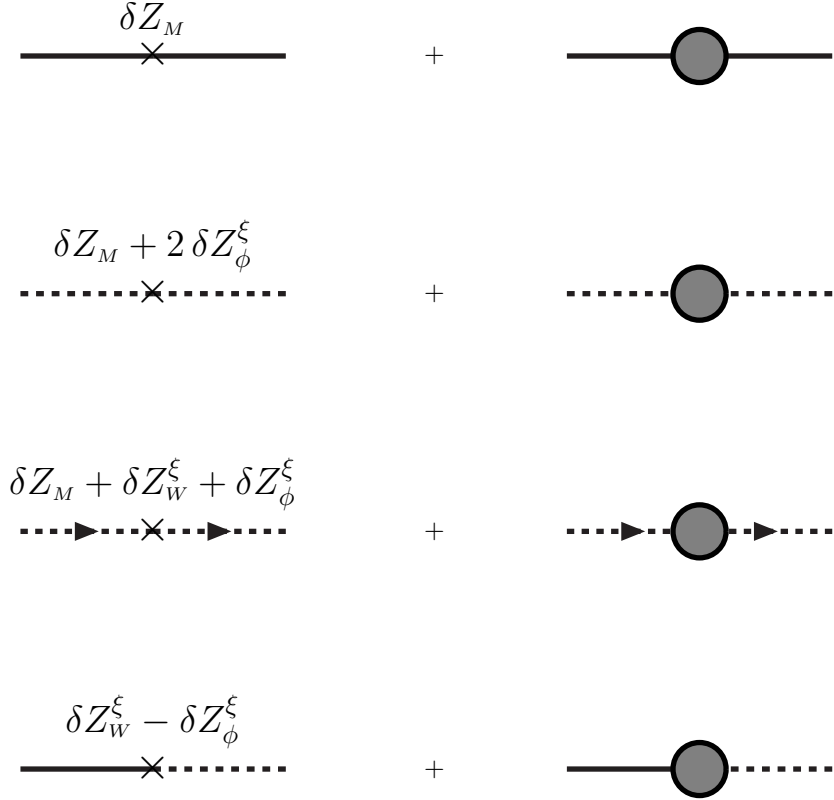
\begin{figure}[th]
\vspace{2.5cm}
\bqas  
  \vcenter{\hbox{
  \begin{picture}(150,0)(0,0)
  \SetScale{1.0}
  \SetWidth{2.}
  \Line(0,0)(50,0)
  \Line(50,0)(100,0)
  \Text(50,8)[cb]{\Large $\delta Z_{\ssM}$}
  \Text(50,-5)[cb]{\LARGE $\times$}
  \end{picture}}}
&+\qquad\qquad&
  \vcenter{\hbox{
  \begin{picture}(150,0)(0,0)
  \SetScale{1.0}
  \SetWidth{2.}
  \Line(0,0)(50,0)
  \Line(50,0)(100,0)
  \GCirc(50,0){10}{0.5}
  \end{picture}}}
\nl\nl\nl\nl\nl\nl
  \vcenter{\hbox{
  \begin{picture}(150,0)(0,0)
  \SetScale{1.0}
  \SetWidth{2.}
  \DashLine(0,0)(50,0){3.}
  \DashLine(50,0)(100,0){3.}
  \Text(50,8)[cb]{\Large $\delta Z_{\ssM} + 2\,\delta Z^{\xi}_{\phi}$}
  \Text(50,-5)[cb]{\LARGE $\times$}
  \end{picture}}}
&+\qquad\qquad&
  \vcenter{\hbox{
  \begin{picture}(150,0)(0,0)
  \SetScale{1.0}
  \SetWidth{2.}
  \DashLine(0,0)(50,0){3.}
  \DashLine(50,0)(100,0){3.}
  \GCirc(50,0){10}{0.5}
  \end{picture}}}
\nl\nl\nl\nl\nl\nl
  \vcenter{\hbox{
  \begin{picture}(150,0)(0,0)
  \SetScale{1.0}
  \SetWidth{2.}
  \DashArrowLine(0,0)(50,0){3.}
  \DashArrowLine(50,0)(100,0){3.}
  \Text(50,8)[cb]{\Large $\delta Z_{\ssM} + \delta Z^{\xi}_{\ssW} + 
\delta Z^{\xi}_{\phi}$}
  \Text(50,-5)[cb]{\LARGE $\times$}
  \end{picture}}}
&+\qquad\qquad&
  \vcenter{\hbox{
  \begin{picture}(150,0)(0,0)
  \SetScale{1.0}
  \SetWidth{2.}
  \DashArrowLine(0,0)(50,0){3.}
  \DashArrowLine(50,0)(100,0){3.}
  \GCirc(50,0){10}{0.5}
  \end{picture}}}
\nl\nl\nl\nl\nl\nl
  \vcenter{\hbox{
  \begin{picture}(150,0)(0,0)
  \SetScale{1.0}
  \SetWidth{2.}
  \Line(0,0)(50,0)
  \DashLine(50,0)(100,0){3.}
  \Text(50,8)[cb]{\Large $\delta Z^{\xi}_{\ssW} - \delta Z^{\xi}_{\phi}$}
  \Text(50,-5)[cb]{\LARGE $\times$}
  \end{picture}}}
&+\qquad\qquad&
  \vcenter{\hbox{
  \begin{picture}(150,0)(0,0)
  \SetScale{1.0}
  \SetWidth{2.}
  \Line(0,0)(50,0)
  \DashLine(50,0)(100,0){3.}
  \GCirc(50,0){10}{0.5}
  \end{picture}}}
\eqas
\vspace{0.5cm}
\caption[]{The $R_{\xi\xi}$ recipe for renormalizing mass dependent ultraviolet
poles in the charged sector.}
\label{wmrenC}
\end{figure}
\section{Running of $\alpha$ beyond one-loop: issues in perspective 
\label{alpharun}}
In the spirit of effective, running, couplings the role played by the running
of $\alpha$ has been crucial in the development of precision tests of the SM.
This is closely related to an hidden thought, universal corrections are the
important ingredient, non-universal ones should be made as small as possible
and represent a (somehow unnecessary) complication to be left for a true 
expert.
Then, universal corrections must be linked to a set of pseudo-observables and
all data should be presented in a way that resembles, as closely as possible,
the language of pseudo-observables. The crucial point is that this language
is intrinsically related to the procedure of resummation, but the latter
comes into conflict with gauge invariance. The whole scheme received further 
boost from precision physics around the $\zb$ mass scale where it is 
relatively easy to perform a discrimination between relevant and (almost) 
irrelevant terms (boxes, for instance, are of little use) paying a very 
little price to gauge invariance.

But what about a definition of the running of $\alpha$ at an arbitrary scale?
What about large energies, where Sudakov logarithms~\cite{Sudakov:1954sw}
come into play? Many words have been spent in order to describe the problem 
and to derive a reasonable solution. The problem is far from trivial since 
large fermionic logarithms must be resummed, according to renormalization 
group arguments.

One idea is to import from QCD the concept of $\MSB$ 
couplings~\cite{Degrassi:2003rw}. QCD is a theory without any obvious 
subtraction point and one defines an $\MSB$ coupling where, 
working in dimensional regularization, poles are thrown away and the arbitrary
unit of mass is promoted to become the relevant scale of the problem. Then, in 
QED, we take advantage of the fact that at $p^2 = 0$ the vacuum polarization 
is gauge invariant, something that can be proved, to all orders, by using 
Nielsen identities~\cite{Gambino:1999ai}. Therefore, the idea is to express 
theoretical predictions by means of a resummed, $\MSB$, 
coupling~\cite{Erler:1998sy}. This choice, is not immune from criticism: 
following common wisdom, it is more physical to use an effective charge
as determined from experiment to define the fundamental coupling.
Nevertheless, we introduce
\bq
\alpha^{-1}_{\MSB}(s) = \alpha^{-1} - 
\frac{1}{4\,\pi}\,\Pi^{\MSB}_{\ssQQ}(0)\bmid_{\mu^2 = s}.
\eq
The gauge parameter independence, at the basis of the $\MSB$ definition,
deserves an additional comment. So long as we are considering the one loop 
case in the $R_{\xi}$ gauge, we have
\bq
\Pi^{(1)}_{\ssQQ\,;\,{\rm bos}}(0) =
(4 - \xi^2)\,\Bigl[\frac{2}{\ep} - \DUV(M^2)\Bigr] - \frac{11}{6}
- \frac{1}{2}\,\xi^2+\Bigl[1 + \frac{3}{\xi^2-1}\Bigr]\,\xi^2\,\ln\xi^2.
\eq
In any gauge where $\Sigma_{3\ssQ}(0) \not= 0$ (for a definition of the LQ
basis see Sect.~9 of I) one has to take into account that the correct factor 
is
\bq
\Pi^{(1)}_{\ssQQ}(0) - \frac{2}{M^2}\,\Sigma^{(1)}_{3\ssQ}(0) =
- \frac{3}{2}\,\Bigl[\frac{2}{\ep} - \DUV(M^2)\Bigr] - \frac{1}{3},
\eq
gauge invariant by inspection. In our case we have
\bq
\Pi_{\ssQQ}(0,\{\xi\}) - \frac{2}{M^2}\,\Sigma_{3\ssQ}(0,\{\xi\}) =
\Pi_{\ssQQ}(0,\{\xi=1\}) - \frac{2}{M^2}\,\Sigma_{3\ssQ}(0,\{\xi=1\}) = 
\Pi_{\ssQQ}(0,\{\xi=1\}),
\eq
due to our $\Gamma$ prescription.

The definition of $\Pi^{\MSB}$ requires some explanation; we introduce the
ultraviolet decomposition
\bq
\Pi_{\ssQQ}(0) =  \sum_{k=-1}^{1}\,
\Pi^{(1)}_{\ssQQ}\lpar 0\,;\,k\rpar\,F^1_k(M^2)+
\sum_{k=-2}^{0}\,\Pi^{(2)}_{\ssQQ}\lpar 0\,;\,k\rpar\,F^2_k(M^2),
\eq
with
\bqa
F^1_{-1}(M^2) &=& \frac{1}{\ep} - \frac{1}{2}\,\DUV + 
\frac{1}{8}\,\DUV^2\,\ep, \quad
F^1_0(M^2) = 1 - \frac{1}{2}\,\DUV\,\ep, \quad
F^1_1(M^2) = \ep,
\nl
F^2_{-2}(M^2) &=& \frac{1}{\ep^2} - \frac{\DUV}{\ep} + \frac{1}{2}\,\DUV^2,
\quad
F^2_{-1}(M^2) = \frac{1}{\ep} - \DUV,
\quad
F^2_0(M^2) = 1,
\eqa
and $\DUV = \gamma + \ln\pi + \ln M^2/\mu^2$. Furthermore, for 
$\Pi^{\rm had}_{\ssQQ}(0)$ we use
\bq
\Pi^{\rm had}_{\ssQQ}(0) = 
\Bigl[ \Pi^{\rm lq}_{\ssQQ}(0) - \Reb\,\Pi^{\rm lq}_{\ssQQ}(s)\Bigr] +
\Reb\,\Pi^{\rm lq}_{\ssQQ}(s),
\eq
where the quantity within brackets is related to
$\Delta \alpha^{5}_{\rm had}(s)$ and the light quark component,
$\Reb\,\Pi^{\rm lq}(s)$, is computed in perturbation theory. The $\MSB$ 
prescription amounts to project $\Pi_{\ssQQ}(0)$ according to
\bq
{\cal P}_{\MSB}\,\ep^{-k} = 0, \quad
{\cal P}_{\MSB}\,\ep = 0, \quad
{\cal P}_{\MSB}\,\DUV = \ln \frac{M^2}{s_{\MSB}}.
\eq
Conventionally we shall use $s_{\MSB} = s$. Our results are shown in 
\tabn{MSBtable}. In our results {\em one-loop} always includes the 
non-perturbative hadronic part; therefore the difference between 
{\em one-loop} and {\em two-loop} is fully perturbative.
We observe mild variations in the range of energies from $\mz$ to $500\,$GeV
induced by a $\pm 5\,$GeV change in $\mt$ an by varying $\mh$ between 
$150\,$GeV and $300\,$GeV.

A possible solution of the puzzle of resummation could be: do the calculation 
in an arbitray gauge, select a gauge parameter independent part of self-energy,
perform resummation while leaving the rest to ensure the same independence
when combined with vertices and boxes. The obvious criticism for this 
procedure is that, although respecting gauge invariance, it violates the 
criterion of uniqueness since any gauge independent quantity can be moved 
back and forward from the resummed part to the non resummed 
one~\cite{Denner:1996gb}. Let us accordingly define
\bq
\frac{\alpha}{\alpha(s)} = 1 + \Delta \alpha(s);
\label{defDalphas}
\eq
our result are shown in \tabn{Fulltable}.
The hadronic, non-perturbative, part has been computed with the help of the 
routine {\tt THADR5} by F.~Jegerlehner~\cite{Jegerlehner:1986vs}; the full
calculation has been performed with $\LB$~\cite{LB}.

Are there ambiguities in the definition of the $\MSB$ parameter? 
There is the question of decoupling of heavy degrees of freedom,
for instance Veltman theorem is violated. One option is to introduce an $\MSB$ 
scheme with decoupling of high degrees of freedom, for instance following
the recipe of~\cite{Marciano:1990dp} according to which the $\ln m_t/\mz$ 
terms in $\Reb\,\Pi_{\gamma\ssZ}(\mzs)$ are subtracted for $m_t > \mz$. More 
generally, the idea is to subtract all contributions that involve particles 
heavier than the $\zb$ boson and do not decouple. Unfortunately, this is 
fully equivalent to shift portions of the result between resummed and 
non-resummed components.

The behavior of $\Reb\,\alpha^{-1}(s)$ for $\sqrt{s} = 200(500)\,\GeV$ and
different values of $\mt, \mh$ is given in \tabn{Compatable}.
In \tabns{Rattablea}{Rattableb} we compare the running of the e.m. coupling 
constant between the $\MSB$-scheme and our option. Sizable differences are 
present for a light $\hb$ boson mass.

To summarize we may say that $\alpha_{\MSB}(s)$ is the only definition of 
the running of the e.m. coupling constant -- beyond one loop -- which is 
anchored to a rigorous, formal, basis; nevertheless, this definition is far 
from physical intuition, a pseudo-observable should always reflect the image 
of a (indirectly-) measured quantity. 

If we disregard the possibility of importing $\MSB$ quantities into the 
electroweak theory only two choices are left: to present the full calculation 
for an arbitrary process, without an attempt to introduce universal corrections
(modifications of the structure of the photon propagator, like in the
Uheling effect) w.r.t. process-dependent corrections, or to introduce
universal corrections according to some convention. The latter choice 
requires agreement in the community.

Any convention should be judged by the quality of the results that can be
predicted. Giving a {\em definition} of the running e.m. coupling constant 
has a link to the idea of introducing some {\em improved Born approximation},
a concept which is also questionable at very high energies.
It is for this reason that we do not give any special emphasis to our numerical
results (if not for the fact they are there, with some degree of novelty).
We observe sizable corrections for a relatively low Higgs mass scenario and
energies well above $\mz$;
the running of $\alpha(s)$ is almost doubled for a Higgs mass below $300\,$Gev
and $\sqrt{s} = 200\,$Gev or higher. At the same time we have studied the
behavior for large values of the Higgs boson mass and $\sqrt{s} < \mh$,
shown in \tabn{mhbe}.

Our definition is also affected by the presence of any threshold, e.g.
the $2\,m_t$ normal threshold (sub-leading Landau singularity) which is
enhanced at two loops by the coupling with a Higgs boson.
Finally and always at two loops we observe a strong correlation between the
range of variability of the top quark and the Higgs boson mass.

A final comment should be devoted to asymptotic limits: for several years 
important results have been obtained for leading and sub-leading results in 
various heavy mass limits, noticeably $\mh$ or $m_t \to \infty$; 
unfortunately, infinity is not always around the corner, and leading effects 
are masked by -- large -- constant terms, as already observed 
by~\cite{Degrassi:2003rw}. 

The observation that asymptopia may affect upper bounds but not central 
values for (pseudo-)observables, should not be confused with a criticism to 
important evolutions in our recent history; only additional progress will 
help in clarifying this issue and we consider our calculation, at non-zero 
momentum, as a tiny step in this direction.
\section{Complex poles: the paradigm \label{CPoles}}
To write additional renormalization equations we also need {\em experimental}
values for gauge boson masses. For the $\wb$ and $\zb$ bosons the IPS is 
defined in terms of pseudo-observables; at first, on-shell quantities are 
derived by fitting the experimental lineshapes with
\bq
\Sigma_{\ssVV}(s) = \frac{N}{(s - M^2_{\OS})^2 + 
s^2\,\Gamma^2_{\OS}/M^2_{\OS}}, \qquad V = W, Z,
\eq
where $N$ is an irrelevant (for our purposes) normalization constant.
Secondly, we define pseudo-observables 
\bq
M_{\ssP} = M_{\OS}\,\cos\psi, \qquad  
\Gamma_{\ssP} = \Gamma_{\OS}\,\sin\psi, \qquad
\psi = \arctan \frac{\Gamma_{\OS}}{M_{\OS}},
\eq
which are inserted in the IPS. At one loop we can use directly the
on-shell masses which are related to the zero of the real part of the inverse
propagator. Beyond one loop this would show a clash with gauge invariance
since only the complex poles
\bq
s_{\ssV} = \mu^2_{\ssV} - i\,\gamma_{\ssV}\,\mu_{\ssV}
\label{moreinit}
\eq
do not depend, to all orders, on gauge parameters.
As a consequence, renormalization equations change their structure.
There is also a change of perspective with respect to old one-loop 
calculations. There one considered the on-shell masses as input parameters
independent of complex poles and {\em derived} the latter in terms of the 
former~\cite{Beenakker:1996kn}. 

Here the situation changes, renormalization equations are written for
real, renormalized, parameters and solved in terms of (among other things) 
experimental complex poles. When we construct a propagator from an IPS that
contains its complex pole, say $s_{\ssV}$, we are left to consider a 
consistency relation between theoretical and experimental values of 
$\gamma_{\ssV}$.
If instead, we derive $\cpw$ from an IPS that contains $\cpz$, this is
a prediction for the full $\wb$ complex pole. Note that there is a conceptual
difference with calculations that relate $\MSB$ and pole masses of gauge 
bosons.

Two type of relations are known in the literature~\cite{Jegerlehner:2003py}:
complex poles in terms of bare (or $\MSB$) masses and their inverse, where 
the $\MSB$-masses are expressed in terms of complex poles. 

Furthermore, consistently with with an order-by-order renormalization 
procedure, renormalized masses in loops and in vertices will be replaced with 
their real solutions of the renormalized equations, truncated to the requested
order. From this point of view there is no problem in our approach with
cutting-equations and unitarity. This scheme can be used for computing
pseudo-observables, including decay widths, but additional complications
arise when we consider processes $2 \to n$ with unstable particles in any
annihilation channel.

Alternatively, one could use Dyson resummed ({\em dressed}) propagators,
\bq
{\bar \Delta}_{\ssV} = 
\frac{\Delta_{\ssV}}{1 - \Delta_{\ssV}\,\Sigma_{\ssVV}},
\eq
also in loops, say two-loop resummed propagators in tree diagrams, one loop 
resummed in one-loop diagrams, tree in two-loop diagrams~\cite{Veltman:1963th}.
Dressed propagators satisfy the K\"allen - Lehmann representation and
processes with external unstable particle should not be considered.
This recipe requires that only skeleton diagrams are included (no insertion
of self-energy sub-loops) and, once again, cutting-equations and unitarity
of the $S\,$-matrix can be proven; we postpone a full discussion to
\sect{UgpIWI} and simply mention that
\bq
{\bar \Delta}^{+}_{\ssV}(p^2) = \theta(p_0)\,
\Bigl[ {\bar \Delta}_{\ssV}(p^2)\Bigr]^2\,
2\,i\,\Reb\,\Sigma_{\ssVV}(p^2),
\eq
while, for a stable particle, the pole term shows up as
\bq
{\bar \Delta}^{+}_{\ssV}(p^2) = \theta(p_0)\,
\Bigl[ {\bar \Delta}_{\ssV}(p^2)\Bigr]^2\,
2\,i\,\Reb\,\Sigma_{\ssVV}(p^2) + 2\,i\,\pi\,\delta(p^2 + m^2_{\ssV}).
\eq
\section{Renormalization equations and their solutions \label{solREe}}
Renormalization with complex poles has more in it than the content of
\eqn{moreinit} and is not confined to prescribe a fixed width for unstable
particles; it allows, al least in principle, for an elegant treatment
of radiative corrections via effective, complex, couplings. The corresponding
formulation, however, cannot be naively extended beyond the fermion loop
approximation~\cite{Beenakker:1996kn}; this is due, once again, to gauge 
parameter independence.
We formulate the next renormalization equation in close resemblance with the 
language of effective couplings and will perform the proper expansions at the
end. At the same time we formulate different choices for IPS.

To proceed further, we also need residual functions defined according to
\bq
\Sigma_{\ssB}(s)= \Sigma_{3\ssQ}(s) + F_{\ssB}(s),
\qquad B= W,Z,\;\;\hbox{and}\;H,
\label{ResF}
\eq
and discuss solutions of the renormalization equations for different IPS.
One of the ingredients in our equations is given by dressed propagators;
for the $W$ boson it has been defined in Eq.(95) of I, whereas propagators
and transitions in the neutral sectors are given in Eqs.(106)--(108) of I.

Furthermore, in Sect.~6 of I we have defined vector boson transitions to all
orders, e.g.
\bq
D_{\ssA\ssA} = \stws\,\Pi_{\ssQ\ssQ\,;\,{\rm ext}}\,p^2 =
\stws\,\sum_{n=1}^{\infty}\,\lpar\frac{g^2}{16\,\pi^2}\rpar^n\,
\Pi^{(n)}_{\ssQ\ssQ\,;\,{\rm ext}}\,p^2,
\eq
etc, distinguishing the $\theta$ dependence originating from external legs and
the one introduced by internal legs. In this section we will drop the
suffix {\em ext}.
\subsection{Running parameters}
As a consequence of introducing higher order corrections the coupling constant 
$g$ will evolve with the scale according to
\bq
\frac{1}{g^2(s)} = \frac{1}{g^2} - \frac{1}{16\,\pi^2}\,\Pi^{(1)}_{3\ssQ}(s) -
\frac{g^2}{(16\,\pi^2)^2}\,\Pi^{(2)}_{3\ssQ}(s).
\label{rung}
\eq
The running of $e^2 = g^2 \stws$ is controlled by
\bq
e^2(s)\,\Bigl[1 - \frac{\alpha}{4\,\pi}\,\Pi_{\ssR}(s)\Bigr] =  
4\,\pi\,\alpha
\label{rune}
\eq
while the running of the weak-mixing angle is defined according to
\bq
s^2(s) = \frac{e^2(s)}{g^2(s)}.
\label{runs}
\eq
\eqns{rung}{runs} still contain renormalized parameters and in the following 
sections we will show how to replace renormalized quantities with 
pseudo-observables of some IPS.

The fact that we use the relation $M_0 = M/\ctw$, valid for bare and 
renormalized parameters, should not bring confusion; as shown in the following
section the renormalization equation for $\ctw$ depends on the IPS.
In \sect{solREe} we will introduce additional, secondary, running parameters.

Obviously our running parameters and the $\MSB$ ones are different objects
and only the former have a physical interpretation, while the latter are
nothing more than a convenient way of expressing the bare parameters of a
renormalizable theory.
\subsection{General structure of self-energies \label{GSSE}}
In this section we clarify the issue of gauge parameter independence,
order by order, of self-energies evaluated at their complex pole.
We now examine more carefully a two-point function to all orders in 
perturbation theory,
\bq
\Sigma_{\ssVV}(s,\xi) = \sum_{n=2}^{\infty}\,
\Sigma^{(n)}_{\ssVV}(s,\xi)\,g^{2n}.
\eq
All one-loop self-energies corresponding to physical particles are 
gauge-parameter independent when put on their, bare or renormalized, 
mass-shell and coincide with the corresponding $\xi = 1$ expression,
i.e.
\bq
\Sigma^{(1)}_{\ssVV}(s,\xi)= 
\Sigma^{(1)}_{\ssVV\,;\,\ssI}(s) + 
(s - M^2_{\ssV})\,\Phi_{\ssVV}(s,\xi).
\label{niI}
\eq 
>From arguments based on Nielsen identities we know that
\bq
\frac{\partial}{\partial \xi}\,\Sigma_{\ssVV}(s_{\ssP},\xi) =  0,
\qquad
s_{\ssP} - M^2_{\ssV} + \Sigma_{\ssVV}(s_{\ssP}) = 0.
\label{invsp}
\eq
To proceed further, we write a decomposition into independent and 
$\xi$-dependent parts,
\bq
\Sigma^{(n)}_{\ssVV}(s,\xi) = \Sigma^{(n)}_{\ssVV\,;\,\ssI}(s) + 
\Sigma^{(n)}_{\ssVV\,;\,\xi}(s,\xi),
\eq
and use the relation between bare (gauge parameter independent) mass and
complex pole,
\bq
M^2_{\ssV} = s_{\ssP} + g^2\,\Sigma^{(1)}_{\ssVV\,;\,\ssI}(s_{\ssP}) + 
g^4\,\Bigl[ \Sigma^{(1)}_{\ssVV\,;\,\ssI}(s_{\ssP})\,
\Sigma^{(1)}_{\ssVV\,;\,\xi}(s_{\ssP},\xi) -
\Sigma^{(2)}_{\ssVV\,;\,\ssI}(s_{\ssP}) -
\Sigma^{(2)}_{\ssVV\,;\,\xi}(s_{\ssP},\xi) \Bigr] + \ord{g^6},
\eq
to derive, as a consequence of \eqn{invsp} and of the fact that $M$ is a
bare quantity,
\bq
\Sigma^{(n)}_{\ssVV\,;\,\xi}(s_{\ssP},\xi) = 
\Sigma^{(n-1)}_{\ssVV\,;\,\ssI}(s_{\ssP})\,
\Phi_{\ssVV}(s_{\ssP},\xi),
\label{niII}
\eq
etc. As a result we can prove that all $\xi$-dependent parts cancel,
\bq
\Sigma_{\ssVV}(s_{\ssP}) = \sum_{n=2}^{\infty}\,
\Sigma^{(n)}_{\ssVV\,;\,\ssI}(s_{\ssP})\,g^{2n}.
\eq
However, this example shows how an all-order relation should be carefully 
interpreted while working at some fixed order.
\subsection{Outline of the calculation}
Here we introduce renormalization equations and their solutions. 
One of our renormalization equations will always be of the type
\bq
\cpv = \mV^2 - \frac{g^2}{16\,\pi^2}\,\Sigma_{\ssVV}(\cpv,\mV^2).
\label{cpi}
\eq
We have three options in using \eqn{cpi} which, we repeat, is gauge
parameter independent if the self-energy is considered to all orders.
\vspace{0.2cm}

\noindent
I) working at $\ord{g^4}$ we use \eqn{cpi} as it stands, i.e. in
$\Sigma_{\ssVV}$ we keep $p^2 = -\cpv$; the result of \sect{GSSE}
guarantees absence of gauge violating terms of $\ord{g^4}$ (violation
is due to the missing $\Sigma^{(3)}_{\ssVV\,;\,\xi}(s_{\ssP},\xi)$ term);

\noindent
II) we replace $\Sigma_{\ssVV}$ with $\Sigma_{\ssVV\,;\,\ssI}$ which
is the correct recipe but requires working in a gauge with arbitrary
(renormalized) gauge parameters; however, up to $\ord{g^4}$ we can use
an explicit calculation for $\Sigma^{(1)}_{\ssVV\,;\,\ssI}(s)$ and 
for $\Phi_{\ssVV}(s,\xi)$ (e.g. from Ref.~\cite{Bardin:1999ak}). Using 
\eqn{niII} we immediately derive
\bq
\Sigma^{(2)}_{\ssVV\,;\,\ssI}(\cpv) = 
\Sigma^{(2)}_{\ssVV}(\cpv,\xi=1) - \Sigma^{(1)}_{\ssVV\,;\,\ssI}(\cpv)\,
\Phi_{\ssVV}(\cpv,\xi=1).
\eq

\noindent
III) we expand, as done in ref.~\cite{Jegerlehner:2003py}
\bq
\cpv= \mV^2 - \frac{g^2}{16\,\pi^2}\,\Sigma^{(1)}_{\ssVV}(\mV^2,\mV^2)
- \lpar\frac{g^2}{16\,\pi^2}\rpar^2\,\Bigl[ \Sigma^{(2)}_{\ssVV}(\mV^2,\mV^2)-
\Sigma^{(1)}_{\ssVV}(\mV^2,\mV^2)\,\Sigma^{(1)}_{\ssVV\,;\,p}(\mV^2,\mV^2)
\Bigr]
\label{sqb}
\eq
where the suffix $p$ denotes derivation,
\bq
\Sigma^{(1)}_{\ssVV\,;\,p}(\mV^2,\mV^2) =
\frac{\partial}{\partial s}\,\Sigma^{(1)}_{\ssVV}(s,\mV^2)\bmid_{s=\mV^2},
\eq
and invert, obtaining $\mV^2$ in terms of $\mu^2_{\ssV} = \Reb\,\cpv$.
The combination within brackets in \eqn{sqb} is gauge invariant
(as shown by explicit calculations~\cite{Jegerlehner:2003py})
while $\Sigma^{(2)}_{\ssVV}(\mV^2,\mV^2)$ is not.
\vspace{0.2cm}

\noindent
Whenever $\mV$ can be reconstructed from other pseudo-observables of the IPS 
not involving $\cpv^{\rm exp}$ (but involving other {\em experimental} 
complex poles),
\bq
\mV^2= \mu^2_{\ssV} + \frac{g^2}{16\,\pi^2}\,\Bigl( m_1 + 
\frac{g^2}{16\,\pi^2}\,m_2\Bigr),
\eq
we derive 
\bq
\cpv^{\rm th} = \mu^2_{\ssV} + \frac{g^2}{16\,\pi^2}\,\Bigl[ x_1 +
\frac{g^2}{16\,\pi^2}\,x_2\Bigr] +
i\,\frac{g^2}{16\,\pi^2}\,\Bigl[ y_1 + \frac{g^2}{16\,\pi^2}\,y_2\Bigr]
\eq
with coefficients
\bq
x_1 = \Reb\,\oSigma^{(1)}_{\ssVV}(\mu^2_{\ssV}), 
\qquad
y_1 = \Imb\,\oSigma^{(1)}_{\ssVV}(\mu^2_{\ssV}), 
\qquad
\oSigma^{(n)}_{\ssVV}(s) = m_n - \Sigma^{(n)}_{\ssVV}(s), 
\eq
\bq
x_2 = \Reb\,\Bigl[ \oSigma^{(1)}_{\ssVV}(\mu^2_{\ssV})\,
        \oSigma^{(1)}_{\ssVV\,;\,p}(\mu^2_{\ssV}) +
        \oSigma^{(2)}_{\ssVV}(\mu^2_{\ssV})\Bigr], 
\quad
y_2 = \Imb\,\Bigl[ \oSigma^{(1)}_{\ssVV}(\mu^2_{\ssV})\,
        \oSigma^{(1)}_{\ssVV\,;\,p}(\mu^2_{\ssV}) +
        \oSigma^{(2)}_{\ssVV}(\mu^2_{\ssV})\Bigr].
\eq
Note, however, that for expanding a function $f(s^{\rm exp})$ around
$\Reb\,s^{\rm exp}$ one has to assume $\Imb\,s^{\rm exp}$ to be $\ord{g^2}$,
where $g^2$ is expressed in terms of pseudo-observables of the same IPS.
This is, for instance, needed in deriving $M$ from an IPS containing
$\cpz^{\rm exp}$.
\subsection{Notations}
Residual functions $F_{\ssB}(s)$ for $B= W,Z$ and $H$ are defined in 
\eqn{ResF}. All functions are expanded up to second order, e.g.
\bq
F_{\ssW} = F^{(1)}_{\ssW} + \frac{g^2}{16\,\pi^2}\,F^{(2)}_{\ssW}.
\eq
Furthermore, we introduce
\bq
\Sigma^{(n)}_{\ssF}(s) = F^{(n)}_{\ssW}(0) - \Reb\,\Sigma^{(n)}_{33}(s) +
\Reb\,\Sigma^{(n)}_{3\ssQ}(s),
\label{SigmaF}
\eq
\bq
\oF^{(n)}_{\ssW}(s) = F^{(n)}_{\ssW}(s) - F^{(n)}_{\ssW}(0),
\qquad
\tF^{(n)}_{\ssW}(s) = \Reb\,\Sigma^{(n)}_{\ssWW}(s) - F^{(n)}_{\ssW}(0),
\label{hF}
\eq
\bq
F^{(n)}_{\ssZ}(s) = \Sigma^{(n)}_{33}(s) - \Sigma^{(n)}_{3\ssQ}(s),
\label{fZ}
\eq
\bq
\oF^{(n)}_{\ssH}(s) = F^{(n)}_{\ssH}(s) - 
\frac{\rph}{\rpz\,\chs}\,F^{(n)}_{\ssW}(0),
\label{fH}
\eq
where $\cpz = \rpz - i\,\gamma_{\ssZ}\,\mu_{\ssZ}$ and
$\Cph = \rph - i\,\gamma_{\ssH}\,\mu_{\ssH}$ are the (input) $Z$ and $H$
boson complex poles. The LQ decomposition has been introduced in Sect.~6 
of I (Eqs.(123)--(125)) and, here, we drop the suffix {\em ext}.
\subsection{Input parameter set I: \boldmath $\alpha, \gf$ and $\mw$}
The starting point for our analysis, as for the solution of renormalization
equations, is to fix an IPS. Our first choice is the $\{\alpha, \gf, \mw\}$ 
IPS; although we still call it $\{\alpha, \gf, \mw\}$, it is clear
from the previous discussion that something slightly different is meant.
We use $\alpha, \gf$ and $\mu_{\ssW}$ and predict, among other things,
$\gamma_{\ssW}$ which, in turn, can be compared with the measured OS
$\Gamma_{\ssW}$. We begin with two equations
\bqa
{}&{}&
G\,\Bigl[ M^2 - \frac{g^2}{16\,\pi^2}\,F_{\ssW}(0)\Bigr] = \frac{g^2}{8}
\label{EQrenGF}
\\
{}&{}&
\rpw = M^2 - \frac{g^2}{16\,\pi^2}\,\Reb\,\Bigl[
\Sigma_{3\ssQ}(\cpw) + F_{\ssW}(\cpw)\Bigr]
\label{EQrenW}.
\eqa
The (finite) mass counterterm of \eqn{EQrenW} is to be contrasted with the
conventional mass renormalization where $\Reb\,\Sigma_{\ssWW}(\mws)$ is
used.

We look for a solution with the following form:
\bq
g^2 = 8\,G\,\rpw\,\Bigl[ 1 + \sum_{n=1}\,C_g(n)\,
\lpar \frac{G}{\pi^2}\rpar^n\Bigr],
\qquad
M^2 = \rpw\,\Bigl[ 1 + \sum_{n=1}\,C_{\ssM}(n)\,
\lpar \frac{G}{\pi^2}\rpar^n\Bigr],
\label{Eqforg}
\eq
where $G$ is the process independent coupling constant of \eqn{defGuniv}.
A straightforward calculation shows that
\bqa
C_g(1) &=& \frac{1}{2}\,
\tF^{(1)}_{\ssW}(\cpw)
\qquad
C_g(2) = C^2_g(1) +
\frac{1}{4}\,\rpw\,
\tF^{(2)}_{\ssW}(\cpw)
\qquad
C_{\ssM}(1) = \frac{1}{2}\,\Reb\,\Sigma^{(1)}_{\ssWW}(\cpw),
\nl 
C_{\ssM}(2) &=& C^2_{\ssM}(1) + \frac{1}{4}\,\Reb\,\Bigl[
\rpw\,\Sigma^{(2)}_{\ssWW}(\cpw) -
F^{(1)}_{\ssW}(0)\,\Sigma^{(1)}_{\ssWW}(\cpw)\Bigr].
\label{solI}
\eqa
Note that there is a special combination of renormalized parameters, 
$M^2/g^2$, which enters into the $W$ propagator; using \eqn{solI} we obtain
\bq
\frac{M^2}{g^2} = \frac{1}{8\,G}\,\Bigl[
1 + \frac{G}{2\,\pi^2}\,F^{(1)}_{\ssW}(0) + \frac{G^2}{4\,\pi^4}\,
\rpw \,F^{(2)}_{\ssW}(0) \Bigr].
\eq
We start with the running of $g$,
\vspace{0.1cm}
\bei
\item[--] {\underline{Renormalized running of $g$}}:
\eei
for this IPS the complete renormalization of the coupling constant $g$ is 
obtained after inserting \eqn{solI} into \eqn{rung},
\bq
\frac{1}{g^2(s)} = \frac{1}{8\,G\,\rpw}
- \frac{1}{16\,\pi^2\,\rpw}\,\delta\,g^{(1)}
- \frac{G}{32\,\pi^4}\,\delta\,g^{(2)},
\qquad
\delta\,g^{(n)} = \rpw\,\Pi^{(n)}_{3\ssQ}(s) + \tF^{(n)}_{\ssW}(\cpw).
\label{rungI}
\eq
To proceed further, we define the
\vspace{0.1cm}
\bei
\item[--] {\underline{Renormalized running of $\stws$}}:
\eei
the renormalization equation for $\stws$ is
\bq
g^2\,\stws = 4\,\pi\,\alpha\,\Bigl[ 1 - \frac{g^2\,\stws}{16\,\pi^2}\,
\Pi_{\ssQQ}(0)\Bigr].
\eq
Using the first of \eqn{Eqforg} we obtain a solution given by
\bqa
\stws &=& \shs\,\Bigl[ 1 + \sum_{n=1}\,C_s(n)\,
\lpar \frac{G}{\pi^2}\rpar^n\Bigr], \qquad
\shs = \frac{1}{2}\,\frac{\pi\,\alpha}{G\,\rpw},
\quad
\delta s^{(n)} = 
\tF^{(n)}_{\ssW}(\cpw) + 2\,\shs\,\rpw\,\Pi^{(n)}_{\ssQQ}(0),
\nl
C_s(1) &=& -\frac{1}{2}\,\delta s^{(1)},
\quad
C_s(2) = -\frac{1}{4}\,\rpw\,\Bigl[ 
\delta s^{(2)} - 2\,\shs\,\Pi^{(n)}_{\ssQQ}(0)\,\delta s^{(1)}\Bigr],
\eqa
Note that we have a residual dependence on $\stws$ in $\delta s^{(2)}$; here 
$\stws$ must be set to the lowest order value $\shs$. 
\vspace{0.1cm}
\bei
\item[--] {\underline{Renormalized $\wb$ propagator}}:
\eei
for the $\wb$ propagator we factorize a $g^2$, insert the solution and write 
its inverse as
\bq
\Bigl[ g^2\,\Delta_{\ssW}(s)\Bigr]^{-1} =  
\frac{s}{g^2(s)} - \frac{1}{8\,G} + \frac{1}{16\,\pi^2}\,\Bigl[
F^{(1)}_{\ssW}(s) - F^{(1)}_{\ssW}(0)\Bigr]
+ \frac{G\,\rpw}{32\,\pi^4}\,\Bigl[
F^{(2)}_{\ssW}(s) - F^{(2)}_{\ssW}(0)\Bigr].
\label{eqnWP}
\eq
Using \eqn{rungI} the same expression can be rewritten as
\bq
\Bigl[ g^2\,\Delta_{\ssW}(s)\Bigr]^{-1} =  
\frac{s}{g^2(s)} - \frac{\rpw}{g^2(\cpw)} + 
\frac{i}{16\,\pi^2}\,R^{(1)}_{\ssW}(\cpw) +
\frac{i\,G\,\rpw}{32\,\pi^4}\,R^{(2)}_{\ssW}(\cpw),
\eq
where the remainders are:
\bq 
R^{(n)}_{\ssW}(\cpw) = \Imb\,\Sigma^{(n)}_{\ssWW}(\cpw) -
\mu_{\ssW}\,\gamma_{\ssW}\,\Pi^{(n)}_{3\ssQ}(\cpw).
\eq
The complex zero of this expression is the theoretical prediction for
the complex pole of the $\wb$ boson. The real part will differ from $\rpw$
only in higher orders, the difference being proportional to
\bq
g^2\,\Reb\,\Bigl[ \Sigma^{(1)}_{\ssWW}(\cpw) - 
\Sigma^{(1)}_{\ssWW}(\rpw)\Bigr],
\eq
which is $\ord{g^2\,\gamma_{\ssW}}$; the solution for the imaginary part is
\bqa
\gamma^{\rm th}_{\ssW} &=& \frac{G\,\mu_{\ssW}}{2\,\pi^2}\,\lpar
   \gamma_1 + \frac{G}{2\,\pi^2}\,\gamma_2\rpar,
\eqa
with coefficients
\bqa
\gamma_1 &=& \Imb\,\Sigma^{(1)}_{\ssWW}(\rpw),
\nl
\gamma_2 &=& \Imb\,\Sigma^{(1)}_{\ssWW}(\rpw)\,\Bigl[
\Reb\,F^{(1)}_{\ssW}(\rpw) - F^{(1)}_{\ssW}(0)\Bigr]
+ \rpw\,\Bigl[
\Imb\,F^{(2)}_{\ssW}(\rpw) -
\Imb\,F^{(1)}_{\ssW}(\rpw)\,
\Reb\,\Sigma^{(1)}_{\ssWW\,;\,p}(\rpw)\Bigr],
\label{wder}
\eqa
where the suffix $p$ denotes derivation. We have one consistency condition
obtained by comparing the derived width of \eqn{wder} with the experimental
input $\gamma_{\ssW}$. The goodness of the comparison is a precision test
of the standard model.

Furthermore, the parameter controlling perturbative (non-resummed) expansion 
is $\gf\,\rpw$ and we derive,
\bq
G = \frac{\gf}{\sqrt{2}}\,
\Bigl\{ 1 - \delta^{(1)}_{\ssG}\,\frac{\gf\,\rpw}{2\,\sqrt{2}\,\pi^2}
+ \Bigl[ 2\,(\delta^{(1)}_{\ssG})^2 - \frac{2}{\rpw}\,
\delta^{(1)}_{\ssG}\,
C_g(1) - \delta^{(2)}_{\ssG}\Bigr]\,\lpar \frac{\gf\,\rpw}{2\,\sqrt{2}\,\pi^2}
\rpar^2 \Bigr\}.
\eq
In other words, we can go from the $G$ option of \subsect{Gopt} to the $\gf$ 
option by replacing 
\bqa
F^{(1)}_{\ssW}(0) \quad &\to& \quad \oF^{(1)}_{\ssW} =
F^{(1)}_{\ssW}(0) + \rpw\,\delta^{(1)}_{\ssG},
\nl
F^{(2)}_{\ssW}(0) \quad &\to& \quad \oF^{(2)}_{\ssW} =
F^{(2)}_{\ssW}(0) + \rpw\,\delta^{(2)}_{\ssG} +
\delta^{(1)}_{\ssG}\,\Bigl[ \rpw\,\delta^{(1)}_{\ssG} + 
\Reb\,F^{(1)}_{\ssW}(\cpw) + \Reb\,\Sigma^{(1)}_{3\ssQ}(\cpw) -
2\,\oF^{(1)}_{\ssW}\Bigr],
\eqa
and $G \to \gf/\sqrt{2}$ in all the results of this section.

All functions appearing in the results depend also on internal masses, $M$ etc.
Therefore we always use, for and arbitrary $f$
\bq
f^{(1)}(s\,;\,M^2\,,\,\dots) = f^{(1)}(s\,;\,\rpw\,,\,\dots) +
\frac{G\,\rpw}{2\,\pi^2}\,\Reb\,
\Sigma^{(1)}_{\ssWW}(\cpw\,;\,\rpw\,,\,\dots)\,
\frac{\partial}{\partial M^2}\,
f^{(1)} (s\,;\,M^2\,,\,\dots)\bmid_{M^2=\rpw}.
\eq
A last subtlety in \eqn{eqnWP} is represented by the residual $\stws$ 
dependence of the $\wb$ self-energy and of $\delta_{\ssG}$; we use 
\bq
\stws = {\bar s}^2\,\Bigl[ 1 - \frac{\gf}{2\,\pi^2}\,\delta s^{(1)}\Bigr] 
\quad
\hbox{in} \quad F^{(1)}_{\ssW},\; \delta^{(1)}_{\ssG}
\qquad
\stws = {\bar s}^2 \quad \hbox{in} \quad F^{(2)}_{\ssW},\; 
\delta^{(2)}_{\ssG}.
\eq
\subsection{Input parameter set II: \boldmath $\alpha, \gf$ and $\mz$}
Within this IPS we use the following three renormalization equations:
\bq
G\,\Bigl[ M^2 - \frac{g^2}{16\,\pi^2}\,F_{\ssW}(0)\Bigr] = \frac{g^2}{8},
\qquad
4\,\pi\,\alpha\,\Bigl[ 1 - \frac{g^2}{16\,\pi^2}\,
\Pi_{\ssQQ}(0)\Bigr] = g^2 \stws\,
\label{renEqIIgs}
\eq
\bq
\Bigl[ 1 - \frac{g^2\,\stws}{16\,\pi^2}\,\Reb\,\Pi_{\ssQQ}(\cpz)
\Bigr]\,\Bigl[ - \rpz\,\ctws + M^2 - \frac{g^2}{16\,\pi^2}\,
\Reb\,\Sigma_{\ssZ\ssZ}(\cpz) \Bigr]
+ \lpar \frac{g^2}{16\,\pi^2}\rpar^2\,\Reb\,
\frac{\Sigma^2_{\ssA\ssZ}(\cpz)}{\cpz}\, = 0.
\label{renEqIIM}
\eq
Within this IPS we look for a formal solution of the renormalization equations 
which improves upon fixed order perturbative expansion. Once this solution is
obtained we will discuss the necessary steps to reinstall gauge parameter
independence. The {\em improved} lowest order solution for $\stws$ is defined 
by
\bq
2\,G\,\rpz = \frac{\pi\,\alpha_{\ssZ}}{\shs\,\chs},
\qquad 4\,\pi\,\alpha(s) = e^2(s), \quad \alpha_{\ssZ} = \Reb\,\alpha(\cpz).  
\label{Is}
\eq
A solution of \eqns{renEqIIgs}{renEqIIM} is written according to the following 
expansion:
\bq
g^2\,\shs = 4\,\pi\,\alpha_{\ssZ}\,\Bigl[
1 + \sum_{n=1}\,C_g(n)\,\alpha^n(\cpz)\Bigr],
\qquad
M^2 = \rpz\,\chs\,\Bigl[ 1 + \sum_{n=1}\,C_{\ssM}(n)\,
\alpha^n(\cpz)\Bigr],
\eq
\bq
\stws = \shs\,\Bigl[ 1 + \sum_{n=1}\,C_s(n)\,\alpha^n(\cpz)\Bigr].
\label{IsolII}
\eq
To explain once again our procedure, we may say the following: despite its
intrinsic simplicity, $\shs$ of \eqn{Is} has problems with gauge 
parameter independence and we can a) expand $\alpha_{\ssZ}$ up to 
$\ord{\alpha^2}$ (improving, in any case, upon one-loop results), b) express
$\alpha_{\ssZ}$ in terms of $\alpha_{\MSB}(0)$ and c) introduce a 
(non-unique) version of $\alpha_{\ssZ}$ where the resummation is performed 
in terms of a gauge parameter independent choice of $\Pi_{\ssR}$ with the rest 
expanded up to second oder. To study gauge boson complex poles it is
important to have a resummation of large fermion logarithms and the use
of $\Pi_{\ssR\,;\,\ssQ\ssE\ssD}$ suffices.

A solution to \eqns{renEqIIgs}{renEqIIM} is provided in terms of the function
$\Sigma^{(n)}_{\ssF}$ of \eqn{SigmaF} and of
\bq
{\hat \Pi}^{(1)}_{\ssA\ssZ}(s) = \Pi^{(n)}_{3\ssQ}(s) -
\shs\,\Pi^{(n)}_{\ssQQ}(s).
\eq
The results are given by
\bqa
C_g(1) &=& \frac{1}{4\,\shs}\,\Reb\,\Bigl[
\frac{\cpz}{\rpz}\,\Pi^{(1)}_{3\ssQ}(\cpz) -
\frac{1}{(\chs - \shs)\,\rpz}\,\Sigma^{(1)}_{\ssF}\Bigr] =
\Delta\ghat^{(1)},
\nl
C_s(1) &=& -\,\frac{1}{4\,\shs}\,\Reb\,\Bigl\{
\frac{\cpz}{\rpz}\,\Bigl[
2\,\Pi^{(1)}_{3\ssQ}(\cpz) + 
{\hat \Pi}^{(1)}_{\ssA\ssZ}(\cpz) \Bigr] +
\frac{1}{(\chs - \shs)\,\rpz}\,\Sigma^{(1)}_{\ssF}\Bigr\} =
\Delta\shat^{(1)},
\nl
C_{\ssM}(1) &=& \frac{1}{4\,\chs}\,\Reb\,\Bigl\{
\frac{1}{\rpz}\,\Bigl[
3\,\cpz\,\Pi^{(1)}_{3\ssQ}(\cpz) + 
\frac{1}{\shs}\,\Sigma^{(1)}_{33}(\cpz) \Bigr] +
\frac{1}{(\chs - \shs)\,\rpz}\,\Sigma^{(1)}_{\ssF}\Bigr\} =
\Delta\Mhat^{(1)}.
\eqa
Furthermore, we introduce
\bqa
\rpz\,\Delta\ghat^{(2)} &=& -\chs\,\Sigma^{(2)}_{\ssF} +
( 1 - 3\,\shs + 2\,\shq )\,\Reb\,\Sigma^{(2)}_{3\ssQ}(\cpz),
\nl
\rpz\,\Delta\shat^{(2)} &=& \chs\,\Sigma^{(2)}_{\ssF}
+ \Reb\,\{\cpz\,\Bigl[
\shs\,( 1 - 3\,\shs + 2\,\shq)\,
{\hat \Pi}^{(2)}_{\ssA\ssZ}(\cpz) + ( 1 - 4\,\shs + 5\,\shq - 2\,\shsix )\,
\Pi^{(2)}_{3\ssQ}(\cpz)\Bigr]\},
\nl
\rpz\,\Delta\Mhat^{(2)} &=& - \shs\,\Sigma^{(2)}_{\ssF} + (\chs - \shs)\,
\Reb\,\Bigl[ \Sigma^{(2)}_{33}(\cpz) - \shs\,
\Sigma^{(2)}_{3\ssQ}(\cpz)\Bigr],
\nl
\Delta {\hat e}^{(n)} &=& \Delta\ghat^{(n)} + \Delta\shat^{(n)},
\eqa
to obtain
\bq
C_g(2)=  \frac{C'_g(2)}{16\,\pi^2\,\shq\,\chs\,(\chs-\shs)},
\quad
C_s(2)=  \frac{C'_s(2)}{16\,\pi^2\,\shq\,\chs\,(\chs-\shs)},
\quad
C_{\ssM}(2)=  \frac{C'_{\ssM}(2)}{16\,\pi^2\,\shq\,\chs\,(\chs-\shs)},
\eq
\bqa
C'_g(2) &=& \Delta\ghat^{(2)} + \shs\,\chs\,\Reb\,\Bigl\{
 16\,\shs\,(1 - 3\,\shs)\,\Bigl[\Delta\ghat^{(1)}\Bigr]^2
+ 16\,\shq\,\lpar 1 - \frac{\rpz}{\cpz}\rpar\,\Bigl[
       \Delta{\hat e}^{(1)}\Bigr]^2
\nl
{}&+& \Pi^{(1)}_{3\ssQ}(\cpz)\,\Bigl[ 
  8\,\shs\,\frac{\cpz}{\rpz}\,\Delta\shat^{(1)} - 
  8\,\shs\, \Delta{\hat e}^{(1)}  +
  \frac{\Sigma^{(1)}_{3\ssQ}(\cpz)}{\rpz}\Bigr]
\Bigr\}.
\eqa
\bqa
C'_s(2) &=& \Delta \shat^{(2)} + \shs\,\chs\,\Reb\,\Bigl\{
16\,\,\shq\,\Bigl[ \Delta\ghat^{(1)}\Bigr]^2 +
16\,\shq\,\frac{\rpz}{\cpz}\,
    \Bigl[ \Delta\,{\hat e}^{(1)} \Bigr]^2 +
16\,\shs\,( 1 - 3\,\shs )\,\Delta\shat^{(1)}\,\Delta{\hat e}^{(1)}  
\nl
{}&+& \Pi^{(1)}_{3\ssQ}(\cpz)\,\Bigl[
8\,\shs\,\Delta \ghat^{(1)} - \frac{\Sigma^{(1)}_{3\ssQ}(\cpz)}{\rpz}
\Bigr]\Bigr\}.
\eqa
\bqa
C'_{\ssM}(2) &=& \Delta \Mhat^{(2)} + \shs\,\chs\,\Reb\,\Bigl\{
16\,\shq\,\Bigl[ \Delta \Mhat^{(1)}\Bigr]^2 -
16\,\shq\,\Bigl[ \Delta \ghat^{(1)} + \Delta \Mhat^{(1)}\Bigr]^2 +
16\,\shs\,\Delta \ghat^{(1)}\,\Delta \Mhat^{(1)} 
\nl
{}&+& 16\,\shq\,\lpar 1 - \frac{\rpz}{\cpz}\rpar\,
   \Bigl[ \Delta {\hat e}^{(1)} \Bigr]^2 -
\Pi^{(1)}_{3\ssQ}(\cpz)\,\Bigl[
    8\,\shs\,\Delta \ghat^{(1)} -
   \frac{\Sigma^{(1)}_{3\ssQ}(\cpz)}{\rpz} \Bigr]\Bigr\}
\eqa
The solution that we have obtained will be interpreted in terms of running 
quantities:
\vspace{0.1cm}
\bei
\item[--] {\underline{Renormalized running of $g$}}:
\eei
The running of $g$, given in \eqn{rung}, is fixed in IPS II by the following 
equation:
\bqa
\frac{1}{g^2(s)} &=& \frac{1}{4}\,\frac{\shs}{\pi\,\alpha_{\ssZ}} -
\frac{1}{16\,\pi^2}\,\Bigl[
 \Pi^{(1)}_{3\ssQ}(s) + \Pi^{(1)}_{3\ssQ}(\cpz) -
\frac{1}{\chs - \shs}\,\frac{\Sigma^{(1)}_{\ssF}}{\rpz}\Bigr]
\nl
{}&+& \frac{\alpha_{\ssZ}}{64\,\pi^3}\,\Bigl\{
\frac{1}{(\chs-\shs)^3}\,\Bigl[
\lpar \frac{\Sigma^{(1)}_{\ssF}}{\rpz} \rpar^2 -
\frac{(\chs-\shs)^2}{\shs}\,
\frac{\Sigma^{(2)}_{\ssF}}{\rpz}\Bigr] 
- \frac{1}{\shs}\,\Pi^{(2)}_{3\ssQ}(s) 
\nl
{}&-& \frac{1}{\shs}\,\Reb\,\Pi^{(2)}_{3\ssQ}(\cpz) +
4\,\Reb\,\Pi^{(1)}_{3\ssQ}(\cpz)\,\Bigl[
   \frac{1}{\chs-\shs}\,\Pi^{(1)}_{3\ssQ}(\cpz) -
   \frac{1}{(\chs-\shs)^2}\,
\frac{\Sigma^{(1)}_{\ssF}}{\rpz}\Bigr]\Bigr\}.
\label{IPSIIrung}
\eqa
At the same time the ratio $M^2/g^2$ becomes
\bqa
\frac{M^2}{g^2} &=&
\frac{\shs\,\chs\,\rpz}{4\,\pi\,\alpha_{\ssZ}} +
\frac{1}{16\,\pi^2}\,\Reb\,\Bigl[ F^{(1)}_{\ssW}(0) + 2\,\chs\,
\Sigma^{(1)}_{3\ssQ}(\cpz)\Bigr]
\nl
{}&+& \frac{\alpha_{\ssZ}}{64\,\pi^3}\,\Reb\,\Bigl\{
\frac{1}{\shs}\,F^{(2)}_{\ssW}(0) -
2\,\frac{\chs}{\shs\,(\chs-\shs)}\,\Sigma^{(2)}_{\ssF} -
2\,\frac{\chs}{\shs}\,\Sigma^{(2)}_{3\ssQ}(\cpz) 
\nl
{}&+& 2\,\frac{\cpz}{\rpz}\,\Pi^{(1)}_{3\ssQ}(\cpz)\,\Bigl[
   \frac{1}{\shs\,(\chs-\shs)}\,\Sigma^{(1)}_{3\ssQ}(\cpz) -
   \frac{1}{\shs}\,F^{(1)}_{\ssW}(0) +
   \lpar \frac{1}{\shs} - \frac{1}{(\chs-\shs)^2}\rpar\,\Sigma^{(1)}_{\ssF}
\Bigr]\Bigr\}.
\label{ratioMg}
\eqa
\vspace{0.1cm}
\bei
\item[--] {\underline{Renormalized running $\wb$ mass}}:
\eei
At this point we introduce a new running parameter, $M(s)$, through the 
relation
\bq
\frac{M^2(s)}{g^2(s)} = \frac{\shs\,\chs\,\rpz}{4\,\pi\,\alpha_{\ssZ}} -
\frac{1}{16\,\pi^2}\,\oF^{(1)}_{\ssW}(s) -
\frac{\alpha_{\ssZ}}{64\,\pi^3\,\shs}\,\,\oF^{(2)}_{\ssW}(s),
\eq
with $\oF^{(n)}_{\ssW}(s)$ given in \eqn{hF}.
At $s= 0$ we have $g^2(0)/M^2(0) = 8\,G$ and, therefore we can describe
the running of $G$ by introducing
\vspace{0.1cm}
\bei
\item[--] {\underline{Renormalized running of $G$}}:
\eei
\bq
G(s)= \frac{1}{8}\,\frac{g^2(s)}{M^2(s)}, \qquad G(0)= G.
\eq
The running parameters introduced so far allow for a simple representation of
the $\wb$ propagator, one of the primary quantities in discussing 
renormalization of the standard model.
\vspace{0.1cm}
\bei
\item[--] {\underline{Renormalized $\wb$ propagator}}:
\eei
Using \eqns{IPSIIrung}{ratioMg} we obtain the following expression for the
inverse $\wb$ propagator (a factor $g^2$ is factorized);
\bqa
\Bigl[ g^2\,\Delta_{\ssW}(s)\Bigr]^{-1} &=&
\frac{s}{g^2(s)} - \frac{M^2(s)}{g^2(s)} + \Reb\,R_{\ssW} =
\frac{s}{g^2(s)} - \frac{\cpw}{g^2(\cpw)} - \frac{M^2(s)}{g^2(s)} + 
\frac{M^2(\cpw)}{g^2(\cpw)}
\nl 
{}&=& \frac{s}{g^2(s)} - \frac{\cpw}{g^2(\cpw)} - \frac{1}{8}\,
\Bigl[ G^{-1}(s) - G^{-1}(\cpw) \Bigr],
\label{virt}
\eqa
where the residual term is
\bqa
R_{\ssW} &=& -\,\frac{1}{16\,\pi}\,\chs\,\Sigma^{(1)}_{3\ssQ}(\cpz)
+ \frac{\alpha_{\ssZ}}{32\,\pi^3}\,\Bigl\{
\frac{\chs}{\shs\,(\chs-\shs)}\,\Sigma^{(2)}_{\ssF} -
\frac{\chs}{\shs}\,\Sigma^{(2)}_{3\ssQ}(\cpz) 
\nl
{}&-& \frac{\Sigma^{(1)}_{3\ssQ}(\cpz)}{\rpz}\,\Bigl[
  \frac{1}{\shs}\,F^{(1)}_{\ssW}(0) +
  \lpar \frac{1}{(\chs-\shs)^2} - \frac{1}{\shs}\rpar\,\Sigma^{(1)}_{\ssF} +
  \frac{\chs}{\shs\,(\chs-\shs)}\,\Sigma^{(1)}_{3\ssQ}(\cpz)\Bigr]
\Bigr\}
\eqa
\eqn{virt} shows the intrinsic beauty of the language; the $W$ propagator is
entirely written in terms of running couplings.
Another useful combination of renormalized parameters is
\bqa
\frac{\ctws}{g^2} &=& \frac{\shs \chs}{4\,\pi\,\alpha_{\ssZ}} +
\frac{1}{16\,\pi^2\,\rpz}\,\Bigl[
\Sigma^{(1)}_{\ssF} + \shs\,\Reb\,\sgh^{(1)}_{\ssA\ssZ}(\cpz) +
(1 - 3\,\shs)\,\Reb\,\Sigma^{(1)}_{3\ssQ}(\cpz)\Bigr]
\nl
{}&+& \frac{\alpha_{\ssZ}}{64\,\pi^3\,\rpz}\,\Reb\,\Bigl\{
\shs\,\sgh^{(2)}_{\ssA\ssZ}(\cpz) +
\frac{1-\shq}{\shs}\,\Sigma^{(2)}_{3\ssQ}(\cpz) -
\frac{1}{\shs (\chs-\shs)}\,\Sigma^{(2)}_{\ssF} 
\nl
{}&+&
\frac{1}{\rpz}\,\sgh^{(1)}_{\ssA\ssZ}(\cpz)\,\Bigl[
4\,\Sigma^{(1)}_{3\ssQ}(\cpz) +
\lpar \frac{\rpz}{\cpz} - 2\rpar\,\sgh^{(1)}_{\ssA\ssZ}(\cpz) +
\frac{2}{\chs-\shs}\,\Sigma^{(1)}_{\ssF}\Bigr]
\Bigr\}.
\eqa
To go from the $G$ option of \subsect{Gopt} to the $\gf$ option we perform 
the following replacements: $\shs$ is now the solution of
\bq
\gf = \frac{\pi\,\alpha_{\ssZ}}{2\,\rpz\,\shs \chs},
\eq
and everywhere we perform the following replacements:
\bqa
F^{(1)}_{\ssW}(0) \quad &\to& \quad \oF^{(1)}_{\ssW} =
F^{(1)}_{\ssW}(0) + \rpz\,\chs\,\delta^{(1)}_{\ssG},
\nl
F^{(2)}_{\ssW}(0) \quad &\to& \quad \oF^{(2)}_{\ssW} =
F^{(2)}_{\ssW}(0) + \rpz \chs\,\delta^{(2)}_{\ssG} +
\delta^{(1)}_{\ssG}\,\Bigl\{
\rpz \chs\,\delta^{(1)}_{\ssG} 
\nl
{}&+& \frac{1}{\chs-\shs}\,\Reb\,\Bigl[
\chs\,\Sigma^{(1)}_{33}(\cpz) -
2\,\shs \chs\,\Sigma^{(1)}_{3\ssQ}(\cpz) -
(2\,\chs - \shs)\,\oF^{(1)}_{\ssW} \Bigr]\Bigr\}.
\eqa
Next, we consider the $\zb$ propagator; however, it is more convenient to
introduce a new parameter which is related to the custodial $SU(2)_V$ 
symmetry:
\vspace{0.1cm}
\bei
\item[--] {\underline{Running $\rho$-parameter}}:
\eei
To write the $\zb$ propagator we introduce a Veltman running parameter,
$\rho(s)$, defined by
\bqa
\frac{1}{\rho(s)} &=& 1 + \frac{\alpha_{\ssZ}}{\pi}\,\Bigl[
\delta \rho^{(1)}(s) + 
\frac{\alpha_{\ssZ}}{4\,\pi\,\shs}\,\delta \rho^{(2)}(s) \Bigr],
\nl
2\,\rpz\,\delta \rho^{(n)}(s) &=& 
  \frac{1}{2}\,\frac{1}{\shs \chs}\,\Bigl[ F^{(n)}_{\ssW}(s)
 - F^{(n)}_{\ssZ}(s)\Bigr]
 - \frac{s}{\chs \rpz}\,\Reb\,\Sigma^{(n)}_{3\ssQ}(\cpz),
\eqa
where we have used \eqn{fZ}. This new parameter has its own relevance insofar it
allows us to simplify the $\zb$ propagator:
\vspace{0.1cm}
\bei
\item[--] {\underline{Renormalized $\zb$ propagator}}:
\eei
With these ingredients it can be shown that
\bq
\frac{\ctws}{g^2}\,\Delta^{-1}_{\ssZ}(s) =
\frac{c^2(s)}{g^2(s)}\,s - \frac{M^(s)}{g^2(s)\,\rho(s)} + R_{\ssZ}(s),
\eq
where $c^2(s)= 1 - e^2(s)/g^2(s)$ and where the residual term is given by
\bqa
R_{\ssZ}(\cpz) &=& 
-\,\frac{c^2(\cpz)}{g^2(\cpz)}\,\cpz + \frac{M^2(\cpz)}{g^2(\cpz)\,\rho(\cpz)}
\nl
R_{\ssZ}(s) &=& - \frac{\chs}{8\,\pi^2}\,
\Reb\,\Sigma^{(1)}_{3\ssQ}(\cpz)
+ \frac{\alpha_{\ssZ}}{16\,\pi^3\,\rpz}\,R^{(2)}_{\ssZ}(s);
\eqa
the second order correction is
\bqa
R^{(2)}_{\ssZ}(s) &=&
\frac{1}{2\,\chs}\,\Bigl[ F^{(1)}_{\ssW}(s) - F^{(1)}_{\ssW}(0)\Bigr]\,
\Bigl[ 
\frac{s}{\rpz}\,\Reb\,\Sigma^{(1)}_{3\ssQ}(\cpz)
- \frac{1}{2\,\shs}\,F^{(1)}_{\ssW}(s)
\nl
{}&+& \frac{1}{2}\,\frac{s}{\rpz}\,\shs\,\Reb\,(\cpz - \rpz)\,
               \Pi^{(1)}_{\ssQQ}(\cpz)
+ \frac{1}{2\,\shs}\,F^{(1)}_{\ssZ}(s) \Bigr]
\nl
{}&+& \frac{1}{2}\,\Bigl[ \frac{1}{\chs - \shs}\,\lpar \frac{s}{\rpz} -
         \frac{\chs}{\shs}\rpar\, + 1\Bigr]\,
   \Bigl[ \Reb\,\Sigma^{(1)}_{3\ssQ}(\cpz)\Bigr]^2
\nl
{}&+& \frac{1}{2}\,\Reb\,\Sigma^{(1)}_{3\ssQ}(\cpz)\,\Bigl[
2\,\sgh^{(1)}_{\ssA\ssZ}(s)
- \Reb\,\sgh^{(1)}_{\ssA\ssZ}(\cpz)
+ \frac{1}{(\chs-\shs)^2}\,\lpar \frac{s}{\rpz} - 1\rpar\,\Sigma^{(1)}_{\ssF}
\nl
{}&-& \shs\,s\,\Reb\,\Pi^{(1)}_{\ssQQ}(\cpz)
- \frac{1}{\shs}\,\Reb\,F^{(1)}_{\ssZ}(\cpz)\Bigr]
\nl
{}&+& \frac{1}{2}\,\shs\,\Bigl[
\Reb\,(\cpz - \rpz)\,\Pi^{(1)}_{\ssQQ}(\cpz)\Bigr]\,
\Bigl[   \frac{1}{\chs-\shs}\,\frac{s}{\rpz}\Sigma^{(1)}_{\ssF} + 
  \sgh^{(1)}_{\ssA\ssZ}(s)\Bigr]
\nl
{}&-& \frac{1}{4}\,\shq\,s\,\Reb\,(\cpz - \rpz)\,
 \lpar\frac{\cpz - \rpz}{\rpz} + 2\rpar\,
  \Bigl[ \Pi^{(1)}_{\ssQQ}(\cpz)\Bigr]^2\,
+ \frac{1}{2}\,\frac{1}{\chs-\shs}\,\lpar \frac{\chs}{\shs}\,\rpz -
      s \rpar\,\Sigma^{(2)}_{\ssF}
\nl
{}&+& \frac{1}{4}\,\shq\,s\,\Reb\,(\cpz - \rpz)\,\Pi^{(2)}_{\ssQQ}(\cpz)
- \frac{1}{2}\,\frac{\chs}{\shs}\,\rpz\,\Reb\,\Sigma^{(2)}_{3\ssQ}(\cpz)
\eqa
One final comment concerns the gauge parameter independence of (finite)
renormalization; \eqnsc{EQrenW}{renEqIIM} are the definition of $W$ and
$Z$ boson complex poles (actually their real part) and therefore 
$\xi$-independent. The second \eqn{renEqIIgs} involves $\Pi(0)$ which is 
gauge parameter independent. As far as \eqn{EQrenGF} is concerned we have to 
understand it as properly expanded to the requested order since only 
$M^2\,\delta^{\ssG} + \Sigma_{\ssWW}(0)$ is $\xi$-independent.
\subsection{Including the Higgs boson}
To describe the Higgs boson lineshape we define the following quantities:
\bq
s_{\ssH} \equiv s^{\rm th}_{\ssH} = 
M^2_{\ssH} - i\,M_{\ssH}\,\Gamma_{\ssH},
\qquad
S_{\ssH} \equiv s^{\rm exp}_{\ssH} = 
\mu^2_{\ssH} - i\,\mu_{\ssH}\,\gamma_{\ssH}.
\eq
To study properties of the Higgs boson a new equation is added
to \eqns{renEqIIgs}{renEqIIM}.
\bq
\rph = \mhs - \frac{g^2}{16\,\pi^2}\,\Reb\,\Bigl[
\Sigma_{3\ssQ}(\Cph) + F_{\ssH}(\Cph)\Bigr]
\label{EQrenH},
\eq
where $\Cph$ is the {\em experimental} $H$ boson complex pole. Expanding $\mhs$
\bq
\mhs = \rph\,\Bigl[ 1 + \sum_{n=1}\,C^h_{\ssM}(n)\,
\lpar\frac{\alpha_{\ssZ}}{\pi\shs}\rpar^n\Bigr],
\eq
we obtain the following solution:
\bqa
\rph\,C^h_{\ssM}(1) &=& \frac{1}{4}\,\Reb\,\Sigma^{(1)}_{\ssHH}(\Cph)
\nl
\rph\,C^h_{\ssM}(2) &=& 
\frac{1}{16}\,\Reb\,\Bigl\{
\Sigma^{(2)}_{\ssHH}(\Cph) + \frac{1}{\rph}\,\Sigma^{(1)}_{\ssHH}(\Cph)\,\Bigl[
\Sigma^{(1)}_{\ssHH}(\Cph) - \frac{\rph}{\chs\,\rpz}\,
F^{(1)}_{\ssW}(0)\Bigr]\Bigr\}.
\eqa
The ratio $\mhs/g^2$ becomes
\bqa
\frac{\mhs}{g^2} &=&
\frac{\rph\,\shs}{4\,\pi\,\alpha_{\ssZ}} +
\frac{1}{16\,\pi^2}\,\frac{\rph}{\rpz\,\chs}\,F^{(1)}_{\ssW}(0)
\nl
{}&+& \frac{\alpha_{\ssZ}}{64\,\pi^3\,\rpz\,\shs\,\chs}\,\Reb\,\Bigl\{
\rph\,F^{(2)}_{\ssW}(0) + F^{(1)}_{\ssW}(0)\,\Bigl[
\frac{1}{\chs-\shs}\,\frac{\rph}{\rpz}\,\Sigma_{\ssH}(\cpz) - 
\Sigma^{(1)}_{\ssHH}(\Cph) \Bigl]\Bigr\},
\label{ratioMhg}
\eqa
\bq
\Sigma_{\ssH}(s) = \frac{1}{\shs}\,\Sigma^{(1)}_{33}(s) -
\frac{1}{\chs}\,F^{(1)}_{\ssW}(0) - 2\,\Sigma^{(1)}_{3\ssQ}(s). 
\eq
\vspace{0.1cm}
\bei
\item[--] {\underline{Renormalized running $\hb$ mass}}:
\eei
After introducing the running Higgs boson mass, $\mh(s)$, through
\bq
\frac{\mhs(s)}{g^2(s)} = \frac{\shs\,\rph}{4\,\pi\,\alpha_{\ssZ}} -
\frac{1}{16\,\pi}\,\oF^{(1)}_{\ssH}(s) -
\frac{\alpha_{\ssZ}}{64\,\pi^3\,\shs}\,\,\oF^{(2)}_{\ssH}(s),
\eq
with $\oF^{(n)}_{\ssH}(s)$ defined in \eqn{fH};
we use \eqns{IPSIIrung}{ratioMhg} to obtain the following expression for the
inverse $H$ propagator (a factor $g^2$ is factorized),
\bq
\Bigl[ g^2\,\Delta_{\ssH}(s)\Bigr]^{-1} =  
\frac{s}{g^2(s)} - \frac{\mhs(s)}{g^2(s)} + \Reb\,R_{\ssH} =
\frac{s}{g^2(s)} - \frac{\Cph}{g^2(\Cph)} - \frac{M^2(s)}{g^2{s}} + 
\frac{M^2(\Cph)}{g^2(\Cph)}, 
\eq
with a residual term
\bqa
R_{\ssH} &=& \frac{\alpha_{\ssZ}\,F^{(1)}_{\ssW}(0)}{64\,\pi^3\,\chs\shs\rpz}\,
\,\Bigl\{
F^{(1)}_{\ssH}(\Cph) -
\frac{\shs}{\chs-\shs}\,\frac{\rph}{\rpz}\,\Sigma_{\ssH}(\cpz) +
\Sigma^{(1)}_{3\ssQ}(\Cph)\Bigr\}.
\eqa
Fuerthermore, if we use input parameter set I we have a mass renormalization 
equation expanded as 
\bq
M^2_{\ssH} = \mu^2_{\ssH}\,\Bigl[ 1 + \sum_{n=1,2}\,
{\overline C}^{(n)}_{\ssM}\,\lpar \frac{G}{2\,\pi^2}\rpar^n\Bigr],
\eq
with coefficients
\bq
{\overline C}^{(1)}_{\ssM} = \frac{\mu^2_{\ssW}}{\mu^2_{\ssH}}\,
\Reb\,\Sigma^{(1)}_{\ssHH}(S_{\ssH}),
\eq
\bq
{\overline C}^{(2)}_{\ssM} = \frac{\mu^2_{\ssW}}{\mu^2_{\ssH}}\,
\Bigl\{ 
\Sigma^{(1)}_{\ssHH}(s_{\ssW})\,\Sigma^{(1)}_{3q}(S_{\ssH}) 
+ \Sigma^{(1)}_{\ssHH}(S_{\ssH})\,\Bigl[
F^{(1)}_{\ssW}(s_{\ssW}) - F^{(1)}_{\ssW}(0) \Bigr] +
\mu^2_{\ssW}\,\Sigma^{(2)}_{\ssHH}(S_{\ssH})\Bigr\}.
\eq
\vspace{0.1cm}
\bei
\item[--] {\underline{Renormalized $\hb$ propagator}}:
\eei
The inverse Higgs propagator becomes
\bqa
\Bigl[ g^2\,\Delta_{\ssH}(s)\Bigr]^{-1} &=&  
\frac{s}{g^2(s)} + 
s\,\Bigl[ \frac{1}{16\,\pi^2}\,\Pi^{(1)}_{3\ssQ}(s) + 
    \frac{g^2}{256\,\pi^4}\,\Pi^{(2)}_{3\ssQ}(s)\Bigr] 
\nl
{}&-& \frac{M^2_{\ssH}}{g^2} +
\frac{1}{16\,\pi^2}\,\Sigma^{(1)}_{\ssHH}(s) + 
  \frac{g^2}{256\,\pi^4}\,\Sigma^{(2)}_{\ssHH}(s).
\eqa
With two-loop accuracy the solution for $s_{\ssH}$ is
\bq
s_{\ssH} = \mu^2_{\ssH} +
\frac{G\,\mu^2_{\ssW}}{2\,\pi^2}\,\Bigl[
\Delta M^{(1)} + \frac{G\,\mu^2_{\ssW}}{2\,\pi^2}\,\Delta M^{(2)}\Bigr] 
- i\,\frac{G\,\mu^2_{\ssW}}{2\,\pi^2}\,\Bigl[
\Delta \Gamma^{(1)} + \frac{G\,\mu^2_{\ssW}}{2\,\pi^2}\,
\Delta \Gamma^{(2)}\Bigr].
\label{effects}
\eq
The coefficients are
\bq 
\Delta{\oSigma}^{(n)} = \Reb\,\Bigl[ \Sigma^{(n)}_{\ssHH}(S_{\ssH}) -
\Sigma^{(n)}_{\ssHH}(\mu^2_{\ssH})\Bigr],
\qquad
\Delta M^{(1)} = \Delta{\oSigma}^{(1)},
\qquad
\Delta \Gamma^{(1)} = \Imb\,\Sigma^{(1)}_{\ssH}(\mu^2_{\ssH}),
\eq
\bqa
\Delta M^{(2)} &=& \frac{1}{\mu^2_{\ssW}}\,\Bigl\{
\mu^2_{\ssW}\,\Delta{\oSigma}^{(2)}
- \Sigma^{(1)}_{\ssWW}(0)\,\Delta{\oSigma}^{(1)}
+ {\oSigma}^{(1)}\,
\Reb\,\Bigl[ 
\Sigma^{(1)}_{\ssWW}(s_{\ssW})
\nl
{}&-&
\mu^2_{\ssW}\,\Sigma^{(1)}_{\ssHH\,\rm p}(\mu^2_{\ssH}) 
\Bigr]
- \mu^2_{\ssW}\,\Imb\,\Sigma^{(1)}_{\ssHH}(\mu^2_{\ssH})\,
\Imb\,\Sigma^{(1)}_{\ssHH\,\rm p}(\mu^2_{\ssH})
\Bigr\}
\nl
\Delta \Gamma^{(2)} &=& \frac{1}{\mu^2_{\ssW}}\,\Bigl\{
\mu^2_{\ssW}\,\Imb\,\Sigma^{(2)}_{\ssHH}(\mu^2_{\ssH})
+ \Bigl[ 
\Reb\,\Sigma^{(1)}_{\ssWW}(s_{\ssW})-\Sigma^{(1)}_{\ssWW}(0) 
\Bigr]\,\Imb\,\Sigma^{(1)}_{\ssHH}(\mu^2_{\ssH})
\nl
{}&+& \mu^2_{\ssW}\,{\oSigma}^{(1)} 
\,\Imb\,\Sigma^{(1)}_{\ssHH\,\rm p}(\mu^2_{\ssH})
- \mu^2_{\ssW}\,\Imb\,\Sigma^{(1)}_{\ssHH}(\mu^2_{\ssH})\,
\Reb\,\Sigma^{(1)}_{\ssHH\,\rm p}(\mu^2_{\ssH})
\Bigr\}
\label{solshipsI}
\eqa
In \eqn{solshipsI} we have introduced the notation $f_{\rm p}(s) =
\partial f/\partial s$ and two-loop contributions include the finite 
renormalization arising from one-loop terms, i.e.
\bqa
F^{(1)}\lpar s\,;\,M^2_{\ssW}\,,\,M^2_{\ssH}\,,\,\dots\rpar &=&
F^{(1)}\lpar s\,;\,\mu^2_{\ssW}\,,\,\mu^2_{\ssH}\,,\,\dots\rpar 
+ \frac{G}{2\,\pi^2}\,
\frac{\partial}{\partial M^2_{\ssW}}\,
       F^{(1)}\lpar s\,;\,M^2_{\ssW}\,,\,M^2_{\ssH}\rpar
       \bmid_{M^2_{\ssW} = \mu^2_{\ssW}\,\dots}\,
       C^{(1)}_{\ssM_{\ssW}}
\nl
{}&+& \frac{G}{2\,\pi^2}\,
\frac{\partial}{\partial M^2_{\ssH}}\,
       F^{(1)}\lpar s\,;\,M^2_{\ssW}\,,\,M^2_{\ssH}\rpar
       \bmid_{M^2_{\ssW} = \mu^2_{\ssW}\,\dots}\,
       C^{(1)}_{\ssM_{\ssH}}
+ \,\dots,
\eqa
with coefficients
\bqa
C^{(1)}_{\ssM_{\ssW}} &=& \Reb\,\Sigma^{(1)}_{\ssWW}(S_{\ssW}),
\qquad
C^{(1)}_{\ssM_{\ssH}} = \frac{\mu^2_{\ssW}}{\mu^2_{\ssH}}\,
\Reb\,\Sigma^{(1)}_{\ssHH}(S_{\ssH}),
\nl
C^{(1)}_s &=&  - \mu^2_{\ssW}\,\shs\,\Pi^{(1)}_{QQ}(0)
- \Reb\,\Sigma^{(1)}_{\ssWW}(s_{\ssW}) + \Sigma^{(1)}_{\ssWW}(0),
\eqa
etc. To give an idea of the impact of radiative corrections in \eqn{effects}
we present few results in \tabn{though}
where it is evident that $\Reb\,(\cph - \Cph)$ is a tiny effect
while anomalous values for the imaginary part are impossible to accommodate
within the standard model.
\subsection{A simple numerical example}
In this subsection we illustrate renormalization equations with a simple 
numerical example.
First we write
\bq
\frac{\gf\,\rpw}{2\,\sqrt{2}\,\pi^2} = x\,( 1 + a_1\,x + a_2\,x^2)
\qquad
x = \frac{g^2}{16\,\pi^2},
\eq
where $g$ is the renormalized coupling constant and the coefficients are
\bqa
a_1 = \delta^{(1)}_{\ssG} + S^{(1)}
&\qquad&
a_2 = S^{(1)}\,\Bigl[ \delta^{(1)}_{\ssG} + S^{(1)}\,\Bigr] 
+ \delta^{(2)}_{\ssG} + S^{(2)},
\nl
S^{(n)} &=& \frac{1}{\rpw}\,\Sigma^{(n)}_{\ssW\ssW}(0).
\eqa
The perturbative solution is 
\bq
x = \frac{g^2}{16\,\pi^2} = X + X^2\,( b_1 + b_2\,X),
\qquad
X = \frac{\gf\,\rpw}{2\,\sqrt{2}\,\pi^2},
\label{numren}
\eq
and the numerical impact of two-loop terms is illustrated in \fig{letut} where
we have shown the percentage one-loop/Born corrections ($b_1\,X$) and
two-loop/one-loop ($b_2\,X/b_1$) for different values of the Higgs boson
mass; note that the large two-loop effect around $\mh = 200\,$GeV is due to 
an accidental cancellation which occurs at one loop level and that for high 
values of $\mh$ the ratio two-loop/one-loop clearly starts to
indicate a questionable regime for the perturbative expansion.
\section{Loop diagrams with dressed propagators \label{LPDP}}
\begin{figure}[th]
\vspace{0.5cm}
\[
  \vcenter{\hbox{
\hspace{-2.cm}
  \begin{picture}(800,0)(0,0)
  \SetWidth{1.2}
  \SetColor{Red}
  \CArc(150,0)(25,0,180)
  \SetColor{Black}
  \DashLine(100,0)(125,0){4.}
  \DashCArc(150,0)(25,180,360){4.}
  \DashLine(175,0)(200,0){4.}
  \Text(125,-40)[cb]{a) skeleton}
  \Text(100,5)[cb]{$\phi$}
  \Text(200,5)[cb]{$\phi$}
  \Text(225,-6)[cb]{{\Large$+$}}
  \SetColor{Red}
  \CArc(300,0)(25,0,25)
  \CArc(300,0)(25,155,180)
  \SetColor{Black}
  \DashLine(250,0)(275,0){4.}
  \DashCArc(300,0)(25,-180,0){4.}
  \DashCArc(300,0)(25,25,155){4.}
  \DashCArc(300,50)(45,-120,-60){4.}
  \DashLine(325,0)(350,0){4.}
  \Text(275,-40)[cb]{b) $\Sigma$ insertion}
  \Text(375,-6)[cb]{{\Large$+$}}
  \SetColor{Red}
  \CArc(450,0)(25,90,180)
  \CArc(450,0)(25,270,360)
  \SetColor{Black}
  \DashLine(400,0)(425,0){4.}
  \DashCArc(450,0)(25,0,90){4.}
  \DashCArc(450,0)(25,180,270){4.}
  \DashLine(450,25)(450,-25){4.}
  \DashLine(475,0)(500,0){4.}
  \Text(425,-40)[cb]{c) skeleton}
  \end{picture}}}
\]
\vspace{1.2cm}
\caption[]{The $\phi$ self-energy with skeleton expansion, diagrams a) and c),
and insertion of a sub-loop $\Sigma_{\Phi\Phi}$, diagram b).}
\label{dressed}
\end{figure}
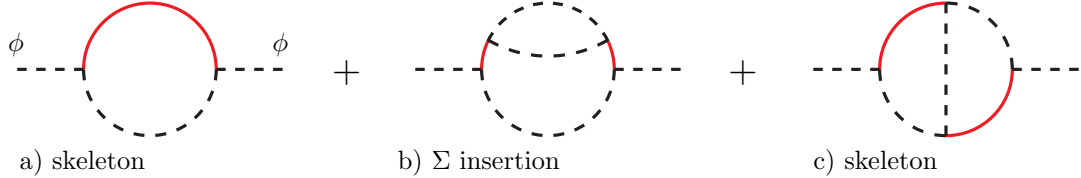

In this section we describe how to reorganize perturbation theory when 
using dressed propagators, the so-called skeleton expansion: 
consider a simple model~\cite{Veltman:1963th} with an interaction Lagrangian
\bq 
L = \frac{g}{2}\,\Phi(x)\,\phi^2(x).
\eq
The mass $M$ of the $\Phi\,$-field and $m$ of the $\phi\,$-field be such that 
the $\Phi\,$-field be unstable. Let $\Delta_i$ be the lowest order propagators
and $\oD_i$ the one-loop dressed propagators, i.e.
\bq
\oD_{\Phi} = 
\frac{\Delta_{\Phi}}{1 - \Delta_{\Phi}\,\Sigma_{\Phi\Phi}},
\qquad
\oD_{\phi} = 
\frac{\Delta_{\phi}}{1 - \Delta_{\phi}\,\Sigma_{\phi\phi}},
\eq
etc. In fixed order perturbation theory, the $\phi$ self-energy is
given ~\cite{Actis:2006xf} in \fig{dressed}.
Note that the imaginary part of $\Sigma_{\phi\phi}$ is non-zero only
for $- p^2 > 9\,m^2$ (the three-particle cut of diagram b) in \fig{dressed}), 
if $m \ll M$.
When we use dressed propagators only diagrams a) and c) are retained in 
\fig{dressed} (for two-loop accuracy) but in a) we use $\oD_{\Phi}$ with 
one-loop accuracy:
\bq
\Sigma^{(a)}_{\phi\phi} = \int\,
\frac{d^n q}{ \lpar q^2 + M^2 - 
\frac{g^2}{16\,\pi^2}\,\Sigma_{\Phi\Phi}(q^2) \rpar\,
\lpar \lpar q + p\rpar^2 + m^2\rpar},
\quad
\Sigma_{\Phi\Phi}(q^2) = B_0(q^2\,;\,m\,,\,m),
\eq
where we assume $p^2 < 0$. Since the complex $\Phi$ pole is defined by
\bq
M^2 - s_{\ssM} - \frac{g^2}{16\,\pi^2}\,\Sigma_{\Phi\Phi}(- s_{\ssM}) = 0,
\eq
we write the inverse (dressed) propagator as
\bq
\Bigl[ 1 - \frac{g^2}{16\,\pi^2}\,
\frac{\Sigma_{\Phi\Phi}(q^2) - \Sigma_{\Phi\Phi}(- s_{\ssM})}
{q^2 + s_{\ssM}} \Bigr]\lpar q^2 + s_{\ssM}\rpar,
\eq
expand in $g$ as if we were in a gauge theory with problems of gauge parameter
independence and obtain
\bqa
\Sigma^{(a)}_{\phi\phi} &=& g^2\,\int\,
\frac{d^n q}{\lpar q^2 + s_{\ssM}\rpar\,
\lpar\lpar q + p\rpar^2 + m^2\rpar}\,\Bigl[
1 + \frac{g^2}{16\,\pi^2}\,
\frac{\Sigma_{\Phi\Phi}(q^2) - \Sigma_{\Phi\Phi}(- s_{\ssM})}
{q^2 + s_{\ssM}} \Bigr]
\nl
{}&=& \frac{i}{2}\,g^2\,\pi^2\,
B_0 \lpar 1,1\,;\,p^2\,;\,s_{\ssM}\,,\,m^2 \rpar
+ i\,\frac{g^4}{16}\,
S^{\ssE}\lpar p^2\,;\,m^2,m^2,s_{\ssM},m^2,s_{\ssM}\rpar
\nl
{}&+&
i\,\frac{g^4}{16}\,B_0 \lpar 2,1\,;\,p^2\,;\,s_{\ssM}\,,\,m^2 \rpar\,
\Bigl[ \Delta_{\ssU\ssV} - \ln\frac{m^2}{\mu^2}  + 2 -
\beta\,\ln\frac{\beta+1}{\beta-1}\Bigr],
\eqa
where $\beta^2 = 1 - 4\,m^2/s_{\ssM}$. Note that there is an interplay between 
using dressed propagators for all internal lines of a diagram and 
combinatorial factors for diagrams with and without dressed propagators.
Note also that the poles in the $q^0\,$ complex plane remain in the same 
quadrants as in the Feynman prescription and Wick rotation can be carried out,
as usual.
Evaluation of diagrams with complex masses does not pose a serious problem;
in the analytical approach one should, however, pay the due attention to
splitting of logarithms. Consider a $B_0$ function,
\bq
B_0(p^2\,;\,M_1\,,\,M_2) = \DUV- \intfx{x}\,\frac{\chi(x)}{\mu^2},
\qquad \chi(x)= - p^2\,x^2 + (p^2 + M^2_2 - M^2_1)\,x + M^2_1 - i\,\delta,
\eq
($\delta \to 0_+$ ) where one usually writes
\bq
\ln \frac{\chi(x)}{\mu^2} = \ln ( - \frac{p^2}{\mu^2} - i\,\delta ) + 
\ln(x - x_-) + \ln(x - x_+).
\eq
Since $\Imb\,\chi(x)$ does not change sign with $x \in [0,1]$ the correct 
recipe for $M^2= m^2 - i\,m\,\gamma$ is
\bq
\ln \frac{\chi(x)}{\mu^2} = \ln |\,p^2\,| + \ln(x - x_-) +
\sum_{\lambda=\pm}\,
\theta(\lambda\,p^2)\,\Bigl[ \ln \lambda\,(x_+ - x) + 
\eta( - x_-\,,\,\lambda x_+)\Bigr],
\eq
Where $\eta$ is the 't Hooft - Veltman function. In the numerical treatment, 
instead, no splitting is performed and no special care is needed. This is 
specially true for higher legs one-loop functions and for two-loop functions.

A $t\,$-channel propagator deserves some additional comment: one should not 
confuse the position of the pole which is always at $\mu^2 - i\,\mu\,\gamma$
with the fact that a dressed propagator function is real in the $t\,$-channel.
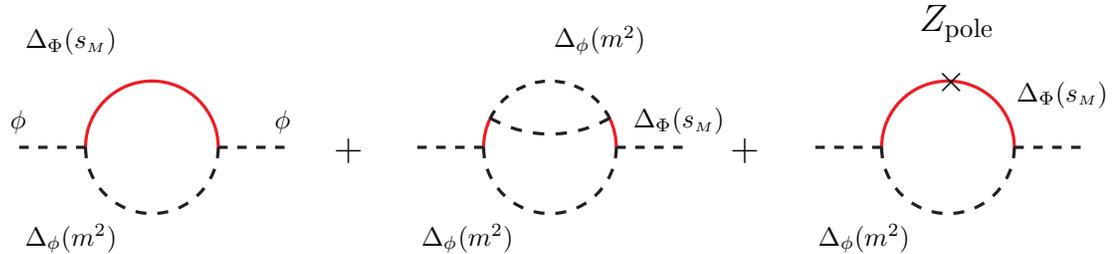
\begin{figure}[th]
\vspace{1.5cm}
\[
  \vcenter{\hbox{
\hspace{-2.cm}
  \begin{picture}(800,0)(0,0)
  \SetWidth{1.2}
  \SetColor{Red}
  \CArc(150,0)(25,0,180)
  \SetColor{Black}
  \DashLine(100,0)(125,0){4.}
  \DashCArc(150,0)(25,180,360){4.}
  \DashLine(175,0)(200,0){4.}
  \Text(100,5)[cb]{$\phi$}
  \Text(200,5)[cb]{$\phi$}
  \Text(120,35)[cb]{$\Delta_{\Phi}(s_{\ssM})$}
  \Text(120,-40)[cb]{$\Delta_{\phi}(m^2)$}
  \Text(225,-6)[cb]{{\Large$+$}}
  \SetColor{Red}
  \CArc(300,0)(25,0,25)
  \CArc(300,0)(25,155,180)
  \SetColor{Black}
  \DashLine(250,0)(275,0){4.}
  \DashCArc(300,0)(25,-180,0){4.}
  \DashCArc(300,0)(25,25,155){4.}
  \DashCArc(300,50)(45,-120,-60){4.}
  \DashLine(325,0)(350,0){4.}
  \Text(270,-40)[cb]{$\Delta_{\phi}(m^2)$}
  \Text(320,35)[cb]{$\Delta_{\phi}(m^2)$}
  \Text(350,5)[cb]{$\Delta_{\Phi}(s_{\ssM})$}
  \Text(375,-6)[cb]{{\Large$+$}}
  \SetColor{Red}
  \CArc(450,0)(25,0,180)
  \SetColor{Black}
  \DashLine(400,0)(425,0){4.}
  \DashCArc(450,0)(25,180,360){4.}
  \DashLine(475,0)(500,0){4.}
  \Text(453,20)[cb]{{\Large$\times$}}
  \Text(455,40)[cb]{{\Large$Z_{\rm pole}$}}
  \Text(420,-40)[cb]{$\Delta_{\phi}(m^2)$}
  \Text(495,15)[cb]{$\Delta_{\Phi}(s_{\ssM})$}
  \end{picture}}}
\]
\vspace{1.2cm}
\caption[]{Diagram b) of \fig{dressed} with one-loop dressed $\Phi$ 
propagators is equivalent, up to $\ord{g^4}$, to the sum of three diagrams
with lowest order propagators with the $\Phi$ mass replaced with
the $\Phi$ complex pole. The $Z_{\rm pole}$ vertex is given in \eqn{Zpole}}
\label{dressedequiv}
\end{figure} 
Therefore, using one-loop diagrams with one-loop dressed $\Phi$ propagators
is equivalent, to $\ord{g^4}$, to introduce the sum of the three diagrams of 
\fig{dressedequiv} where $\Phi$ propagators are at lowest order but with
complex mass $s_{\ssM}$ and where the vertex $Z_{\rm pole}$ is defined by
\bq
Z_{\rm pole} = \frac{g^2}{16\,\pi^2}\,
B_0 \lpar - s_{\ssM}\,;\,m\,,\,m\rpar.
\label{Zpole}
\eq
In the following sections we will discuss an extension of this simple scalar 
theory.
\section{Unitarity, gauge parameter independence and WST identities
\label{UgpIWI}}
The simple scalar model of \sect{LPDP} is not adequate for describing the 
complexity of a gauge theory.
A critical question is how to construct a scheme for a gauge theory with
unstable particles which allows to deal with the calculation of physical 
processes at one and two loops. This scheme must satisfy a certain number
of requisites, in particular we seek for a scheme that
\begin{itemize}

\item[a)] respects the unitarity of the $S-$matrix;

\item[b)] gives results that are gauge-parameter independent;

\item[c)] satisfies the whole set of WST identities.

\end{itemize}
Resummation will be part of any scheme, a fact that introduces additional
subtleties if $a)-c)$ are to be respected. Consider in more details the
definition of dressed propagator: we consider a skeleton expansion of
the self-energy $\Sigma$ with propagators that are resummed up to $\ord{n}$
and define
\bq
\Delta^{(n)}(p^2) = \Delta^{(0)}(p^2)\,\Bigl[
\Delta^{(0)}(p^2) - \Sigma^{(n)}\lpar p^2\,,\,
\Delta^{(n-1)}(p^2)\rpar\Bigr]^{-1},
\eq
where the Born propagator (tensor structures are easily included) is
\bq
\Delta^{(0)}(p^2) = \frac{1}{p^2 + m^2}.
\eq
If it exists, we define a dressed propagator as the formal limit
\bq
\oD(p^2) = \lim_{n \to \infty}\,\Sigma^{(n)}(p^2),
\qquad
\oD(p^2) =
\Delta^{(0)}(p^2)\,\Bigl[
\Delta^{(0)}(p^2) - \Sigma\lpar p^2\,,\,\oD(p^2)\rpar\Bigr]^{-1},
\label{dr}
\eq 
which is not equivalent to a {\em rainbow} approximation and coincides with 
the Schwinger - Dyson solution for the propagator.
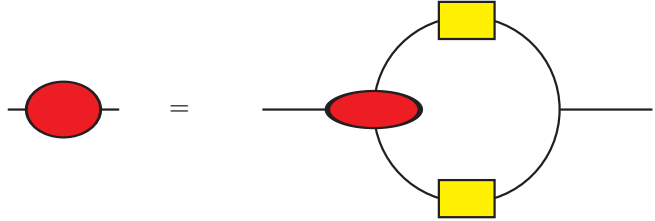
\begin{figure}[th]
\vspace{1.cm}
\bqas  
{}&{}&
  \vcenter{\hbox{
  \begin{picture}(0,0)(100,0)
  \SetScale{0.7}
  \SetWidth{1.2}
  \Line(0,0)(60,0)
  \COval(30,0)(15,20)(0){Black}{Red}
  \Text(65,-2)[cb]{$=$}
  \end{picture}}}
\qquad\qquad 
  \vcenter{\hbox{
  \begin{picture}(0,0)(40,0)
  \SetScale{0.7}
  \SetWidth{1.2}
  \Line(-10,0)(50,0)
  \CArc(100,0)(50,0,360)
  \Line(150,0)(200,0)
  \COval(50,0)(10,25)(0){Black}{Red}
  \CBoxc(100,-48)(30,20){Black}{Yellow}
  \CBoxc(100,48)(30,20){Black}{Yellow}
  \end{picture}}}
\eqas
\vspace{1.cm}
\caption[]{Schwinger - Dyson equation for the self-energy (l.h.s). In the 
r.h.s. the red oval is the SD vertex and the yellow box is the dressed 
propagator.}
\label{SDEse}
\end{figure}
\vspace{1.cm}
\begin{figure}[th]
\vspace{1.cm}
\bqas  
{}&{}&
  \vcenter{\hbox{
  \begin{picture}(0,0)(100,0)
  \SetScale{0.7}
  \SetWidth{1.2}
  \Line(0,0)(25,0)
  \CBox(25,-10)(50,10){Black}{Yellow}
  \Line(50,0)(75,0)
  \Text(65,-2)[cb]{$=$}
  \Line(115,0)(190,0)
  \Text(160,-3)[cb]{$+$}
  \end{picture}}}
\qquad\qquad 
\qquad\qquad
\qquad\qquad
  \vcenter{\hbox{
  \begin{picture}(0,0)(40,0)
  \SetScale{0.7}
  \SetWidth{1.2}
  \Line(0,0)(150,0)
  \COval(50,0)(15,20)(0){Black}{Red}
  \CBox(95,-10)(120,10){Black}{Yellow}
  \end{picture}}}
\eqas
\vspace{0.3cm}
\caption[]{Schwinger-Dyson equation for a dressed propagator (yellow box);
the red oval is the SD self-energy.}
\label{SDEdp}
\end{figure}
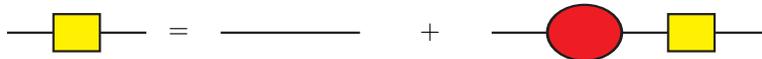
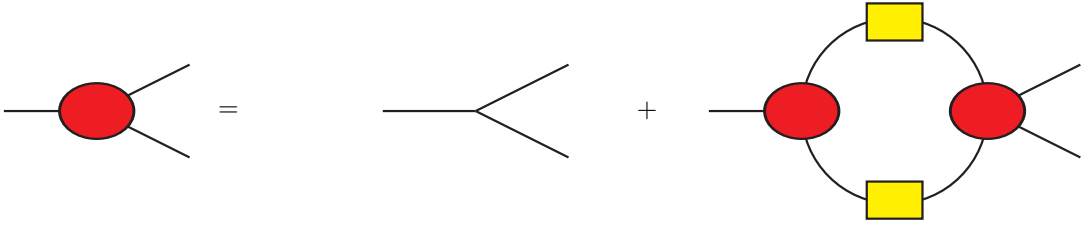
\begin{figure}[th]
\vspace{1.cm}
\bqas  
{}&{}&
  \vcenter{\hbox{
  \begin{picture}(0,0)(100,0)
  \SetScale{0.7}
  \SetWidth{1.2}
  \Line(0,0)(50,0)
  \Line(50,0)(100,25)
  \Line(50,0)(100,-25)
  \COval(50,0)(15,20)(0){Black}{Red}
  \Text(85,-2)[cb]{$=$}
  \end{picture}}}
\qquad\qquad 
\qquad\qquad
  \vcenter{\hbox{
  \begin{picture}(0,0)(40,0)
  \SetScale{0.7}
  \SetWidth{1.2}
  \Line(0,0)(50,0)
  \Line(50,0)(100,25)
  \Line(50,0)(100,-25)
  \Text(100,-3)[cb]{$+$}
  \end{picture}}}
\qquad\qquad 
\qquad\qquad
\qquad\qquad
  \vcenter{\hbox{
  \begin{picture}(0,0)(40,0)
  \SetScale{0.7}
  \SetWidth{1.2}
  \Line(0,0)(50,0)
  \Line(150,0)(200,25)
  \Line(150,0)(200,-25)
  \CArc(100,0)(50,0,360)
  \COval(50,0)(15,20)(0){Black}{Red}
  \COval(150,0)(15,20)(0){Black}{Red}
  \CBoxc(100,48)(30,20){Black}{Yellow}
  \CBoxc(100,-48)(30,20){Black}{Yellow}
  \end{picture}}}
\eqas
\vspace{1.cm}
\caption[]{Schwinger-Dyson equation for a dressed vertex (l.h.s); in the
r.h.s. we have SD three- and four-point vertices (red ovals) and dressed
propagators (yellow boxes).}
\label{SDEdv}
\end{figure}

Technically speaking, one should also introduce an integral equation for the 
four-point functions, as they appear in \fig{SDEdv}. However, this is well
beyond the scope of the present discussion and its structure shows a 
somewhat greater complexity. In discussing unitarity an essential tool is
representend by the so-called cutting rules.
\begin{itemize}
\item[--] {\bf Cutting rules}
\end{itemize}
We assume that \eqn{dr} has a solution that obeys K\"allen - Lehmann 
representation,
\bq
\Reb\,\oD(p^2) = \Imb\,\Sigma(p^2)\,
\Bigl[ \lpar p^2 + m^2 - \Reb\,\Sigma(p^2)\rpar^2 + 
\lpar \Imb\,\Sigma(p^2)\rpar^2\Bigr]^{-1} = \pi\,\rho(-\,p^2).
\eq
A dressed propagator, being the result of an infinite number of iterations,
\bq
\oD(p^2) = \int_0^{\infty}\,ds\,
\frac{\rho(s)}{p^2 + s - i\,\delta},
\label{KLrep}
\eq
is a formal object which is difficult to handle for all practical purposes.

Unitarity follows if we add all possible ways in which a diagram with
given topology can be cut in two. The shaded line separates $S$ from 
$S^{\dagger}$. For a stable particle the cut line, proportional to 
$\oD^{+}$, contains a pole term
\bq
\oD^{+} = 2\,i\,\pi\,\theta(p_0)\,\delta(p^2+m^2),
\eq
whereas there is no such contribution for an unstable particle. We express
$\Imb\,\Sigma$ in terms of cut self-energy diagrams and repeat the procedure
ad libidum and prove that cut unstable lines are left with no contribution, 
i.e. unstable particles contribute to the unitarity of the $S-$matrix via
their stable decay products. 
\begin{figure}[th]
\vspace{1.cm}
\bqas  
{}&{}&
  \vcenter{\hbox{
  \begin{picture}(0,0)(100,0)
  \SetWidth{1.2}
  \Line(0,0)(25,0)
  \CBox(25,-10)(50,10){Black}{Yellow}
  \Line(50,0)(75,0)
  \SetColor{Blue}
  \LinAxis(32,-25)(45,25)(25,1,-1,0,1)
  \SetColor{Black}
  \Text(120,-2)[cb]{$=$}
  \end{picture}}}
\qquad\qquad 
\qquad\qquad
\qquad\qquad
\qquad\qquad
  \vcenter{\hbox{
  \begin{picture}(0,0)(100,0)
  \SetWidth{1.2}
  \Line(0,0)(25,0)
  \CBox(25,-10)(50,10){Black}{Yellow}
  \Line(50,0)(75,0)
  \COval(100,0)(25,25)(0){Black}{Red}
  \Line(125,0)(150,0)
  \CBox(150,-10)(175,10){Black}{Yellow}
  \Line(175,0)(200,0)
  \end{picture}}}
  \SetColor{Blue}
  \LinAxis(-10,-35)(10,35)(20,1,-1,0,1)
  \SetColor{Black}
\eqas
\vspace{0.5cm}
\caption[]{Cutting equation for dressed propagator.}
\label{cutdp}
\end{figure}
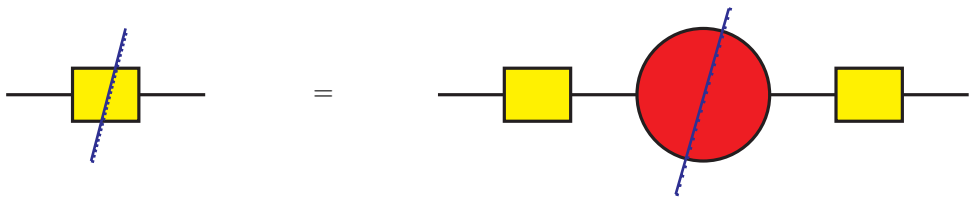
\vspace{0.5cm}
The consistent use of dressed propagators gives a general scheme where
unitarity is satisfied which is essentially a statement on the imaginary
parts of the diagrams. Approximated, or truncated, schemes (e.g.
resummation of one-loop self energies, or {\em rainbow} approximation
without further resummation of the vertex functions) usually lead to
gauge dependent results.
\begin{itemize}
\item[--] {\bf WST identities}
\end{itemize}
We assume that WST identities hold at any fixed order in perturbation theory
for diagrams that contain bare propagators and vertices; they again form 
dressed propagators and vertices when summed.
We expect that an arbitrary truncation that preferentially resums specific 
topologies will lead to violations of WST identities. These violations, of 
course, are not present in the exact calculations.
\begin{itemize}
\item[--] {\bf Gauge parameter dependence}
\end{itemize}
A truncated approximation, e.g. simple resummation of two-point functions, 
necessarily leads to gauge dependent results. A convenient tool is to
analyze the gauge invariance of the effective action where one can show
that on-shell gauge dependence always occurs at higher order than the order
of truncation.
\begin{itemize}
\item[--] {\bf Complex pole}
\end{itemize}
A property of the $S-$matrix is the complex pole 
$\oD^{-1}(p^2\,=\,-s_{\ssP}) = 0$,
which is gauge parameter independent as shown by a study of Nielsen identities,
see \sect{GSSE}.

\begin{figure}[th]
\vspace{2.cm}
\bqas  
{}&{}&
  \vcenter{\hbox{
  \begin{picture}(0,0)(100,0)
  \SetWidth{1.2}
  \Line(0,0)(50,0)
  \CArc(75,0)(25,0,360)
  \Line(100,0)(150,0)
  \COval(75,25)(5,5)(0){Black}{Red}
  \COval(75,-25)(5,5)(0){Black}{Red}
  \SetColor{Blue}
  \LinAxis(60,-35)(90,35)(25,1,-1,0,1)
  \SetColor{Black}
  \Text(160,-2)[cb]{$\approx$}
  \end{picture}}}
\qquad\qquad 
\qquad\qquad
\qquad\quad
  \vcenter{\hbox{
  \begin{picture}(0,0)(40,0)
  \SetWidth{1.2}
  \Line(0,0)(50,0)
  \Line(150,0)(200,0)
  \CArc(100,0)(50,0,67)
  \CArc(100,0)(50,113,180)
  \CArc(100,0)(50,180,247)
  \CArc(100,0)(50,293,360)
  \ArrowArc(100,50)(20,0,180)
  \ArrowArc(100,50)(20,180,360)
  \ArrowArc(100,-50)(20,0,180)
  \ArrowArc(100,-50)(20,180,360)
  \SetColor{Blue}
  \LinAxis(90,-80)(110,80)(40,1,-1,0,1)
  \SetColor{Black}
  \end{picture}}}
\eqas
\vspace{1.5cm}
\caption[]{Cutting equation for a contribution to the $Z$ self-energy
using $W$ propagators of \eqn{cpapp}.}
\label{CEcpapp}
\end{figure}
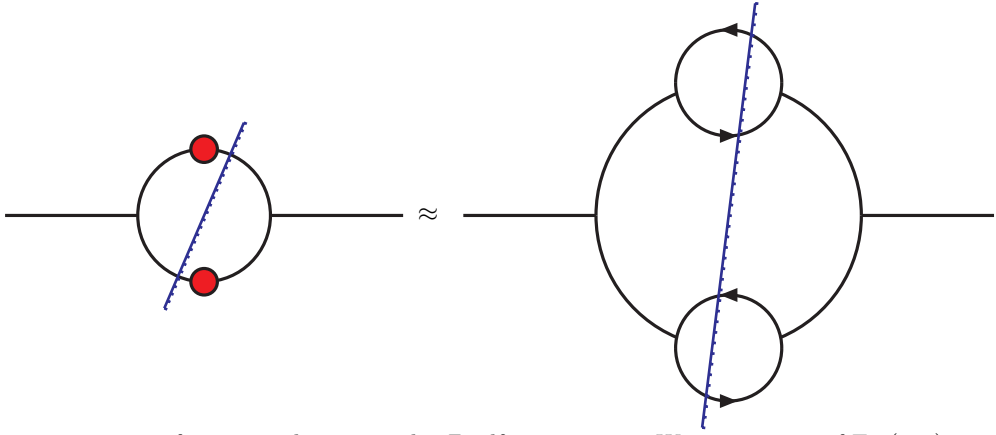

An approximate solution of the unitarity constraint is as follows:
\bq
2\,\Imb\,T_{ii} = \sum_n\,\bmid T_{ni}\bmid^2,
\qquad
\sum_n\,\bmid T_{ni}\bmid^2 = \bmid D(p^2)\bmid^2\,\sum_n\,\int\,dPS_n\,
\bmid M_{1\,\to\,n}\bmid^2,
\eq
where, $S = 1 +i\,T$ and where $D(p^2)$ is the unknown form of the propagator.
Making the approximation,
\bq
\sum_n\,\int\,dPS_n\,\bmid M^{1\,\to\,n}\bmid^2 \equiv m\,\Gamma_{\rm tot},
\label{wapp}
\eq
we derive $\Imb\,D(p^2) = m\,\Gamma_{\rm tot}$.
A simple but, once again, approximate solution is 
\bq
D(p^2) = \lpar p^2 + m^2 - i\,m\Gamma_{\rm tot} \rpar^{-1},
\label{fapp}
\eq
which is valid far from the mass shell and where the invariant mass at which 
the decay is evaluated is identified with $m^2$. We can improve upon this 
solution by writing instead
\bq
D(p^2) = \lpar p^2 - s_{\ssP} \rpar^{-1},
\label{cpapp}
\eq
which is equivalent to resum only the self-energy (up to some fixed order),
and to use
\bq
m^2 = s_{\ssP} + \Sigma(s_{\ssP}),
\qquad
D(p^2) = -\,\Bigl[ s - s_{\ssP} - \Sigma(s) + \Sigma(s_{\ssP})\Bigr]^{-1} =
-\,\lpar p^2 - s_{\ssP} \rpar^{-1} + \mbox{h.o.},
\eq
where higher order terms are neglected. Another way to see that \eqn{cpapp}
is an improvement of \eqn{fapp} is to observe that
\bq
p^2 + m^2 + i\,\frac{\Gamma_{\rm tot}}{m}\,p^2 = 
\lpar 1 + i\,\frac{\Gamma_{\rm tot}}{m}\rpar\,( p^2 + s_{\ssP} ) +
\mbox{h.o.} \approx p^2 + s_{\ssP}.
\eq
A propagator with the correct analytical structure, $p^2 - s_{\ssP}$, will be 
represented with a thick dot. The approximation of \eqn{cpapp} allows us to 
write the cutting equation of \fig{CEcpapp}.
One can see that using truncated propagators with complex poles (at the 
one-loop level of accuracy) is still respecting unitarity of the $S-$matrix
within the approximation of \eqn{wapp} if the complex pole is computed from
fermions only; however, this scheme violates gauge invariance since vertices
are not included. There is a solution to this problem, namely replacing
everywhere the (real) masses with the complex poles, couplings included;
this is known in the literature as complex mass scheme~\cite{Denner:2005fg}

\begin{itemize}
\item[--] {\bf The complex mass scheme}
\end{itemize}
Since WST identities are algebraic relations satisfied separately by the real
and the imaginary part one starts from WST identities with real masses,
satisfied at any given order, replaces everywhere $m^2 \to s_{\ssP}$ without
violating the invariance.
In this scheme, using for instance IPS I(II), also $\stws$ is expanded around
a complex value,
\bq
\shs =  \frac{1}{2}\,\frac{\pi\,\alpha}{G\,\cpw}, \quad \mbox{or} \quad
2\,G\,\cpz = \frac{\pi\,\alpha_{\ssZ}}{\shs\,\chs}.
\eq
In turns, this scheme violates unitarity i.e. we cannot identify the two 
sides of any cut diagram with $T$ and $T^{\dagger}$ respectively. 

To summarize, the analytical structure of the $S-$matrix is correctly 
reproduced when we use propagator factors $p^2 + s_{\ssP}$ but unitarity of
$S$ requires more, a dressed propagator
\renewcommand{\arraystretch}{1.3}
\renewcommand{\tabcolsep}{15pt}
\begin{center}
\begin{tabular}[t]{|c|c|}   
\hline
$p^2 + s_{\ssP}$ & $p^2 + s_{\ssP} - \Sigma(p^2) + \Sigma(-\,s_{\ssP})$ \\
\hline
analyticity & unitarity \\
\hline
\end{tabular}
\end{center}
\renewcommand{\tabcolsep}{6pt}
\renewcommand{\arraystretch}{1}
Another drawback of the scheme is that all propagators for unstable particles
will have the same functional form both in the time-like and in the 
space-like region while, for a dressed propagator the presence of a pole
on the second Riemann sheet does not change the real character of the function
if we are in a $t-$channel.

In some sense the scheme becomes more appealing when we go beyond one loop.
As a preliminary step we have verified that WST identities are satisfied with
bare (i.e. non-dressed) propagators and vertices up to two loops; we may assume
that they are verified order by order to all orders, 
\bq
W^{(1)}\lpar \{\Gamma\} \rpar = W^{(2)}\lpar \{\Gamma\} \rpar =
\,\cdots\, = 0,
\eq
where $\{\Gamma\}$ is a set of (off-shell) Green function and $W = 0$ is the
WST identity.

Next we write the same set of WST identities but using a skeleton expansion
with one-loop dressed propagators. Calling the scheme {\em complex mass
scheme} is somehow misleading; to the requested order we replace everywhere
$m^2$ with $s_{\ssP} + \Sigma(s_{\ssP})$ which is real by construction.
If only one-loop is needed then $m^2 \to s_{\ssP}$ everywhere (therefore
justifying the name {\em complex mass}) and
\bq
W^{(1)}\lpar \{\Gamma\}\rpar\bmid_{m^2\,=\,s_{\ssP}} = 0,
\label{wstol}
\eq
is trivially true. Also,
\bq
W^{(2)}\lpar \{\Gamma\}\rpar\bmid_{m^2\,=\,s_{\ssP}} = 0.
\label{wsttl}
\eq
We have two-loop diagrams with no self-energy insertions where 
$m^2 = s_{\ssP}$ and one-loop diagrams where $m^2 = s_{\ssP} + 
\Sigma(s_{\ssP})$ and the factor
\bq
\frac{\Sigma(p^2) - \Sigma(s_{\ssP})}{p^2 + s_{\ssP}},
\label{fexp}
\eq
expanded to first order with $\Sigma = \Sigma^{(1)}$. Furthermore, in vertices
we use $m^2 = s_{\ssP}$ in two-loop diagrams and $m^2 = s_{\ssP} + 
\Sigma(s_{\ssP})$ in one-loop diagrams. Expanding the factor of \eqn{fexp}
generates two-loop diagrams with insertion of one-loop self-energies plus
one-loop diagrams with one more propagator and a vertex proportional to
$\Sigma(s_{\ssP})$; furthermore one-loop diagrams with $m^2$ dependent
vertices get multiplied by $\Sigma(s_{\ssP})$; it follows that
\bq
W^{(1+2)}\lpar \{\Gamma\}_{\rm skeleton/expanded} 
\rpar\bmid_{m^2\,=\,s_{\ssP}+\Sigma(s_{\ssP})} = 0,
\eq
as a consequence of \eqns{wstol}{wsttl}.  

In this scheme, one replaces \eqn{EQrenW} by
\bq
\cpw = M^2 - \frac{g^2}{16\,\pi^2}\,\Bigl[
\Sigma_{3\ssQ}(\cpw) + F_{\ssW}(\cpw)\Bigr]
\eq
and organizes the perturbative expansion according to strategy outlined in 
this section, namely, skeleton expansion expanded around complex poles. Note
that $M$ is always a real quantity.
v
\section{Beyond the complex mass scheme: outlook \label{giveitatry}}
A possible improvement of the complex mass scheme which also respects the 
unitarity of the $S-$matrix can be constructed when we observe that
unstable particles contribute to the unitarity of the $S-$matrix via their 
stable decay products, i.e. light fermions. Furthermore, at one loop, 
the so called fermion-loop scheme respects all WST identities.

The birth of a complex pole is full of subtleties; the origin is the
instability of the corresponding quantum state, reflected by the fact that
the self-energy develops an imaginary part. Several statements that are
usually made in this context are questionable and should be understood with
the due caution. We have a few ingredients, a bare (renormalized) mass
and the corresponding complex pole; we observe that $\Reb\,\cpv$ and 
$M^2_{\ssV}$ differ because of $\Imb\,\Sigma_{\ssVV} \not= 0$. A criterion
of naturalness (which also should be taken with the due caution) requires
a small difference, fully accountable in perturbation theory.

Consider the toy model of \sect{LPDP}; the usual statement that {\em
the mass $M$ of the $\Phi\,$-field and $m$ of the $\phi\,$-field be such that 
the $\Phi\,$-field be unstable} is based on the assumption of naturalness
since both masses are not input data but should be derived by using some IPS,
i.e. no assumption can be made a priori on Lagrangian parameters.
Indeed one could introduce the following paradox: assume that $m$ and $M$
have been derived with some IPS and that they respect the condition $m \ll M$;
consider the $\phi$ self-energy given by diagram a) of \fig{dressed}; look 
for a complex solution of
\bq
m^2 - s_m - \frac{g^2}{16\,\pi^2}\,\Sigma(-s_m) = 0,
\label{parad}
\eq
where, for $\Reb\,s_m$ large enough, $\Imb\,\Sigma(s_m) \not= 0$. The paradox
is that we end up with $\Reb\,s_m \gg m^2$ and the notion of stable particle
is lost. The resolution of the paradox lies in the observation that $\Phi$
is unstable and must be removed from the in/out bases of the Hilbert space;
in other words we cannot derive anything by naively cutting diagram a)
of \fig{dressed}. This diagram, with a bare $\Phi$ propagator, should not
be there and only its dressed version should be considered; therefore,
the imaginary part starts with the three-particle cut of diagram b) which
enters whenever a virtual $\phi$ appears in a diagram.
The solution of \eqn{parad} is real and $\phi$ is stable.

Consider now a gauge theory: how is the complex pole affected by bosonic 
corrections? Assume that $\mw$ and $\mz$ (both renormalized parameters) have
been derived by some IPS and that, as expected, $\mz < 2\,\mw$. Due to
gauge invariance the one-loop $Z$ self-energy cannot develop an imaginary 
part, for $\Reb\,s_{\ssZ}$ close to $\mzs$, in diagrams with $W, \phi$ or
FP ghosts so that bosonic corrections only modify $\Reb\,s_{\ssZ}$ and
instability is a consequence of (light) fermion corrections. Even this
statement should be slightly modified: bosonic states enter through their
dressed propagators and when we cut the sequence stops only when cutting
(at some higher order in perturbation theory) (light) fermion lines.
Therefore, bosonic contributions change $\Imb\,s_{\ssZ}$ only at higher
orders through their stable decay products. Only light fermions matter and
this represents the basis for our proposal.

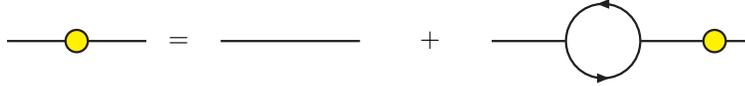
\begin{figure}[th]
\vspace{1.cm}
\bqas  
{}&{}&
  \vcenter{\hbox{
  \begin{picture}(0,0)(100,0)
  \SetScale{0.7}
  \SetWidth{1.2}
  \Line(0,0)(75,0)
  \CCirc(37.5,0){6}{Black}{Yellow}
  \Text(65,-2)[cb]{$=$}
  \Line(115,0)(190,0)
  \Text(160,-3)[cb]{$+$}
  \end{picture}}}
\qquad\qquad 
\qquad\qquad
\qquad\qquad
  \vcenter{\hbox{
  \begin{picture}(0,0)(40,0)
  \SetScale{0.7}
  \SetWidth{1.2}
  \Line(0,0)(40,0)
  \Line(80,0)(120,0)
  \Line(120,0)(140,0)
  \ArrowArc(60,0)(20,0,180)
  \ArrowArc(60,0)(20,180,360)
  \CCirc(120,0){6}{Black}{Yellow}
  \end{picture}}}
\eqas
\caption[]{One-loop dressed propagator in the fermion loop approximation.}
\label{FLdrP}
\end{figure}

\begin{figure}[th]
\vspace{1.cm}
\bqas  
{}&{}&
  \vcenter{\hbox{
  \begin{picture}(0,0)(100,0)
  \SetScale{0.7}
  \SetWidth{1.2}
  \Line(0,0)(40,0)
  \Line(40,0)(60,30)
  \Line(40,0)(60,-30)
  \CCirc(40,0){6}{Black}{Red}
  \Text(65,-2)[cb]{$=$}
  \end{picture}}}
\qquad\qquad 
  \vcenter{\hbox{
  \begin{picture}(0,0)(40,0)
  \SetScale{0.7}
  \SetWidth{1.2}
  \Line(0,0)(20,0)
  \ArrowArc(40,0)(20,-45,45)
  \ArrowArc(40,0)(20,45,180)
  \ArrowArc(40,0)(20,180,315)
  \Line(52,15)(60,30)
  \Line(52,-15)(60,-30)
  \Text(120,-2)[cb]{$+\;\;\;$ permutations}
  \end{picture}}}
\eqas
\vspace{0.5cm}
\caption[]{One-loop triple gauge boson vertex in the fermion loop 
approximation.}
\label{FLVVV}
\end{figure}
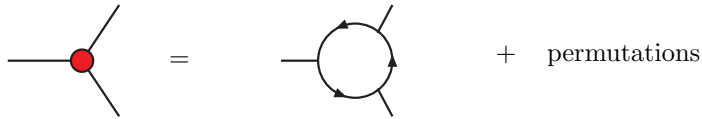

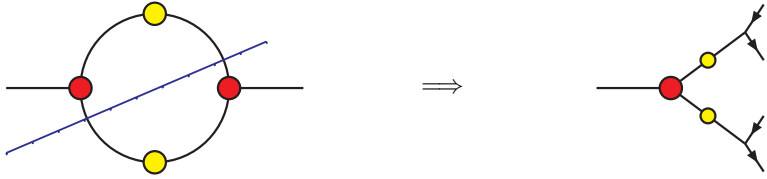
\begin{figure}[th]
\vspace{1.cm}
\bqas  
{}&{}&
  \vcenter{\hbox{
  \begin{picture}(0,0)(100,0)
  \SetScale{0.7}
  \SetWidth{1.2}
  \Line(0,0)(40,0)
  \Line(120,0)(160,0)
  \CArc(80,0)(40,0,90)
  \CArc(80,0)(40,90,180)
  \CArc(80,0)(40,180,270)
  \CArc(80,0)(40,270,360)
  \CCirc(40,0){6}{Black}{Red}
  \CCirc(120,0){6}{Black}{Red}
  \CCirc(80,40){6}{Black}{Yellow}
  \CCirc(80,-40){6}{Black}{Yellow}
  \SetColor{Blue}
  \LinAxis(0,-35)(140,25)(10,1,-1,0,1)
  \SetColor{Black}
  \Text(165,-2)[cb]{$\Longrightarrow$}
  \end{picture}}}
\qquad\qquad 
\qquad\qquad 
\qquad\qquad 
\qquad\qquad 
  \vcenter{\hbox{
  \begin{picture}(0,0)(40,0)
  \SetScale{0.7}
  \SetWidth{1.2}
  \Line(0,0)(40,0)
  \Line(40,0)(80,30)
  \Line(40,0)(80,-30)
  \ArrowLine(90,45)(80,30)
  \ArrowLine(80,30)(90,15)
  \ArrowLine(80,-30)(90,-45)
  \ArrowLine(90,-15)(80,-30)
  \CCirc(40,0){6}{Black}{Red}
  \CCirc(60,15){4}{Black}{Yellow}
  \CCirc(60,-15){4}{Black}{Yellow}
  \end{picture}}}
\eqas
\vspace{1.cm}
\caption[]{One-loop dressed propagator in the fermion loop approximation.}
\label{unitarity}
\end{figure}

\begin{figure}[th]
\vspace{1.cm}
\bqas  
{}&{}&
  \vcenter{\hbox{
  \begin{picture}(0,0)(100,0)
  \SetScale{0.7}
  \SetWidth{1.2}
  \Line(0,0)(40,0)
  \Line(40,0)(80,30)
  \Line(40,0)(80,-30)
  \ArrowLine(90,45)(80,30)
  \ArrowLine(80,30)(90,15)
  \ArrowLine(80,-30)(90,-45)
  \ArrowLine(90,-15)(80,-30)
  \CCirc(40,0){6}{Black}{Red}
  \CCirc(60,15){4}{Black}{Yellow}
  \CCirc(60,-15){4}{Black}{Yellow}
  \Text(120,-3)[cb]{$+$}
  \end{picture}}}
\qquad\qquad 
\qquad\qquad 
  \vcenter{\hbox{
  \begin{picture}(0,0)(40,0)
  \SetScale{0.7}
  \SetWidth{1.2}
  \Line(0,0)(40,0)
  \ArrowLine(90,45)(40,0)
  \ArrowLine(40,0)(90,-45)
  \Line(65,20)(75,10)
  \ArrowLine(90,25)(75,10)
  \ArrowLine(75,10)(90,-5)
  \CCirc(70,15){2}{Black}{Yellow}
  \Text(140,-2)[cb]{$+\;\;\;$ permutations}
  \end{picture}}}
\eqas
\vspace{1.cm}
\caption[]{Fermion-loop improved tree diagrams for $V \to 4\,f$.}
\label{All4f}
\end{figure}
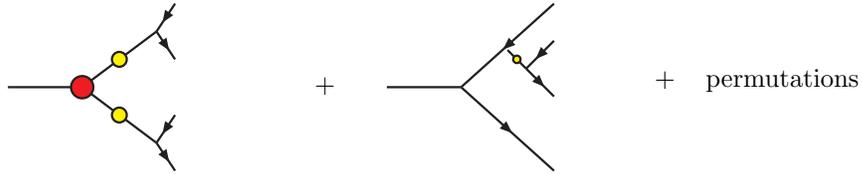

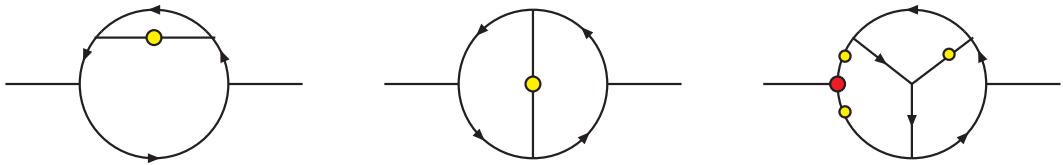
\begin{figure}[b]
\vspace{1.cm}
\bqas  
{}&{}&
  \vcenter{\hbox{
  \begin{picture}(0,0)(100,0)
  \SetScale{0.7}
  \SetWidth{1.2}
  \Line(0,0)(40,0)
  \ArrowArc(80,0)(40,180,360)
  \ArrowArc(80,0)(40,0,45)
  \ArrowArc(80,0)(40,45,135)
  \ArrowArc(80,0)(40,135,180)
  \Line(48,25)(113,25)
  \Line(120,0)(160,0)
  \CCirc(80,25){4}{Black}{Yellow}
  \end{picture}}}
\qquad\qquad 
\qquad\qquad 
  \vcenter{\hbox{
  \begin{picture}(0,0)(40,0)
  \SetScale{0.7}
  \SetWidth{1.2}
  \Line(0,0)(40,0)
  \ArrowArc(80,0)(40,180,270)
  \ArrowArc(80,0)(40,270,360)
  \ArrowArc(80,0)(40,0,90)
  \ArrowArc(80,0)(40,90,180)
  \Line(80,40)(80,-40)
  \Line(120,0)(160,0)
  \CCirc(80,0){4}{Black}{Yellow}
  \end{picture}}}
\qquad\qquad 
\qquad\qquad 
\qquad\qquad 
\qquad
  \vcenter{\hbox{
  \begin{picture}(0,0)(40,0)
  \SetScale{0.7}
  \SetWidth{1.2}
  \Line(0,0)(40,0)
  \Line(120,0)(160,0)
  \ArrowArc(80,0)(40,0,45)
  \ArrowArc(80,0)(40,45,135)
  \CArc(80,0)(40,135,180)
  \CArc(80,0)(40,180,270)
  \ArrowArc(80,0)(40,270,360)
  \ArrowLine(80,0)(80,-40)
  \ArrowLine(48,25)(80,0)
  \Line(113,25)(80,0)
  \CCirc(40,0){4}{Black}{Red}
  \CCirc(44,15){3}{Black}{Yellow}
  \CCirc(44,-15){3}{Black}{Yellow}
  \CCirc(100,16){3}{Black}{Yellow}
  \end{picture}}}
\eqas
\vspace{1.cm}
\caption[]{Examples of the multi-loop {\em bosonic} self-energy replacement.}
\label{urepl}
\end{figure}

Consider any unstable vector boson $V$ and construct the corresponding 
one-loop self-energy; a rearrangement of the perturbative expansion is 
required. First we define the $V$ dressed propagator according to \fig{FLdrP},
where only the one-loop fermion self-energy is resummed. Any triple gauge
boson vertex is also computed in the fermion-loop approximation of
\fig{FLVVV}.
In constructing $\Sigma{\ssVV}$ at one loop we only include 
$\Sigma^{\rm f}_{\ssVV}$. What about the bosonic part? Consider the loop
diagram in \fig{unitarity}; cutting dressed propagators gives stable
intermediate states with the result shown in the r.h.s of \fig{unitarity}.
Thus, the imaginary part of the diagram is one of the terms contributing to
$V \to 4\,f$ in the fermion loop approximation. Clearly this fermion-loop
improved tree diagram cannot respect gauge invariance by itself.
Consider all fermion-loop improved tree diagrams contributing to $V \to 4\,f$, 
as illustrated in \fig{All4f}; their sum is clearly gauge invariant and 
respects all WST identities. Using this total as the imaginary part of a 
self-energy we can reconstruct a self-energy which has the right properties 
wrt unitarity and gauge invariance.
The reorganization of perturbation theory replaces the simple one-loop
diagram in the $VV$ self-energy with internal unstable bosons with a series
of multi-loop self-energies where all internal boson lines represent dressed
propagators as in \fig{FLdrP}. An example is given in \fig{urepl}.

The next step in the calculation of the bosonic $VV$ self-energy starts with
the set of all tree diagrams contributing to $V \to 6\,f$ written with
the fermion loop improvement; after identifying this set with $T(1 \to 6)$
we can reconstruct another piece of the $VV$ self-energy, $T\,T^{\dagger}$,
using the relation of \fig{cutdp} and removing the cut on the dressed 
propagator. 

The same algorithm can be generalized to deal with $2 \to 2$ (or $2 \to n$) 
processes (no unstable particles in the initial and final state) by 
constructing the fermion-loop improved $T(2 \to 2(n))$, by writing all the 
cut diagrams with the cut separating $T$ from $T^{\dagger}$, using the 
relation of \fig{cutdp} and removing the cut on the dressed propagator. 

Computing loop integrals with dressed propagators is not the difficult part
in this extension of the familiar scheme; they satisfy \eqn{KLrep}, so that
we can compute loop integrals with masses $s_1,...,s_n$ and fold the
result with spectral functions, $\rho(s_i)$. The intrinsic obstacle, for
all practical purposes, is represented by the presence of multi-loops.
\section{Conclusions}
\label{fin}
In this paper, we begin with the (two-loop) renormalized Lagrangian of the 
Standard Model and show how to construct (pseudo-)observables up to two 
loops. For this we have written a set of (finite) renormalization equations 
(RE) and shown how to solve them. These equations are based on the choice of 
some input parameter set; in this case we use the Fermi coupling constant 
$\gf$, the fine structure constant $\alpha$ and complex poles for unstable 
vector bosons. The structure of our renormalization equations at two loops 
generalizes one-loop results that are well known in the 
literature~\cite{Bardin:1999ak}. 

At any order in perturbation theory quantities in the IPS -- like $\gf$ -- are
experimental data points and, for the specific example, we do not make any 
prediction for the two-loop decay rate, no more than we do for the two-loop
$Z$ complex pole. These quantities are input for the procedure and allow
for the determination of other observables. Of course, if we are really 
interested in the muon lifetime, then another data point instead of $\gf$ 
should be used and the lifetime would follow consistently. 
It is, however, an important observation that the correction $\Delta g$ of 
\eqn{deg}, which is known to be ultraviolet and infrared finite at one loop,
only needs one-loop counterterms to show the same property at two loops.

For the IPS specified by $\alpha, \gf$ and the $Z$ complex pole we have also 
considered a solution of the RE improved by resummation.
Note that the transition from real to pseudo-observables involves certain
assumptions but, within these assumptions, there is a well defined
mathematical procedure.

We have discussed the change of perspectives with respect to old one-loop 
calculations where one considers the on-shell masses, derived from a fit
to pseudo-observables (line-shapes etc.), as input parameters and 
derives complex poles in terms of the former. Since only complex poles are 
gauge parameter independent to all orders it follows that on-shell masses 
loose their central position in any renormalization procedure which goes 
beyond one loop.

We have also discussed how to generalize the concept of running couplings
and presented numerical results for the value of of an effective e.m,
coupling $\alpha$ at an arbitrary scale; this shows that not only the 
algorithmic part of our scheme is at work but also that its parallel 
implementation in a numerical code is complete~\cite{LB}.

It is an unfortunate consequence of gauge invariance that we cannot have a 
full extension of the running, resummed, couplings to the bosonic sector of the
theory; in any case, even the concept of a fermionic sector is meaningless at 
two loops.

We have devoted the last three sections to discuss renormalization of a gauge
theory in the presence of unstable particles, objects which should be removed
from the asymptotic states of the Hilbert space (in/out bases). Although 
there is a way of formulating the general treatment of the theory, as shown 
by Veltman in the sixties in his seminal paper, one has to admit that all 
practical approaches have loopholes since they, necessarily, rely on 
truncation of the Schwinger-Dyson equations.

By a suitable modification of the form of the solution of our renormalization 
equations we recover the complex mass scheme of Ref.~
(the {\em less inconsistent} and, so far, the only available scheme) and 
discuss its extension to two loops as well as a possible improvement aimed to 
satisfy the unitarity of the $S$-matrix.
\Acknowledgments
We gratefully acknowledge several discussions with Dima Bardin, Ansgar Denner, 
Stefan Dittmaier, Ayres Freitas and Mikhail Kalmykov as well as the unvaluable 
assistance of Sandro Uccirati in several steps of this project. S.~A. is
indebted to Chiara Arina for useful collaboration in an early stage of this 
paper. S.~A. is indebted to Nigel Glover for hospitality at the Institute for 
Particle Physics Phenomenology of the University of Durham where part of the 
manuscript was written. G.~P. is indebted to Dima Bardin for the invitation
to the International School-Seminar CALC2006, Dubna, 15-25 July 2006, where
parts of this work were presented. 
\clearpage
\appendix
\section{Tables}
\vspace{1.cm}
\begin{table}[ht]\centering
\setlength{\arraycolsep}{\tabcolsep}
\renewcommand\arraystretch{1.2}
\begin{tabular}{|l|l|l|l|}
\hline 
&&& \\
$\mh^{\ssO\ssS}\,$ [GeV] & $150$ & $300$ & $500$ \\
&&& \\
\hline
&&& \\
$\frac{\gf\,\rpw}{2\,\pi^2}\,
\frac{\delta^{(2)}_{\ssG}}{\delta^{(1)}_{\ssG}}$ &
$18.29\,\%$ & $8.89\,\%$ & $-24.62\%$ \\
&&& \\
&&& \\
\hline
\end{tabular}
\caption[]{The ultraviolet, infrared finite remainder for $\gf$ of
\eqn{defDG} for different values of the on-shell Higgs mass. Here $\rpw$ is
the real part of the $W$ boson complex pole, see \eqn{cpi}.}
\label{nideltag}
\end{table}
\vspace{2.cm}
\begin{table}[ht]\centering
\setlength{\arraycolsep}{\tabcolsep}
\renewcommand\arraystretch{1.2}
\begin{tabular}{|l|l|l|l|l|l|}
\hline 
\hline 
$m_t = 174.3\,\GeV$ & $\mh = 150\,\GeV$ &&&& \\
\hline 
{$\sqrt{s}\,\,[\GeV]$} & $\mz$ & $120$ & $160$ & $200$ & $500$ \\
\hline 
one-loop & $128.104$ & $127.974$ & $127.839$ & $127.734$ & $127.305$ \\
two-loop & $128.040$ & $127.967$ & $127.891$ & $127.831$ & $127.586$ \\
$\%$     & $-0.05$   & $-0.01$   & $+0.04$   & $+0.08$   & $+0.22$    \\
\hline 
$m_t = 179.3\,\GeV$ & $\mh = 150\,\GeV$ &&&& \\
\hline 
one-loop & $128.113$ & $127.982$ & $127.847$ & $127.742$ & $127.313$ \\
two-loop & $128.048$ & $127.980$ & $127.911$ & $127.857$ & $127.636$ \\
$\%$     & $-0.05$   & $-$       & $+0.05$   & $+0.09$   & $+0.25$    \\
\hline 
$m_t = 174.3\,\GeV$ & $\mh = 300\,\GeV$ &&&& \\
\hline 
{$\sqrt{s}\,\,[\GeV]$} & $\mz$ & $120$ & $160$ & $200$ & $500$ \\
\hline 
one-loop & $128.104$ & $127.974$ & $127.839$ & $127.734$ & $127.305$ \\
two-loop & $128.046$ & $127.921$ & $127.790$ & $127.689$ & $127.272$ \\
$\%$     & $-0.05$   & $-0.04$   & $-0.04$   & $-0.04$   & $+0.03$    \\
\hline 
\hline
\end{tabular}
\caption[]{$\alpha_{\MSB}(s)$ for different values of $\mt, \mh$ and
$\mw = 80.380\,$Gev, $\mz = 91.1875\,$GeV; $\alpha_s(\mz) = 0.11$. The
percentage effect of two-loop corrections is shown.}
\label{MSBtable}
\end{table}
\begin{table}[ht]\centering
\setlength{\arraycolsep}{\tabcolsep}
\renewcommand\arraystretch{1.2}
\begin{tabular}{|l|l|}
\hline 
$\Delta\alpha$ & value at $\sqrt{s} = 200\,$GeV \\
\hline
$2L\;\Reb\,$EW    &  $-0.00645(1)$ \\
$2L\;\Imb\,$EW    &  $+0.00075(1)$ \\
$2L\;\Reb\,$p-QCD &  $+0.0005512(5)$ \\
$2L\;\Imb\,$p-QCD &  $+0.0001174(5)$ \\
fin ren           &  $-0.0000977 - 0.0000998\,i$ \\
$1L\;\Reb\,$EW    &  $-0.06953$ \\ 
$1L\;\Imb\,$EW    &  $+0.00334$ \\ 
non-pert          &  $-0.03266$      \\
\hline
Full $\Reb\,\alpha(s)$    &   $0.0078934(1)$ \\
\hline
$1L + {\rm non-pert}\;\Reb\,\alpha^{-1}(s)$    &   $127.509$ \\
Full $\Reb\,\alpha^{-1}(s)$                    &   $126.688(2)$ \\
\hline 
\hline
\end{tabular}
\caption[]{$\Delta \alpha(s)$ of \eqn{defDalphas} at $\sqrt{s} = 200\,$GeV;
$\mw = 80.380\,$Gev, $\mz = 91.1875\,$GeV, $\mt = 174.3\,$GeV,
$\mh = 150\,$GeV and $\alpha_s(\mz) = 0.11$. p-QCD is perturbative QCD.
Only the error for the perturbative calculation is reported.}
\label{Fulltable}
\end{table}
\begin{table}[ht]\centering
\setlength{\arraycolsep}{\tabcolsep}
\renewcommand\arraystretch{1.2}
\begin{tabular}{|l|l|l|}
\hline 
$\sqrt{s}= 200\,\GeV$ && \\
\hline
$\mt, \mh\,$[Gev] & $1L + 2L$ & $2L/1L$ perturbative only \\
$169.3,\; 150$ &  $126.774(3)$ & $14.64\,\%$ \\
$174.3,\; 150$ &  $126.688(2)$ & $16.27\,\%$ \\
$179.3,\; 150$ &  $126.598(3)$ & $17.97\,\%$ \\
$169.3,\; 300$ &  $127.300(3)$ & $4.35\,\%$ \\
$174.3,\; 300$ &  $127.313(2)$ & $1.51\,\%$ \\
$179.3,\; 300$ &  $127.122(3)$ & $7.73\,\%$ \\
\hline
$\sqrt{s}= 500\,\GeV$ && \\
\hline
$\mt, \mh\,$[Gev] & $1L + 2L$ & $2L/1L$ perturbative only \\
$169.3,\; 150$ &  $125.430(2)$  & $29.97\,\%$ \\
$174.3,\; 150$ &  $125.259(2)$  & $33.92\,\%$ \\
$179.3,\; 150$ &  $125.070(2)$  & $38.29\,\%$ \\
$169.3,\; 300$ &  $127.121(2)$  & $4.19\,\%$ \\
$174.3,\; 300$ &  $126.960(2)$  & $0.57\,\%$ \\
$179.3,\; 300$ &  $126.781(2)$  & $3.44\,\%$ \\
\hline
\hline 
\end{tabular}
\caption[]{$\alpha(s)$ at $\sqrt{s} = 200(500)\,$GeV;
$\mw = 80.380\,$Gev, $\mz = 91.1875\,$GeV, $\alpha_s(\mz) = 0.11$ and
different values of $\mt, \mh$. The ratio between two-loop and one-loop 
perturbative corrections is shown. Only the error for the perturbative 
calculation is reported.}
\label{Compatable}
\end{table}
\begin{table}[ht]\centering
\setlength{\arraycolsep}{\tabcolsep}
\renewcommand\arraystretch{1.2}
\begin{tabular}{|l|l|l|l|l|l|}
\hline 
&&&&& \\
$\sqrt{s}\,$[GeV] & $100(\sqrt{s_0})$ & $200$ & $300$ & $500$ & $1000$ \\
&&&&& \\
\hline
&&&&& \\
$\Reb\,\alpha^{-1}(s)/\Reb\,\alpha^{-1}(s_0)$ & $1$ & $0.984$ & $0.981$
& $0.972$ & $0.957$ \\ 
&&&&& \\
\hline
&&&&& \\
$\alpha^{-1}_{\MSB}(s)/\alpha^{-1}_{\MSB}(\mz)$ & $1$ & $0.999$ & $0.998$
& $0.997$ & $0.995$ \\
&&&&& \\
\hline
&&&&& \\
$\Reb\,\alpha^{-1}(s)/\alpha^{-1}_{\MSB}(s)$ & $1.006$ & $0.991$ & $0.989$
& $0.982$ & $0.967$ \\
&&&&& \\
\hline
\end{tabular}
\caption[]{Comparing the running of $alpha$ with the running of
$\alpha_{\MSB}$ for $\mt= 174.3\,$GeV and $\mh= 150\,$GeV.}
\label{Rattablea}
\end{table}
\begin{table}[ht]\centering
\setlength{\arraycolsep}{\tabcolsep}
\renewcommand\arraystretch{1.2}
\begin{tabular}{|l|l|l|l|l|l|}
\hline 
&&&&& \\
$\sqrt{s}\,$[GeV] & $100(\sqrt{s_0})$ & $200$ & $300$ & $500$ & $1000$ \\
&&&&& \\
\hline
&&&&& \\
$\Reb\,\alpha^{-1}(s)/\Reb\,\alpha^{-1}(s_0)$ & $1$ & $0.987$ & $0.988$ 
& $0.985$ & $0.977$ \\ 
&&&&& \\
\hline
&&&&& \\
$\alpha^{-1}_{\MSB}(s)/\alpha^{-1}_{\MSB}(\mz)$ & $1$ & $0.998$ & $0.996$
& $0.994$ & $0.990$ \\
&&&&& \\
\hline
&&&&& \\
$\Reb\,\alpha^{-1}(s)/\alpha^{-1}_{\MSB}(s)$ & $1.006$ & $0.996$ & $0.999$
& $0.998$ & $0.992$ \\
&&&&& \\
\hline
\end{tabular}
\caption[]{Comparing the running of $alpha$ with the running of
$\alpha_{\MSB}$ for $\mt= 174.3\,$GeV and $\mh= 300\,$GeV.}
\label{Rattableb}
\end{table}
\clearpage
\begin{table}[ht]\centering
\setlength{\arraycolsep}{\tabcolsep}
\renewcommand\arraystretch{1.2}
\begin{tabular}{|l|l|l|l|l|}
\hline 
&&&& \\
$\sqrt{s}\,/\,\mh\,$[GeV] & $150$ & $300$ & $400$ & $450$  \\
&&&& \\
\hline
$100$  & $128.806(2)\;(2.79\%)$  &  $128.831(3)\;(3.03\%)$ 
       & $128.382(8)\;(9.55\%)$  &  $127.893(13)\;(24.96\%)$  \\
\hline
$200$  & $126.688(2)\;(16.27\%)$  & $127.213(3)\;(5.99\%)$
       & $127.910(4)\;(6.85\%)$   & $128.285(6)\;(12.34\%)$ \\
&&&& \\
\hline
\end{tabular}
\caption[]{$\alpha(s)$ for $\mt= 174.3\,$GeV increasing values of $\mh$.
The second entry gives the ratio between the perturbative two-loop contribution
and the perturbative one-loop.}
\label{mhbe}
\end{table}
\vspace{2.cm}
\begin{table}[ht]\centering
\setlength{\arraycolsep}{\tabcolsep}
\renewcommand\arraystretch{1.2}
\begin{tabular}{|l|l|l|l|}
\hline 
&&& \\
$\mu_{\ssH}$ & $\gamma_{\ssH}$ & $M_{\ssH}$ & $\Gamma_{\ssH}$ \\
&&& \\
\hline
$300$ & $4$ & $299.96$ & $8.374$ \\
\hline
$300$ & $12$ & $299.87$ & $8.376$ \\
\hline
$500$ & $40$ & $500.17$ & $63.37$ \\
\hline
$500$ & $80$ & $500.42$ & $63.34$ \\
\hline
\end{tabular}
\caption[] {Values for $\cph$ derived from $\Cph$, \eqn{effects}.}
\label{though}
\end{table}
\vspace{1.5cm}
\begin{table}[ht]\centering
\setlength{\arraycolsep}{\tabcolsep}
\renewcommand\arraystretch{1.2}
\begin{tabular}{|l|l|l|l|l|l|}
\hline 
&&&&& \\
$\mh^{\ssO\ssS}\,$ [GeV] & $150$ & $200$ & $250$ & $300$ & $350$ \\ 
&&&&& \\
\hline
$b_1\,X\,(\%)$ & $+3.31$ & $+0.13$ & $-2.30$ & $-4.84$ & $-7.85$ \\
\hline
$b_1$ & $+12.28$ & $+0.47$ & $-8.51$ & $-17.95$ & $-29.07$ \\
\hline
$b_2\,X$ & $+0.25$ & $-1.31$ & $-1.38$ & $-2.58$ & $-9.26$ \\
\hline
&&&&& \\
$\frac{b_2}{b_1}\,X\,(\%)$ & $+2.06$ & $-277.31$ & $+16.16$ & 
$+14.35$ & $+31.85$ \\
&&&&& \\
\hline
\hline
\end{tabular}
\caption[]{Numerical solution for the renormalized $SU(2)$ coupling
constant, \eqn{numren} corresponding to $\mw = 80.380\,$Gev, 
$\mz = 91.1875\,$GeV, $\mt = 174.3\,$GeV.}
\label{letut}
\end{table}
\clearpage

\end{document}